%% file: ms.tex
\documentclass[apj]{emulateapj}
\usepackage{side}
\usepackage[rotateright]{rotating}

\newcommand{\bea}{\begin{eqnarray}}
\newcommand{\eea}{\end{eqnarray}}

\newcommand{\Lx}{L_{\rm X}}
\newcommand{\Tx}{T_{\rm X}}
\newcommand{\Mx}{M_{\rm X}}
\newcommand{\Rx}{R_{\rm X}}
\newcommand{\rmag}{\>^{0.1}{\rm M}_r-5\log h}
\newcommand{\mpch}{\>h^{-1}{\rm {Mpc}}}
\newcommand{\msunh}{\>h^{-1}\rm M_\odot}
\newcommand{\msunhh}{\>h^{-2}\rm M_\odot}
\newcommand{\kms}{\>{\rm km}\,{\rm s}^{-1}}
\newcommand{\etal}{{et al.~}}

\shorttitle{Cross-identification between X-ray and optical clusters}
\shortauthors{Wang et al.}

\begin{document}


\title{Cross identification between X-ray and Optical Clusters of Galaxies in
  the SDSS DR7 Field}

\author{Lei Wang\altaffilmark{1}, Xiaohu Yang \altaffilmark{1}, Wentao
  Luo\altaffilmark{1}, Erwin T. Lau\altaffilmark{1,2}, Yu
  Wang\altaffilmark{3}, H.J. Mo\altaffilmark{4}, Frank C. van den
  Bosch\altaffilmark{5}, Q.D. ~Wang\altaffilmark{4} }

\altaffiltext{1}{Key Laboratory for Research in Galaxies and Cosmology,
  Shanghai Astronomical Observatory,  Nandan Road 80,
  Shanghai 200030, China; E-mail: leiwang@shao.ac.cn}

\altaffiltext{2} {Department of Physics, Yale University, New
      Haven, CT 06520-8120, USA}

\altaffiltext{3}{Key Laboratory for Research in Galaxies and Cosmology, Center
  for Astrophysics, University of Science and Technology of China, Hefei,
  230026, China}

\altaffiltext{4}{Department of Astronomy, University of Massachusetts, Amherst
  MA 01003-9305}

\altaffiltext{5} {Department of Astronomy, Yale University, P.O. Box 208101, New
      Haven, CT 06520-8101, USA}


\begin{abstract} 
  We use the ROSAT all sky  survey X-ray cluster catalogs and the optical SDSS
  DR7 galaxy  and group catalogs  to cross-identify X-ray clusters  with their
  optical counterparts, resulting in a sample of 201 X-ray clusters in the sky
  coverage  of SDSS  DR7.   We investigate  various  correlations between  the
  optical  and X-ray properties  of these  X-ray clusters,  and find  that the
  following optical  properties are correlated with the  X-ray luminosity: the
  central galaxy luminosity, the central galaxy mass, the characteristic group
  luminosity  ($\propto   \Lx^{0.43}$),  the  group   stellar  mass  ($\propto
  \Lx^{0.46}$),  with  typical 1-$\sigma$  scatter  of  $\sim  0.67$ in  $\log
  \Lx$. Using the observed number distribution of X-ray clusters, we obtain an
  unbiased scaling  relation between the X-ray luminosity,  the central galaxy
  stellar mass and the characteristic  satellite stellar mass as ${\log L_X} =
  -0.26 + 2.90 [\log (M_{\ast, c} + 0.26 M_{\rm sat}) -12.0]$ (and in terms of
  luminosities, as ${\log L_X} = -0.15 + 2.38 [\log (L_{c} + 0.72 L_{\rm sat})
    -12.0]$). We  find that the  systematic difference between  different halo
  mass estimations,  e.g., using the  ranking of characteristic  group stellar
  mass or using the X-ray luminosity scaling relation can be used to constrain
  cosmology.  Comparing the  properties of groups of similar  stellar mass (or
  optical   luminosities)   and   redshift   that  are   X-ray   luminous   or
  under-luminous, we find that X-ray luminous groups have more faint satellite
  galaxies and higher red  fraction in their satellites.  The cross-identified
  X-ray  clusters  together with  their  optical  properties  are provided  in
  Appendix B.
\end{abstract}





\section{Introduction}


Clusters  of  galaxies  are  the   most  massive  virialized  objects  in  the
universe. Their  abundance and spatial distribution  are powerful cosmological
probes (e.g.,  Majumdar \& Mohr  2004; Vikhlinin et  al.  2009b; Mantz  et al.
2010a). In addition, galaxy clusters provide extreme environments for studying
the  formation  and  evolution  of   galaxies  within  the  framework  of  the
hierarchical build-up  of the most  massive halos.  One important  property of
clusters is that both their stellar and gas components are readily observable:
their gravitational wells are deep  enough to retain any energetic gas ejected
from  their  member  galaxies  which  can  be  observed  in  the  optical  and
infrared. The intracluster  medium (ICM) are also hot  enough to be observable
in X-ray.   The observed thermodynamic state  of the ICM is  determined by the
combined  effects  of  shock  heating  during  accretion,  radiative  cooling,
feedback  from stellar  evolution (stellar  winds and  supernovae)  and active
galactic nuclei.   The density, temperature,  and entropy profiles of  the ICM
therefore carry important information  regarding the entire thermal history of
cluster  formation.   The  hot  ICM,  with temperatures  between  $10^7$K  and
$10^8$K, emits  X-rays in the form  of thermal bremsstrahlung  and atomic line
emissions  (e.g., Kellogg  \etal 1971;  Forman \etal  1971).  Since  the X-ray
emission is proportional  to the gas density squared,  X-ray selected clusters
are  more suitable than  optically-selected clusters  for mapping  the spatial
distribution of clusters as they  suffer less from projection effects (Ebeling
\etal  1998;  Jones \&  Forman  1999).   By  assuming hydrostatic  equilibrium
between the  intracluster gas and the  cluster potential, one  can also derive
the  gravitational   mass  of  the  cluster  using   density  and  temperature
measurements provided by X-ray data.

Clusters have also been observed by other means in addition to X-ray: optical,
infrared, radio,  Sunyaev-Zel'dovich effect and  gravitational lensing.  Among
these, the most complete cluster  samples to date are optically-selected (e.g.
Abell et al.  1989; Zwicky et  al.  1961-68) and X-ray selected (e.g.  Ebeling
et al.   1998; B\"ohringer et al.   2000). To synthesize the  benefits of both
X-ray and optical  observations of galaxy systems, it is  useful to relate the
X-ray  systems to optically  selected groups  and clusters.   Numerous studies
have cross-identified optical groups or  clusters with X-ray clusters, or vice
versa, in order  to compare their optical and  X-ray properties (e.g., Bahcall
1977; Edge \&  Stewart 1991; Donahue \etal 2001, 2002;  Yee \& Ellingson 2003;
Mulchaey  \etal 2003;  Gilbank \etal  2004).  Most of  these earlier  studies,
however,  were severely hampered  by the  lack of  large samples  with uniform
observations in  both passbands. The situation improved  dramatically with the
completion of  a number  of large  surveys.  In recent  years, with  the great
advance in  optical surveys, especially with  the advent of  the Sloan Digital
Sky Survey  (SDSS), more  and more  effort has been  made to  characterize the
X-ray properties  of optically selected  clusters (e.g., Mulchaey  \etal 2003;
Rykoff \etal 2008a,b; Rozo \etal  2009a,b; Hansen \etal 2009; Hao \etal 2010).
In particular, Popesso  \etal (2004) cross correlated the  X-ray clusters from
the ROSAT All-Sky  Survey (RASS; Voges \etal 1999) with  optical data from the
SDSS data  release 1 (DR1),  resulting in a  sample of 114 clusters  with both
X-ray and optical data.  

Although  X-ray selection  is  arguably  the most  reliable  method to  select
clusters, X-ray selection typically has  a low efficiency. 
In fact, a  significant fraction of optically detected clusters
falls on  the general  scaling relation between  optical luminosity  and virial
mass  (inferred from,  for  example,  the velocity  dispersion  of the  member
galaxies), but is  undetected in the X-ray (i.e., does  not follow the scaling
relation between  X-ray luminosity and virial  mass).  This has  given rise to
the notion that  there exists a genuine population of  clusters that are X-ray
under-luminous (e.g.,  Castander \etal 1994;  Lubin \etal 2004;  Popesso \etal
2007; Castellano et  al. 2011;  Balogh et al.  2011).  In  addition to
simply cross-correlating optical and X-ray catalogs, one can also use stacking
techniques.   Dai, Kochanek \&  Morgan (2007)  used a  NIR selected  sample of
$\sim 4000$  nearby ($\langle z  \rangle \sim 0.02$) galaxy  clusters selected
from the  Two Micron All Sky  Survey (2MASS) using a  matched filter algorithm
(Kochanek \etal  2003), and  probed their X-ray  properties by  stacking X-ray
data from  the RASS. A similar  approach was taken by  Rykoff \etal (2008a,b),
who used  as their input catalog  the large maxBCG sample  of ~14,000 clusters
($\langle  z \rangle  \simeq 0.23$)  selected from  the photometric  SDSS data
(Koester  \etal 2007a,b).   Although these  stacking techniques  are extremely
powerful for  determining the {\it average} scaling  relations between optical
and  X-ray properties,  they contain  little to  no information  regarding the
corresponding scatter.

Note  that  the  optical  clusters  so  far  are  mainly  extracted  from  the
photometric  data.  These photometric  samples  are  quite  complete for  most
massive  clusters, (which  have the  most constraining  power  as cosmological
probes),  and they are  much deeper  than those  based on  spectroscopic data.
However, their galaxy members are not well constrained.  In this paper, we use
the SDSS DR7  group catalogs of Yang \etal (2007),  which are constructed from
the SDSS spectroscopic data (Abazajian \etal 2009).  These catalogs provide us
with galaxy groups  that have reliable galaxy memberships  which are important
in  probing  the halo  occupation  distribution  (HOD)  statistics and  galaxy
formation models. (e.g. Yang et al.  2008; 2009).  The SDSS DR7 group catalogs
also  span a  large halo  mass  range, from  rich clusters  to isolated  faint
galaxies,  allowing  us  to  investigate  the X-ray  luminosity  and  hot  gas
distribution not only in massive clusters but also in relatively small halos.

As the first  paper in a series, we focus  on the cross-identification between
the optical galaxy groups with existing X-ray cluster catalogs, e.g. the ROSAT
X-ray  clusters from the  NASA/IPAC Extragalactic  Database (NED;  see Section
\ref{sec:X-ray}   for  their   original  references).    Some  straightforward
comparisons between  the optical  and X-ray properties  of these  clusters are
investigated.  We  will  address  the  more  specific  probes  of  the  galaxy
properties in the  X-ray clusters and the X-ray  properties around the optical
groups in forthcoming papers.

This paper  is organized  as follows.  In  Section \ref{sec_data},  we briefly
describe  the group  samples  of Yang  et al.   (2007)  for SDSS  DR7 and  our
extraction  and   treatment  of  the   X-ray  cluster  samples.    In  section
\ref{sec:match}, we  present the selection  criteria for matching  groups with
X-ray clusters.  The correlation between  the X-ray and optical properties are
investigated in section \ref{sec_correlation}.   The properties of groups with
and  without strong X-ray  emissions are  compared in  Section \ref{sec_diff}.
Finally,  we present  our conclusions  in section  \ref{sec_summ}.  Throughout
this paper, we use the  $\Lambda$CDM cosmology whose parameters are consistent
with  the 7-year  data release  of the  WMAP mission:  $\Omega_{\rm m}=0.275$,
$\Omega_{\rm  \Lambda}=0.725$,  $h=0.702$,  and  $\sigma_8=0.816$,  where  the
reduced  Hubble constant,  $h$,  is  defined through  the  Hubble constant  as
$H_0=100h~{\rm km~s^{-1}~Mpc^{-1}}$ (Komatsu et al.  2011).


\section{DATA}
\label{sec_data}


\subsection{The SDSS DR7 Galaxy and Group catalogs}
\label{sec:group}

The optical  data used  in our analysis  is taken  from the SDSS  galaxy group
catalogs of Yang  \etal (2007; hereafter Y07), constructed  using the adaptive
halo-based group finder of Yang \etal  (2005a), here updated to Data Release 7
(DR7).   The related  galaxy catalog  is the  New York  University Value-Added
Galaxy catalog  (NYU-VAGC; Blanton \etal  2005b) based on SDSS  DR7 (Abazajian
\etal  2009), which  contains  an independent  set  of significantly  improved
reductions. DR7 marks the completion of  the survey phase known as SDSS-II. It
features a spectroscopic  sample that is now complete  over a large contiguous
area  of the  Northern Galactic  cap,  closing the  gap which  was present  in
previous data releases.  From the NYU-VAGC, we select all galaxies in the Main
Galaxy Sample  with an  extinction-corrected apparent magnitude  brighter than
$r=17.72$, with  redshifts in  the range $0.01  \leq z  \leq 0.20$ and  with a
redshift completeness ${\cal  C}_z > 0.7$.  The resulting  SDSS galaxy catalog
contains a  total of $639,359$  galaxies, with a  sky coverage of  7748 square
degrees. Note  that a  very small  fraction of galaxies  in this  catalog have
redshifts taken from the Korea Institute for Advanced Study (KIAS) Value-Added
Galaxy  Catalog  (VAGC) (e.g.   Choi  \etal  2010)\footnote{These were  kindly
  provided to us by Yun-Young Choi and Changbom Park.}.

Following  Y07, three  group samples  are constructed  from  the corresponding
galaxy samples: Sample I, which only uses the $599,301$ galaxies with measured
$r$-band magnitudes and redshifts from  the SDSS; Sample II, which includes in
addition  $3269$ galaxies  with SDSS  $r$-band magnitudes  but  with redshifts
taken from alternative  surveys; and Sample III, which  includes an additional
$36,759$  galaxies  that  do  not  have redshift  measurements  due  to  fiber
collisions,  but  are  assigned  the  redshifts of  their  nearest  neighbors.
Although the  fiber-collision correction works  well in roughly 60  percent of
the cases,  the assigned  redshifts of  the remaining 40  percent can  be very
different from their true values (Zehavi \etal 2002).  In this study, in order
not to miss  any potential group members for  cross-identification, we use the
group catalogs  of Sample III.  For  completeness, two sets  of group catalogs
were  constructed:  one  in which  we  use  the  Petrosian magnitudes  of  the
galaxies, and the other in which  we use the model magnitudes (Yang \etal 2012
in  preparation).  In  total there  are  $474,085$ groups  based on  Petrosian
magnitudes and $472,673$ groups based  on model magnitudes. Among these groups
about $23,700$  have three member galaxies or  more. In this paper  we use the
group catalog  based on  the model magnitudes.   We have tested,  though, that
using the group catalog based on  the Petrosian magnitudes does not affect any
of our results in any significant way.

Following Y07,  for each group in  the catalog, we  estimate the corresponding
halo mass using the ranking of its characteristic stellar mass, defined as the
total stellar  mass of all  group members with  $\rmag \leq -19.5$.   Here the
halo mass  function obtained by Tinker  et al. (2008) for  WMAP7 cosmology and
$\Delta=200$ is  used in our calculation,  where $\Delta$ is  the average mass
density  contrast in  the spherical  halo.  We  indicate the  group  mass thus
obtained by $M_{\rm G}$, where the letter  'G' is used to indicate that it has
been obtained  from the optical group  catalog. Note that  groups whose member
galaxies are all  fainter than $\rmag = -19.5$ cannot be  assigned a halo mass
with this method.  For these systems,  one could in principle use the relation
between halo mass and the stellar  mass of the central galaxy obtained by Yang
\etal (2011) to  estimate their halo masses. However, since  our main focus is
on the cross  identification of X-ray clusters with  optical groups, which are
in general quite massive, we do not require halo masses for these groups.

The halo  masses $M_{\rm  G}$ thus  assigned to the  groups are  calibrated to
correspond to  $M_{200}$, the  mass of the  halo defined  so that it  has an
average overdensity of 200. Along similar lines, we define the `virial radius'
of the group as $r_{200}$, which is given by 
\begin{equation}\label{eq:r200}
r_{\rm 200}=\left[\frac{M_{200}}{\frac{4\pi}{3}\times200\Omega_{\rm m}\times
\frac{3H_0^2}{8\pi G}} \right]^{1/3} (1+z_{\rm G})^{-1}\,,
\end{equation}
where $z_{\rm G}$ is the redshift  of the group (i.e., the average redshift of
the group  members).  Tests  with detailed mock  galaxy redshift  surveys have
shown that  the statistical error on  $M_{\rm G}$ is  of the order of  0.3 dex
(see Y07 for details).

\subsection{The X-ray Cluster Catalogs}
\label{sec:X-ray}

\begin{figure*}
\plotone{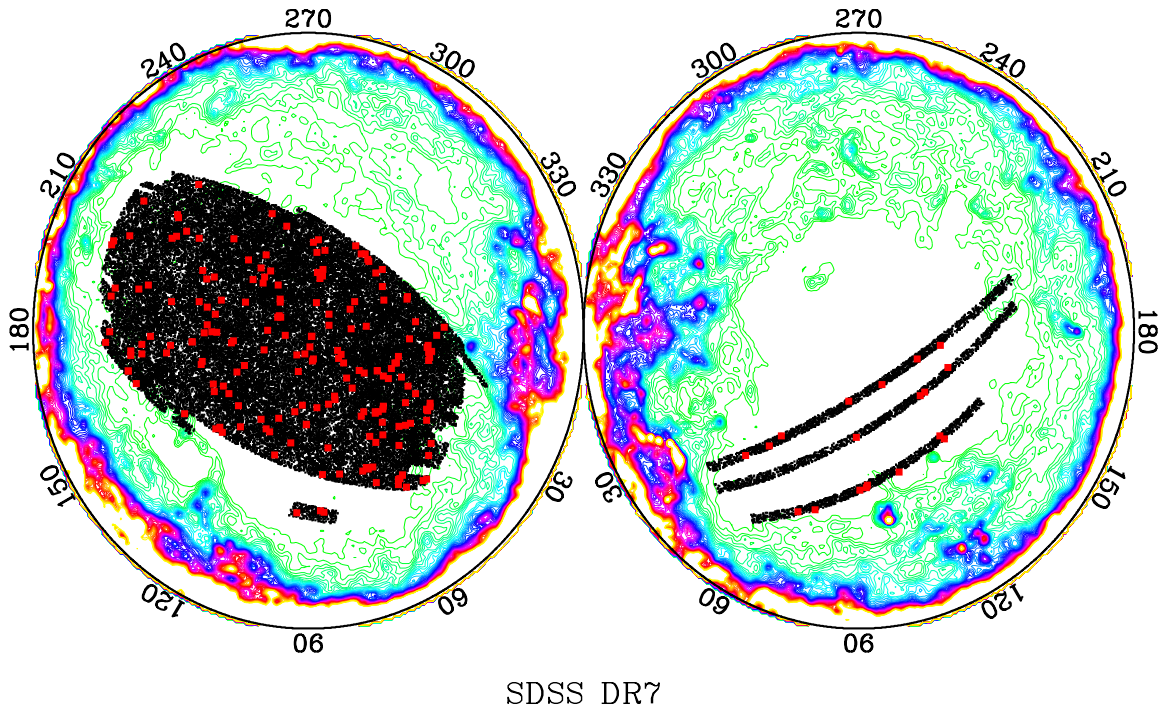}
\caption{The distribution  of X-ray clusters (squares) that  coincide with the
  sky coverage of the SDSS DR7 galaxies (black area), overlaid on the galactic
  extinction contours of Schlegel, Finkbeiner \& Davis (1998).}
\label{fig:skycover}
\end{figure*}

The main aim of this paper  is to cross-identify the optically selected groups
and  clusters (described  above) with  X-ray selected  cluster samples  and to
study the correlations between X-ray and optical properties.  
For this purpose,  we use the ROSAT
catalogs at  the broad  band $0.1$-$2.4$  keV as our  primary input  sample of
X-ray clusters. In particular, we combine the following ROSAT cluster samples:
the  ROSAT  Brightest  Cluster  Sample  (BCS) and  their  low-flux  extensions
compiled by Ebeling \etal (1998, 2000), and the Northern ROSAT All-Sky (NORAS)
and  ROSAT-ESO Flux  Limited X-ray  (REFLEX) samples  compiled  by B\"ohringer
\etal (2000, 2004).  Within these catalogs, the BCS (Ebeling \etal 1998) has a
flux limit  $F_X \ge  4.4 \times 10^{-12}  {\rm erg~cm^{-2}~s^{-1}}$  and flux
completeness  $f_X\simeq 90$  percent in  the northern  hemisphere ($\delta\ge
0^{\circ}$),  at high Galactic  latitudes ($|b|\ge20^{\circ}$).   Its low-flux
extension   (Ebeling  \etal  2000)   has  flux   limits  $2.8\times10^{-12}\le
F_X\le4.4\times10^{-12}{\rm  erg~cm^{-2}~s^{-1}}$ and  completeness $f_X\simeq
75$  percent. The NORAS  cluster sample  (B\"ohringer \etal  2000) has  a flux
limit  $F_X  \sim  1.0  \times  10^{-12} {\rm  erg~cm^{-2}~s^{-1}}$  and  flux
completeness $f_X\simeq 50$ percent  with respect to REFLEX\footnote{Note that
  this  value is  estimated in  the  ${\rm 9^h-14^h}$  region.} at  $\delta\ge
0^{\circ}$   and   $|b|\ge20^{\circ}$.   And   finally,   the  REFLEX   sample
(B\"ohringer \etal  2004), which covers  4.24 steradians in the  southern sky,
has   flux   limits   $F_X\ge3.0\times10^{-12}{\rm  erg~cm^{-2}~s^{-1}}$   and
completeness $f_X\ge 90$ percent.

These four catalogs  combined contain a total of 1138  unique sources (see the
NED  website\footnote{http://nedwww.ipac.caltech.edu/}),  with information  on
the rest-frame X-ray luminosity (K-correction applied; e.g.  B\"ohringer \etal
2004) and gas temperature (mostly  estimated from the X-ray luminosity) listed
for each  of the sources.   However, 213 entries  in this raw,  combined X-ray
cluster sample are duplicates, implying  a total sample of 925 unique sources,
many of which have already been cross-identified with Abell or Zwicky clusters
in NED.   For duplicate clusters,  we take the  ones with the  most up-to-date
information  for their  characteristic quantities.  Throughout, we  use $\Lx$,
$\Tx$, $\Mx$,  $\Rx$, to  denote the X-ray  luminosity, gas  temperature, halo
mass and halo radius for each X-ray cluster, where $\Mx$ and $\Rx$ are defined
later  in Eq.  \ref{eq:convert}.   These  quantities are  quoted  in units  of
$10^{44} {\rm erg~s^{-1}}$, ${\rm keV}$, $\msunh$ and $\mpch$, respectively.

As a  first step of  our cross identification  we remove those  X-ray clusters
that are located  outside the sky area and redshift range  covered by the SDSS
DR7. For  each X-ray  cluster we adopt  the right ascension,  declination, and
redshift, $z$, recommended by the NED website, unless the information provided
by NED  is incomplete,  in which  case we use  the data  from the  most recent
documentation. Only  clusters with  $0.01 \leq  z \leq 0.20$  and with  a SDSS
redshift completeness of ${\cal C}_z >  0.7$ at the cluster's center are kept.
This results in  a sample of 217 unique X-ray clusters.   As a final selection
criterion,  we  follow   Y07  and  remove  all  X-ray   clusters  that  suffer
significantly from  an edge  effect (i.e.,  are located close  to one  or more
boundaries of  the SDSS survey  volume), which leaves  a sample of  207 unique
sources.

Throughout this paper, if not  specified otherwise, we use the subscripts `g',
`G' and  `X' to  refer to  quantities for galaxies,  optical groups  and X-ray
clusters, respectively.

\section{Matching Optical Groups with X-ray Clusters via Central Galaxies}
\label{sec:match}


Since we have only 207 X-ray  clusters, we decide to use simple eyeball checks
to  make  the cross-identification  with  optical  groups.   Our criterion  to
cross-identify X-ray  clusters with  optical groups is  based on  their common
central galaxies.

To find the central galaxies for our sample of 207 X-ray clusters, we make use
of the SDSS  skyserver to extract an optical image  around each X-ray cluster.
In each  of these images we first  find the brightest galaxy  according to its
apparent $r$-band magnitude provided by the skyserver within a 7 arcmin radius
from the  center of the  X-ray cluster \footnote{Since different  X-ray source
  characterization techniques were used by  Ebeling et al.  (1996, 1998, 2000;
  Voronoi  tessellation and percolation,  VTP) and  B\"ohringer et  al. (2000,
  2004; Growth  curve analysis, GCA),  the position difference  from different
  references can be as large as $\sim 5$ arcmins for a given source}.  If this
galaxy  is red  [with  $^{0.1}(g-r)>0.8$] and  has  an offset  smaller than  2
arcmins from  the X-ray cluster position  obtained from Ebeling  et al. (1998,
2000) and  B\"ohringer et  al.  (2000,  2004), it is  regarded as  the central
galaxy of the cluster.  This criterion follows that of Allen et al. (1992) and
Crawford et  al. (1995,  1999) who  showed that the  bright central  galaxy is
usually found within  1-2 arcmin of the centroid of the  X-ray emission of the
cluster.   About $170$  clusters  in our  final  catalogue are  found in  this
catagory.  For all the other X-ray clusters, where the brightest galaxies have
offsets $\ga 2$ arcmins from the ROSAT X-ray cluster positions, we make use of
high-resolution X-ray  images from e.g. ROSAT/PSPC,  ROSAT/HRI, XMM-Newton and
Chandra when  available, or from  previous identifications (e.g.   Crawford et
al.   1995,  1999).  About  $20$  central  galaxies  are identified  with  the
high-resolution  X-ray  images, and  the  remaining  $\sim  10$ are  based  on
previous   identifications   (see  the   notes   on   individual  sources   in
Table~\ref{tab}).  An exception  is RXC J1554.2 +3237 for  which no bright and
red galaxy  is found within  10 arcmins from  the X-ray cluster  center.  This
cluster is  therefore removed  from our sample.   Close inspection  shows that
large  offsets mainly  come  from (i)  multiple  maxima in  the X-ray  images,
e.g. the X-ray  cluster position is located between the 2  maxima of the X-ray
emission; (ii) very extended sources for which the uncertainty in the position
of the X-ray maximum is very large; and (iii) low resolution of RASS for which
the 2 pixel offset along pixel's diagonal line is larger than 2 arcmins.

As a further check of the reliability of our central galaxy identification, we
overlay the X-ray contour of each  cluster on the optical image.  We find that
the central galaxy  is in general located close to the  X-ray flux maximum. In
fact,  for each of  the 90  clusters where  high-resolution (with  pixel sizes
about one arcsecond and position error  about a few arcseconds) X-ray data are
available from the Chandra and/or  XMM-Newton databases, we find that the peak
of  the X-ray  emission is  almost exactly  ($\le 5$  arcsec) centered  on the
`central galaxy' that we have identified using the method described above.  In
summary,  among  all the  X-ray  clusters  that  are identified  with  central
galaxies,  about  $170$  have  $<2$  arcmin offsets  from  the  X-ray  cluster
positions, about $30$ have $>2$ arcmin offsets, and only 2 have offsets larger
than 7  acrmins.  For  the last  two cases, the  X-ray distributions  are very
extended.   The offsets  between the  central galaxies  and the  X-ray cluster
positions are provided in Appendix~B.

Although NED  provides redshifts for all  X-ray clusters in  our sample, these
are not always reliable. For  example, for clusters for which no spectroscopic
redshift information is  available, Ebeling \etal (1996; 1998;  2000) assign a
redshift to  the X-ray cluster based  on the magnitude of  the tenth brightest
galaxy (cf., Abell 1958; Corwin 1974; Abell \etal 1989; Peacock \& West 1992).
Since each of the X-ray clusters in  our sample is linked to a central galaxy,
we can use the spectroscopic  SDSS galaxy catalog to obtain improved redshifts
for these  X-ray clusters.  Unfortunately, because of  fiber collisions, which
are relatively frequent for galaxies in high-density regions such as clusters,
spectroscopic  redshifts are  only  available  for $\sim  75$  percent of  the
central galaxies in our sample of X-ray clusters.  For the remaining $\sim 25$
percent the redshifts are obtained using  the redshifts of the nearest (or the
second,    third,   ...,    nearest)   galaxies    close   to    the   central
galaxies\footnote{For these central  galaxies without spectroscopic redshifts,
  the redshifts  provided in the  NYU-VAGC according to the  nearest neighbors
  may  not be  always  correct.  Since  we  have made  eyeball  check of  each
  `central' galaxy  with respect to  its neighboring galaxies, if  its nearest
  neighbor  is  an isolated  blue  galaxy, we  suspect  this  galaxy has  been
  assigned a wrong redshift. In that  case we update its redshift with that of
  the second (or  third, etc.) nearest galaxy with red colors,  and with a few
  more galaxies at the same redshift, i.e.  the most possible redshift for the
  cluster.  }.  Both the original  and updated redshift for each X-ray cluster
are  provided in  Appendix~B.  For  the updated  one we  indicated  whether it
corresponds  to  the  spectroscopic  redshift  of  the  central  galaxy  ({\tt
  ztype}$=1$) or whether  it has been estimated from  the neighboring galaxies
({\tt ztype}$=2$).  In  204 of the 206 cases the  updated redshift agrees with
the original redshift to better than  $\Delta z = 0.02$. Throughout this paper
we  use our  updated redshifts,  and all  related quantities,  such  as $\Lx$,
$\Mx$, and $\Rx$ have been updated accordingly.

Starting from the central galaxies associated with the X-ray clusters, we look
into the  SDSS DR7  group catalog for  the cross-identified galaxy  groups. We
find that  not all  the central galaxies  in the  X-ray clusters are  the most
massive galaxies  (MMG) in  their respective groups.   About 20 (10\%)  of the
X-ray  clusters have galaxies  more massive  than the  centrals and  which are
offset from the X-ray center by more than 7 arcmins. This is in agreement with
other studies;  for example, Zhang \etal  (2011), using 62  galaxy clusters in
the  HIghest   X-ray  FLUx  Galaxy  Cluster  Sample   (HIFLUGCS;  Reiprich  \&
B\"ohringer 2002), have  shown that the brightest galaxy in  the cluster has a
lognormal  offset from the  X-ray flux-weighted  center with  a mean  value of
about $10  {\rm kpc}$  and a  10-based logarithmic scatter  of 0.55  (see also
Skibba  \etal 2011  and references  therein).  In  what follows,  we  will use
subscripts $1$ and  $c$ to refer to the most massive  galaxies and the central
galaxies, respectively  (e.g., in case of  the luminosities we  will use $L_1$
and $L_c$).

As a final step of our cross-identification, we check for duplicates (clusters
that are cross-identified  with the same central galaxy)  and pairs of merging
clusters (clusters that are cross-identified  with the same optical group, but
with different central galaxies), as  outlined in Appendix~A. In both cases we
remove the  smaller of  the two clusters  from our  sample. This results  in a
final sample of  $201$ X-ray clusters with an  optical cross-identification in
our SDSS DR7 group catalog.  As an illustration, Fig.~\ref{fig:skycover} shows
the  distribution  of  these  X-ray  clusters  on  the  sky  overlaid  on  the
distribution of galaxies in the SDSS  DR7. Before we proceed with studying the
correlations of their X-ray and optical  properties, we point out that many of
the rich optical clusters/groups in  our group catalog are not associated with
any existing X-ray  cluster entries.  We will address this  issue in detail in
Section \ref{sec_diff}.

\begin{figure*}
\center{ \includegraphics[height=14.0cm,width=12.0cm,angle=270]{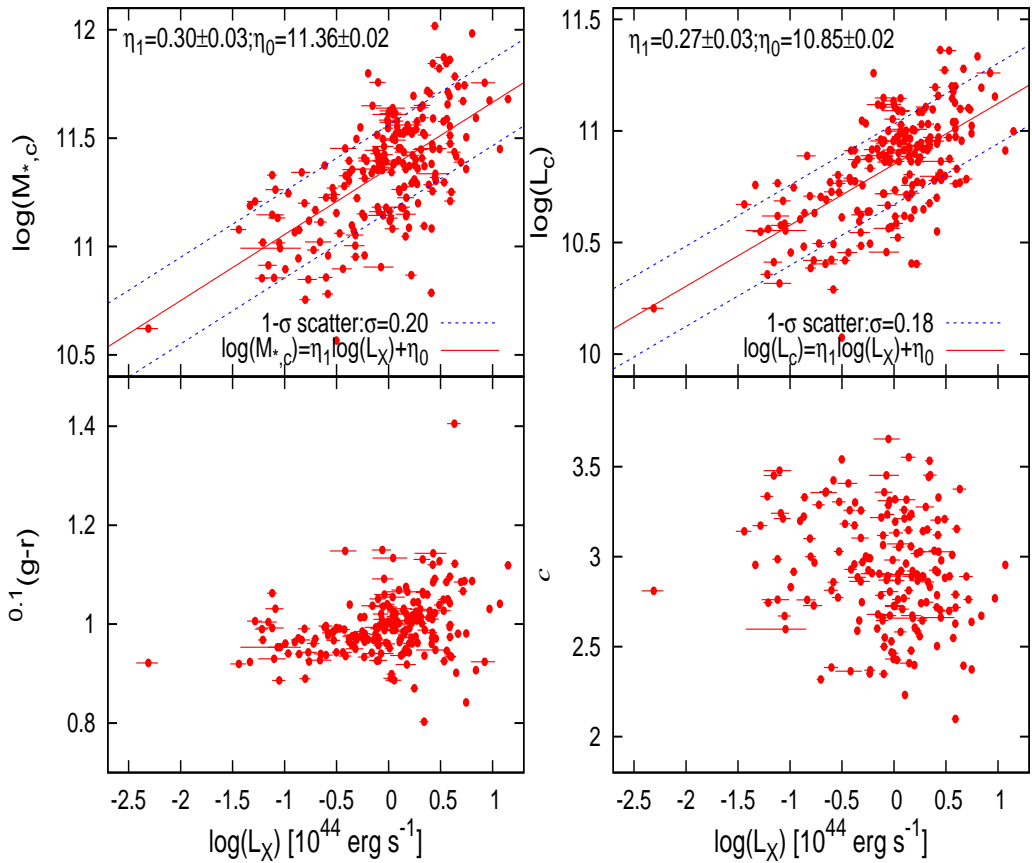}}
\caption{The distributions  of various optical properties  of central galaxies
  as  a  function  of the  X-ray  luminosity  of  their host  cluster,  $\Lx$.
  Clockwise  from the  upper left-hand  panel, the  optical properties  of the
  central  galaxies  are the  stellar  mass,  $M_{\ast,  c}$, the  luminosity,
  $L_{c}$, the $^{0.1}(g-r)$ color,  and the concentration parameter $c$.  The
  solid  and dashed  lines  in  the upper  two  panels are  the  best fit  and
  1-$\sigma$ deviations  of the  $M_{\ast, c}-\Lx$ and  $L_{c}-\Lx$ relations,
  respectively.}
\label{fig:central-Lx}
\end{figure*}

\begin{figure*}
\center{ \includegraphics[height=15.0cm,width=12.0cm,angle=270]{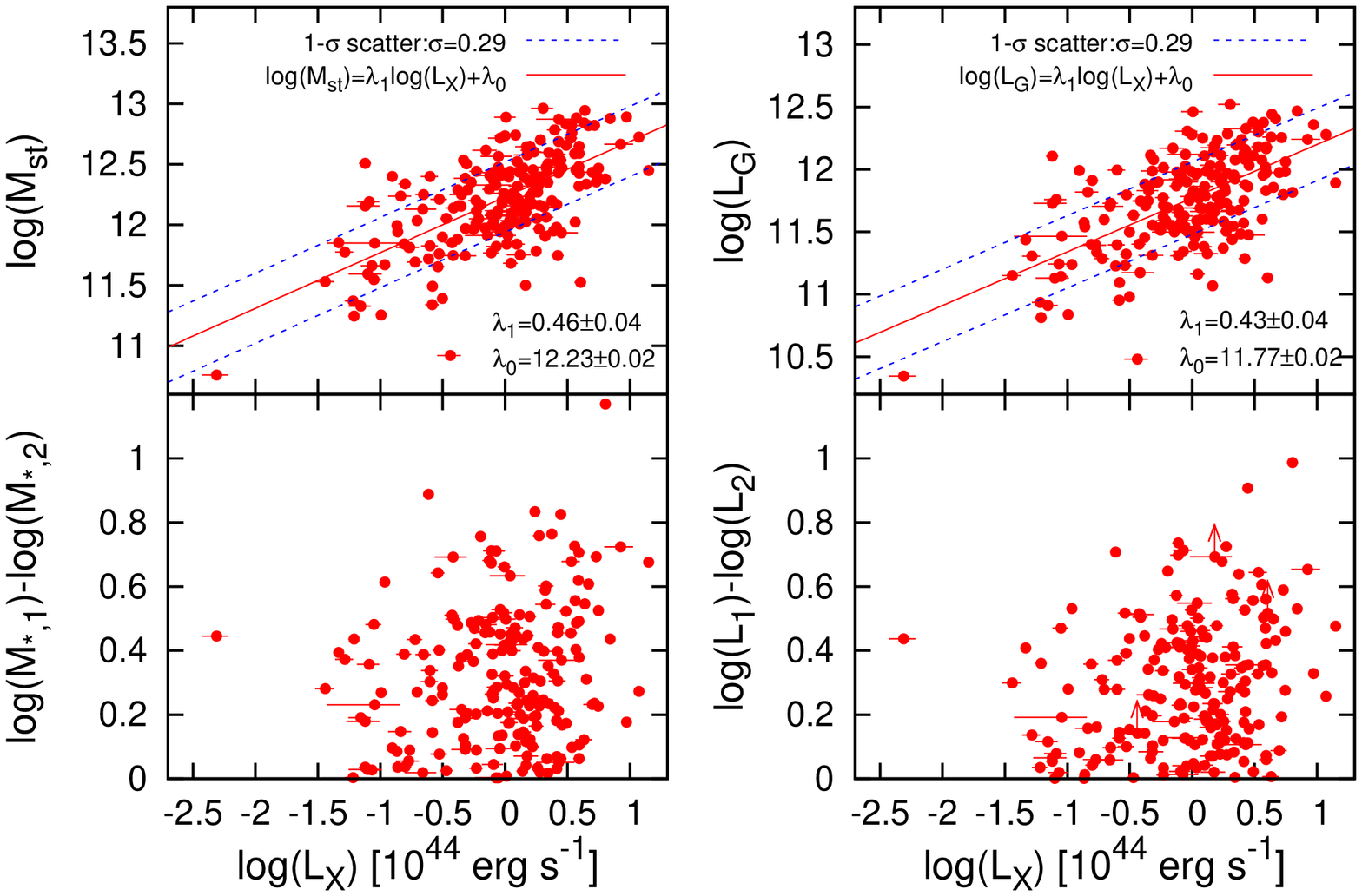}}
\caption{Upper  panels:  the distributions  of  characteristic stellar  masses
  (left) and  luminosities (right)  of groups as  a function of  X-ray cluster
  luminosity  $\log \Lx$.  Lower  panels: the  distributions of  stellar mass
  gaps  between the  first and  second most  massive galaxies  (left)  and the
  luminosity gaps between the first and second brightest galaxies (right) as a
  function of $\log \Lx$.  The solid and dashed lines in the upper row panels
  are  the best  fit and  1-$\sigma$ deviations  of the  $M_{\rm  st}-\Lx$ and
  $L_{G}-\Lx$ relations, respectively.  }
\label{fig:group-Lx}
\end{figure*}

\section{Correlation between the X-ray and optical properties} 
\label{sec_correlation}

Now that we have cross-identified the X-ray clusters with optical groups, we
proceed by examining various (possible) correlations between the X-ray and
optical properties of the X-ray clusters.

\subsection{The General Correlations}
\label{sec:general}

The  optical properties  to be  investigated  in this  subsection include  the
characteristic  luminosity  $L_G$  and   stellar  mass  $M_{\rm  st}$  of  the
group/cluster, defined as  the total luminosity and total  stellar mass of all
member  galaxies  with  $\rmag\le  -19.5$,  respectively  (see  Y07  for  more
detail). In addition, we will  also consider the following properties of their
central   galaxies:  the   $r$-band   luminosity  $L_c$,   the  stellar   mass
$M_{\ast,c}$,   the   $^{0.1}(g-r)$  color,   and   the  concentration   index
$c=r_{90}/r_{50}$\footnote{  $r_{50}$ and  $r_{90}$ are  the  radii containing
  $50\%$  and   $90\%$  of  the   Petrosian  flux  (Blanton  et   al.   2005),
  respectively.}.   We examine  if there  are any  correlations  between these
properties and the X-ray cluster luminosity $\Lx$.

We first examine the distributions, as a function of X-ray cluster luminosity,
of the stellar  mass and luminosity of the central  galaxies.  The results are
shown in the upper row  panels of Fig.~\ref{fig:central-Lx}: left panel is for
the stellar  mass and right panel for  the luminosity. 
There are clear correlations between these quantities.   
To quantify these correlations, we use a least square  
linear regression  method in  the log-space to  obtain the
regression lines.  Here we did  not take into account measurement errors
in $\Lx$ and $M_{\ast, c}$ (or $L_c$) as they are much smaller 
than the scatter among different clusters. The  same weight  
is assigned to each cluster in the fitting. The best-fit lines are
shown as  the solid lines, together with their parameters, 
in the corresponding panels. For both  relations  the 
correlation  coefficient  is about $0.63$.   As  an
illustration,  we also show  as  the  short-dashed  lines   
the  $\pm  1\sigma$ scatter of the distributions with respect to 
the best fit lines. For fixed $L_X$, the $1\sigma$ scatter 
is $\sim 0.20$ dex and $\sim 0.18$ dex in $M_{\ast, c}$ and $L_c$, 
respectively. For fixed $M_{\ast,c}$ ($L_c$), the scatter 
in $L_X$ is $\sim 0.20/0.30=0.67$ dex ($0.18/0.27=0.67$ dex).  
Despite the  relatively large scatter, there is a clear trend 
that clusters with brighter X-ray luminosities have central  
galaxies that are more massive and more luminous.  The correlation  
slope between the stellar mass (luminosity) of the central galaxies 
and the cluster X-ray luminosity is $\sim 3.5$. 
However, since the  X-ray cluster sample is flux limited,
the correlations may be affected by the Malmquist bias.  We will 
come back to this issue in Section \ref{sec:conditional}.

Next  we  check   the  distributions  of  the  $^{0.1}(g-r)$   color  and  the
concentration  index, $c$, of  the central  galaxies, again  as a  function of
X-ray cluster luminosity.  The distributions  are shown in the lower panels of
Fig.~\ref{fig:central-Lx}.  Clearly, the majority of central galaxies in X-ray
clusters   are  red   and  of   early-type  (relatively   large  concentration
parameters).  Here, we  did not see any obvious  correlation between the color
(or concentration) and the X-ray luminosity of the cluster.

Apart from those properties of central galaxies, we proceed to investigate the
properties of groups.  We show in the upper  panels of Fig.~\ref{fig:group-Lx}
the  distributions of  the characteristic  stellar mass  ($M_{\rm  st}$; upper
left-hand  panel)  and  the  characteristic  luminosity  ($L_{\rm  G}$;  upper
right-hand panel) as a function of  $\Lx$.  Similar to the case of the central
galaxy, we  see a  clear positive correlation  between the group  stellar mass
(luminosity) and its  X-ray luminosity.  Using the same  algorithm for central
galaxies, we fit  the regression lines for the groups.   The results are shown
in the upper panels of  Fig.~\ref{fig:group-Lx} as solid lines.  The slopes of
the best-fit lines  are somewhat smaller ($\sim 2.5$) than in  the case of the
stellar mass/luminosity of the {\it central} galaxy. Here again, we show using
the  short-dashed   lines  the  $\pm  1\sigma$  statistical   scatter  of  the
distributions with respect  to the best fit lines.  The $1\sigma$ scatters are
about  $0.29$dex   in  $M_{\rm   st}$  and  $L_G$   for  a  given   $L_X$,  or
$0.29/0.46=0.63$dex and $0.29/0.43=0.67$dex in  $L_X$ for a fixed $M_{\rm st}$
and $L_{\rm G}$.  And we will also check if  these relations are significantly
affected by the Malmquist bias in Section \ref{sec:conditional}.

Finally, we  check the distributions of  the stellar mass  and luminosity gaps
between the  first and second  most massive (luminous) galaxies.   The results
are  shown in  the lower-left  and  lower-right panels  in the  middle row  of
Fig.~\ref{fig:group-Lx}, respectively.  As  discussed in D'Onghia \etal (2005)
and Milosavljevi\'c  \etal (2006) this gap statistic  quantifies the dynamical
age of a system of galaxies: haloes with a small gap must be relatively young,
as dynamical friction  will cause multiple luminous galaxies  in the same halo
to merge  on a relatively  short time scale.   Evidently, there is  no obvious
correlation between the  stellar mass gap or luminosity  gap and the cluster's
X-ray  luminosity.  Furthermore, a  comparison with  the distributions  of the
luminosity and  stellar mass  gaps in massive  groups presented in  Yang \etal
(2008),  shows  that  the X-ray  clusters  have  gaps  that are  in  excellent
agreement with  those expected for haloes  with $M_h \ga  10^{14} \msunh$ (see
also  van den  Bosch \etal  2007).   Hence, there  is no  indication that  the
magnitude of the luminosity and/or stellar mass gaps are in any way correlated
with X-ray luminosity.

In  the  literature, the  luminosity  gap  has often  been  used  to define  a
``special population''  of galaxy groups, called ``fossil  groups'', which are
defined as  having an $R$-band luminosity  gap $\log L_1-\log  L_2 >0.8$ (e.g.
Ponman \etal 1994).  These fossil  groups are usually dominated by one central
early-type galaxy, and  a bright extended X-ray halo with  a cooling time that
is long enough for its bright satellite galaxies to have merged away (i.e., to
have been  cannibalized or  disrupted by the  central galaxy).  Among  our 201
X-ray clusters, only  2 fall in this category (the  clusters with the sequence
numbers 46  and 73 in Appendix~B).  In  addition, there are 3  clusters in our
sample  that have  no satellite  galaxies above  the apparent  magnitude limit
$r=17.72$ (not  observed since  $L_2< L_{\rm limit}$),  which might  be fossil
groups  as well.   Their gaps  are plotted  as lower  limits  (upward pointing
arrows)  using $L_2 =  L_{\rm limit}$.   Note that  all five  potential fossil
groups  in our  sample  have relatively  high  X-ray luminosities,  suggesting
either (i) that  they reside in relatively massive haloes,  or (ii) that their
X-ray luminosity  is a poor indicator  of their halo  mass.  Unfortunately the
sample is too  small to draw any meaningful  conclusions. Nevertheless, as the
X-ray clusters are of quite different gap distributions, we will check in more
detail the  galaxy properties (e.g., the  star formation rate,  etc.) in X-ray
clusters within different gap regions in a future probe.

\begin{figure}
\center{\includegraphics[height=7.0cm,width=7.0cm,angle=270]{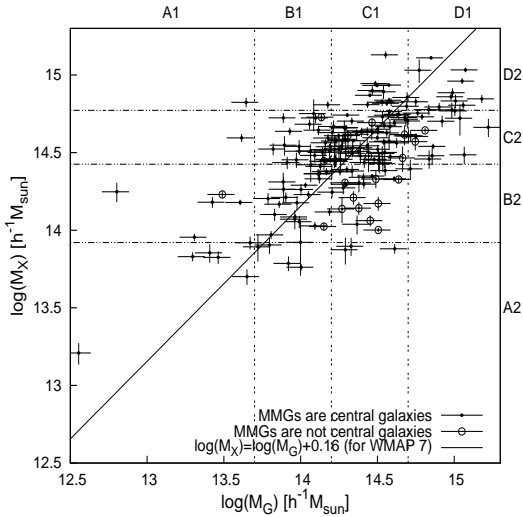}}
\caption{The X-ray  cluster mass $\log  M_{\rm X}$ v.s.  the  cross identified
  group mass, $\log M_{\rm G}$.  Here  results are shown for X-ray clusters in
  which  the  most massive  galaxies  are (solid  symbols)  or  are not  (open
  symbols)  central galaxies.  To  check their  large scale  environments, the
  X-ray  clusters are  divided into  4 subsamples  (A1-D1) according  to $\log
  M_{\rm G}$  using the three  vertical dashed lines  shown in the  plot.  For
  comparison,  we also  divide the  X-ray clusters  into 4  subsamples (A2-D2)
  according to  $\log M_{\rm X}$, each  containing exactly the  same number as
  the corresponding subsample in A1-D1.}
\label{fig:MX-MG}
\end{figure}

\subsection{The Halo Masses of the X-ray Clusters}

In this section  we compare two different methods to  estimate the halo masses
of the X-ray clusters in our sample. The first method is the one presented and
tested in Y07, and uses abundance matching to infer a halo mass for each group
in the  SDSS group catalog, under  the assumption of a  one-to-one (i.e., zero
scatter) relation between halo mass and either $L_{\rm G}$ or $M_{\rm st}$. As
shown in Yang \etal (2005a) and  Y07, the typical uncertainty in the resulting
halo mass  (hereafter $M_G$) is at  the level of  $\sigma_{\log M_G}\sim 0.3$.
However, since the majority of the  groups contain only one or two members, if
we restrict to  groups with at least 3  members\footnote{Among 201 groups that
  are matched  with the X-ray clusters, 197  have at least 3  members.  }, the
resulting  uncertainty  is about  $\sigma_{\log  M_G}\sim  0.25$.  The  second
method that  we consider  in this section  is the hydrodynamical  mass, $M_X$,
inferred from the  X-ray emission under the assumption that  the cluster is in
hydrostatic equilibrium.   In what follows we  convert all halo  masses to the
same  definition,  namely  the  mass  inside  a  spherical  volume  of  radius
$r_{200}$, inside  of which  the average density  is 200 times  the background
density  of the  Universe.   By  construction, the  masses  $M_G$ are  already
consistent with this definition.

The hydrodynamical  mass, $M_X$, requires accurate measurements  of the radial
temperature profile of  the ICM. Since such data is only  available for a tiny
fraction of the 201 X-ray clusters  in our sample, we use a statistical method
instead,  based on  the {\it  average} $\Lx-M_X$  relation (e.g.   Reiprich \&
B\"ohringer  2002; Stanek \etal  2006; Vikhlinin  \etal 2009;  Leauthaud \etal
2010; Arnaud  \etal 2010)  .  Since we  focus only on  low-redshift ($0.01\leq
z\leq0.2$) X-ray clusters, with  rest-frame X-ray luminosities measured in the
broad ROSAT  passband ($0.1$-$2.4$  keV), we use  the $\Lx$-$M_X$  relation of
Arnaud  \etal (2010).   By investigating  the regularity  of  cluster pressure
profiles with REXCESS (B\"ohringer et al. 2007) for a representative sample of
33 local  ($z < 0.2$) clusters,  and with the  help of $N$-body/hydrodynamical
simulations, Arnaud \etal (2010)  obtained the following X-ray luminosity-mass
scaling relation,
\begin{equation}\label{eq:LMz}
\frac{L_{500\rm  c}}{10^{44} {\rm erg~ s^{-1}}}=C\Big(\frac{M_{500\rm  c}}
{3\times10^{14} M_{\odot}}\Big)^{\alpha}\left[\Omega_{\rm m} (1 + z)^3 + 
\Omega_{\rm \Lambda}\right]^{\frac{7}{6}}\,,
\end{equation}
where $\log(C) = 0.193$, and $\alpha = 1.76$.  $M_{500\rm c}$ is the halo mass
of the X-ray  cluster within radius $r_{500\rm c}$  whose average mass density
is $500$ times  the critical mass density of the  Universe, and $L_{500\rm c}$
is the total X-ray luminosity  within $r_{500\rm c}$. This fitting formula has
an  intrinsic  scatter  in   the  log-log  plane  of  $\sigma_{\log  M_{500\rm
    c}}=0.199$.  Note  that the X-ray luminosity  $\Lx$ used in  this paper is
the total luminosity without  cluster core exclusion.  Piffaretti \etal (2010)
employed an iterative  algorithm to calculate $L_{500\rm c}$  for sources with
available   aperture   luminosities  $L_{\rm   ap}$,   and  found   $L_{500\rm
  c}/\Lx=0.91$ for the total X-ray luminosities.  With this transformation, we
can obtain $M_{500\rm  c}$ and $r_{500\rm c}$ for an  X-ray cluster with given
$\Lx$. The  final step  is then  to convert $M_{500\rm  c}$ to  $M_{200}$, for
which we use the relations
\begin{eqnarray} \label{eq:convert} 
r_{200} &\simeq& 2.70\times r_{500\rm c} \,, \nonumber \\ 
M_{200} &=& M_{500\rm c}\times \frac{200}{500}\times\Omega_{\rm
  m}\times \left( \frac{r_{200}}{r_{500\rm c}}\right)^3 \,,
\end{eqnarray}
where we  have assumed that  dark matter have  a NFW density  profile (Navarro
\etal  1997) with  concentration  parameters given  by the  concentration-mass
relation of  Maccio \etal  (2007). Note  that we have  not made  a distinction
between cool-core and non cool-core  systems.  As shown in Pratt \etal (2009),
the $L_X$-$M_X$  relation for cool-core  clusters has a  systematically higher
normalization  than non  cool-core systems.  We will  come back  to this  in a
forthcoming paper by  probing the optical properties of  galaxies in cool-core
and non cool-core systems.

Fig.~\ref{fig:MX-MG} plots  the hydrodynamic mass $M_X$ versus  the group mass
$M_G$. There  is a clear  correlation between these  two sets of  halo masses,
with a lognormal  scatter at the level of  $\sigma_{\log M_{\rm X}}\sim 0.25$.
This is perhaps  due to mixing of cool-core and non  cool-core clusters in the
sample.  For comparison,  results are shown separately for  X-ray clusters for
which  the most  massive galaxies   are  (solid dots)  and are  not (open
circles)  central galaxies.  There  is a  hint that  X-ray clusters  with none
central MMGs are slightly more massive in terms of group mass $M_G$.   

To check if there is any  systematic difference between these two sets of halo
mass  measurements, we  again fit  using  the least  square linear  regression
method with  only 1 free parameter,  $ \Delta \log  M$ to obtain the  best fit
$\log M_X=\Delta  \log M +  \log M_G$ relation.   The result is shown  in Fig.
~\ref{fig:MX-MG}  as  the  solid  line.   Here we  see  small  but  noticeable
systematic difference between the two sets  of halo masses. Note that the halo
masses obtained  in Y07 are based  on the abundance matching  method where the
halo mass  function of  given cosmology  (WMAP7 in this  paper) is  used. Thus
obtained halo  masses are quite  sensitive to the cosmological  parameters. In
case  the  $M_X$ provided  by  the X-ray  scaling  relation  is reliable,  the
systematic difference  can be straightforwardly  used to probe  the cosmology.
For instance, here the slightly underestimated systematic difference for $M_G$
may  indicate that  the data  require  the slightly  larger $\Omega_m$  and/or
$\sigma_8$ than WMAP7.  And of cause, to carry  out reliable constraints along
this line, more detailed error analyses are needed.

\begin{figure*}
\center{\includegraphics[height=8.0cm,width=8.0cm,angle=270]{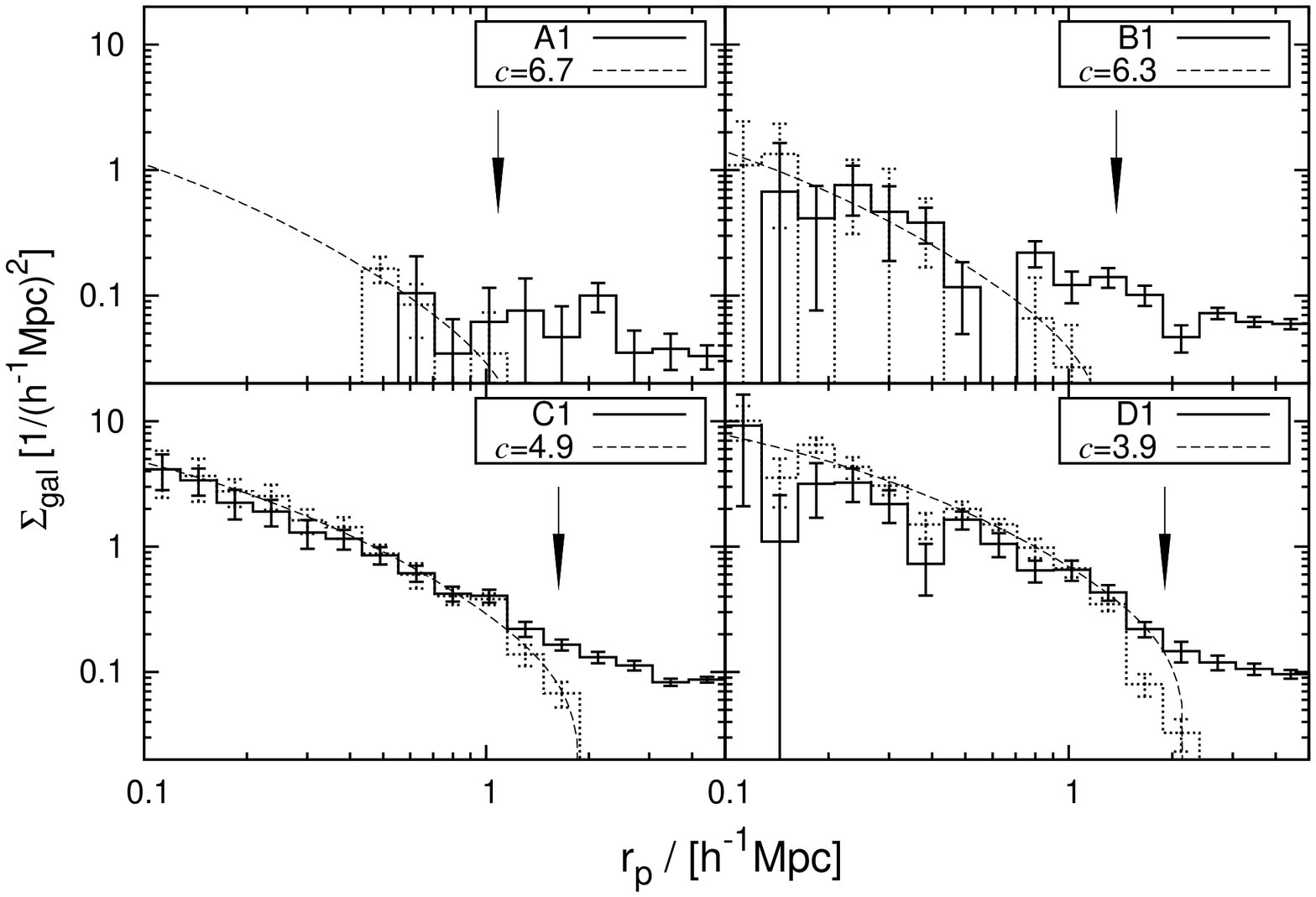}
\includegraphics[height=8.0cm,width=8.0cm,angle=270]{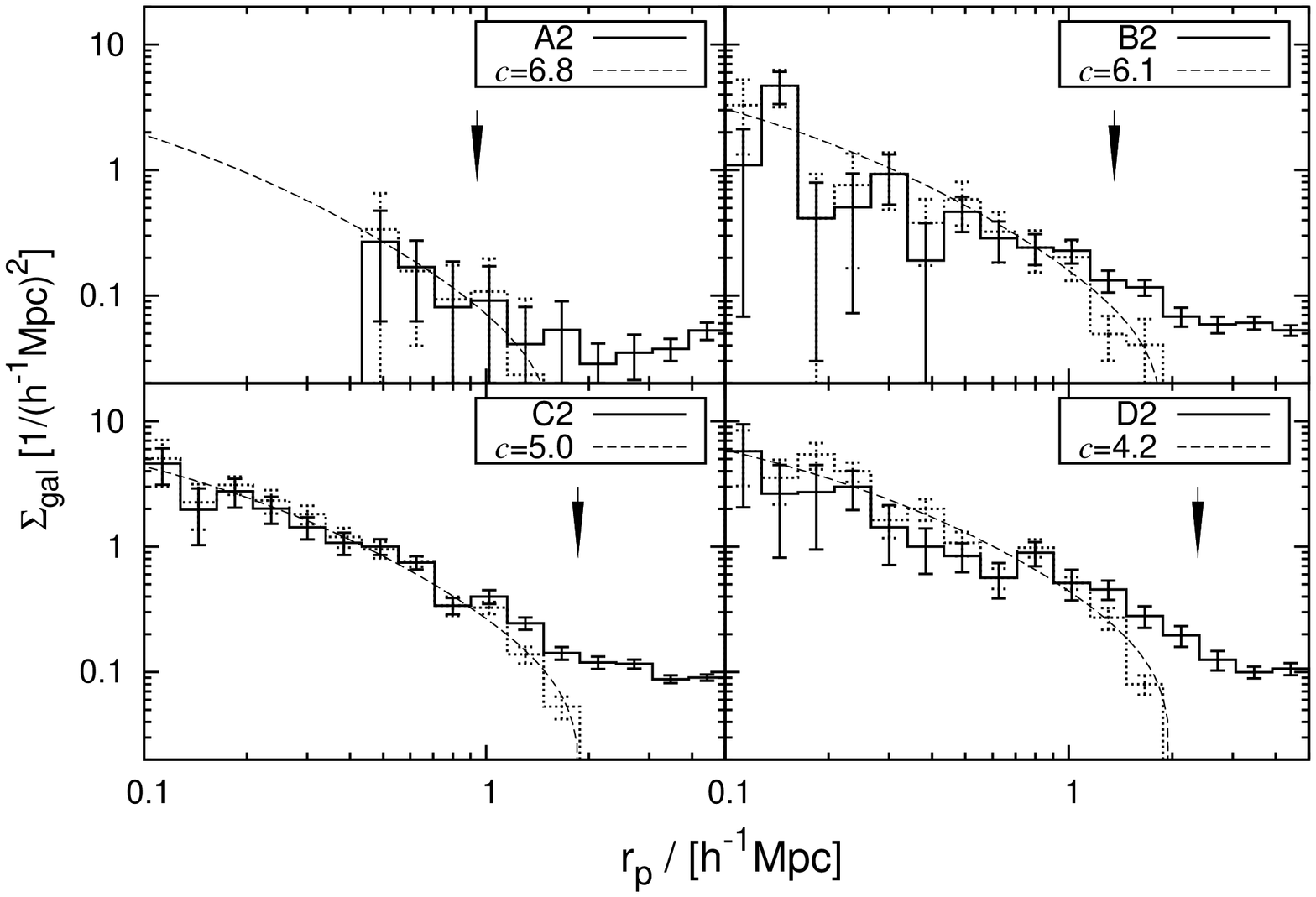}}
\caption{Surface number  density of galaxies  in and around X-ray  clusters in
  units  of $1/(\mpch)^2$.   Different  panels correspond  to different  X-ray
  cluster subsamples. The  error bars are $1\sigma$ scatter  obtained from 200
  bootstrap  re-samplings.  The  vertical arrow  in each  panel  indicates the
  average halo radius, $r_{200}$, of  the X-ray clusters in consideration. The
  dashed lines are the best fit 2-D NFW profiles. }
\label{fig:rho_g}
\end{figure*}

\begin{figure}
\center {\includegraphics[height=7.0cm,width=6.5cm,angle=270]{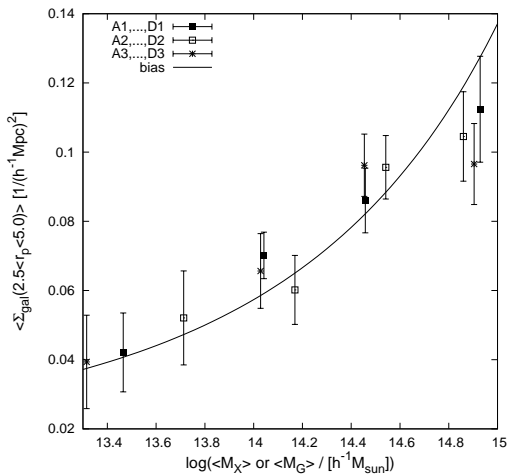}}
\caption{The average galaxy surface number density within $2.5\mpch <r_{\rm p}
  <  5\mpch$ as a  function of  halo mass  $M_X$ or  $M_G$. The  errorbars are
  obtained  from   200  bootstrap  re-samplings  of   the  clusters/groups  in
  consideration.   For comparison, we  show as  the solid  line the  halo bias
  predicted by Sheth et al. (2001), properly shifted to match most of the data
  points. }
\label{fig:ratios}
\end{figure}

\subsection{Distribution of Galaxies inside and around X-ray Clusters}
\label{sec:sbgal}

As a general check of the X-ray  cluster masses $M_{\rm G}$ and $M_X$, we make
use  of the  fact  that, in  CDM  cosmologies, more  massive  haloes are  more
strongly clustered  (e.g., Mo \& White  1996).  Hence, if  our mass indicators
are  reliable, we  should find  that  haloes with  larger $M_X$  or $M_G$  are
located in denser  environments.  We can check this  using the distribution of
galaxies in the  cluster outskirts. To do so, we proceed  as follows. We first
divide our sample  of 201 X-ray clusters in  four subsamples (A1-D1) according
to their  assigned mass $M_G$, and  in another set of  four subsamples (A2-D2)
according  to  their  assigned  mass  $M_X$.  The  samples  are  indicated  in
Fig.~\ref{fig:MX-MG} as horizontal and vertical dot-dashed lines.

Since the typical velocity dispersion  of cluster galaxies is $\sim 1000\kms$,
we measure the surface number density of galaxies in and around X-ray clusters
as a function of the projected distance 
\begin{equation}\label{eq:drp_X}
r_{\rm p} = \sqrt{|{\bf r}_{\rm X} - {\bf r}_{\rm g}|^2-(d_{\rm  X} - d_{\rm
    g})^2} \,,
\end{equation}
using the following criterion:
\begin{equation}\label{eq:dz_X}
c\Delta z = c|z_{\rm X}-z_{\rm g}| \le 1000\kms \,.
\end{equation}  
Here $c$ is the speed of light, while ($z_{\rm g}$, ${\bf r}_{\rm g}$, $d_{\rm
  g}$) and  ($z_{\rm X}$,  ${\bf r}_{\rm X}$,$d_{\rm  X}$) are  the redshifts,
co-moving coordinates,  and co-moving radial  distances from the  observer, of
the galaxy and  the X-ray cluster (i.e., its central  galaxy) in question.  To
avoid  potential   inhomogeneities  caused  by   Malmquist  bias,  we   use  a
volume-limited galaxy sample with $\rmag\le -21.27$.

Fig.~\ref{fig:rho_g}  shows  the resulting  galaxy  surface number  densities,
$\Sigma_{\rm gal}(r_{\rm  p})$ for the  X-ray clusters in the  four subsamples
(A1-D1) of  mass $M_G$  (left-hand panel) and  the four subsamples  (A2-D2) of
mass $M_X$ (right-hand panel). In  each panel the vertical arrow indicates the
average halo  radius, $r_{200}$,  of the X-ray  clusters in  consideration. On
small scales  ($r_{\rm p} \la r_{200}$),  the signal is  dominated by galaxies
that reside in the dark matter halo  of the cluster. To show this we determine
the `1-halo'  term by  simply computing the  average projected  surface number
density of  all galaxies that  belong to the  X-ray clusters in each  bin {\it
  according  to  the  Y07  group  catalog}.  These are  shown  as  the  dotted
histograms  in Fig.~\ref{fig:rho_g},  and,  as expected,  nicely overlap  with
$\Sigma_{\rm  gal}(r_{\rm p})$ on  small scales.   Note that  in the  two most
massive  bins  (D1  and  D2)  these  1-halo terms  are  somewhat  larger  than
$\Sigma_{\rm gal}(r_{\rm p})$, which is a  result of our cut in redshift space
(criterion~[\ref{eq:dz_X}]). We fit the  1-halo term profiles with a projected
NFW model (Eq.~7 in  Yang et al.  2005a), and the results  are also plotted in
Fig. \ref{fig:rho_g}  as the dashed curves.  The  concentration parameters $c$
thus  obtained   are  indicated  in  each  panel.    Compared  to  theoretical
predictions for the concentration parameters of dark matter haloes (e.g., Zhao
\etal 2009), these best fit  concentrations $c$ are somewhat lower, suggesting
that  satellite galaxies  have  a  number density  distribution  that is  less
centrally  concentrated  than  the   dark  matter.   Although  in  qualitative
agreement with other  studies (e.g., Lin \etal 2004;  Collister \& Lahav 2005;
Yang \etal  2005b; Chen 2008;  More \etal 2009),  we caution that,  because of
interlopers and  other selection effects,  a more quantitative measure  of the
true concentration  of the number  density distribution of  satellite galaxies
requires a more careful analysis (e.g.  Yang et al. 2005b; Chen 2008).

At large projected  radii ($r_{\rm p} \ga r_{200}$)  the galaxy surface number
densities,  $\Sigma_{\rm gal}(r_{\rm  p})$, flatten  over to  roughly constant
values.  A comparison  with  the 1-halo  terms  shows that  this reflects  the
distribution of galaxies in the direct surroundings of the clusters. Since all
surface number  density profiles  are obtained using  the same  volume limited
sample  of  galaxies,  the  ratios  between  the  large-scale  surface  number
densities are directly  proportional to the ratios of the  biases of the X-ray
clusters  in the different  subsamples. We  show in  Fig.~\ref{fig:ratios} the
average galaxy surface number densities  within $2.5\mpch <r_{\rm p} < 5\mpch$
as a  function of halo  mass, $M_G$ (solid  squares) or $M_X$  (open squares).
Clearly,  more massive  clusters have  higher galaxy  surface  number density,
indicating that  they reside in  denser environments (i.e., are  more strongly
biased). For comparison,  the solid line in Fig.~\ref{fig:ratios}  is the halo
bias  predicted  by Sheth  \etal  (2001),  properly  shifted in  the  vertical
direction  to match  most of  the data  points.  Clearly,  the data  and model
prediction agree  remarkably well, for  both $M_G$ and $M_X$.

\begin{figure*}
\plotone{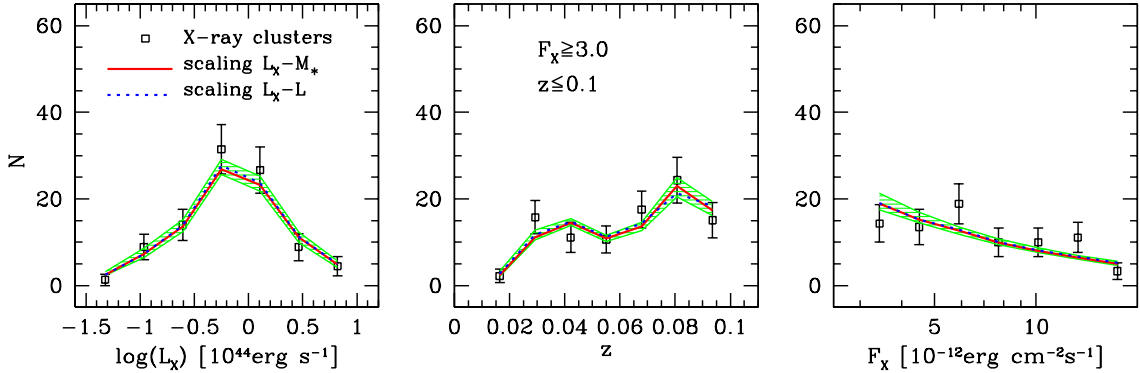}
\caption{The differential  number distribution of  X-ray clusters in  the SDSS
  DR7  region with  $F_X\ge  3.0\times10^{-12}\,{\rm erg~cm^{-2}~s^{-1}}$  and
  redshift  $z\le  0.1$ (squares  with  error bars)  as  a  function of  X-ray
  luminosity  (left  panel), redshift (middle  panel)  and  X-ray flux  (right
  panel), respectively. The solid and dashed  lines in each panel are the best
  fit model  predictions by the  scaling relations using group  stellar masses
  and luminosities,  respectively. The shaded areas represent  the 68\% ranges
  of the model predictions. }
\label{fig:for_FP}
\end{figure*}

\begin{figure*}
\plottwo{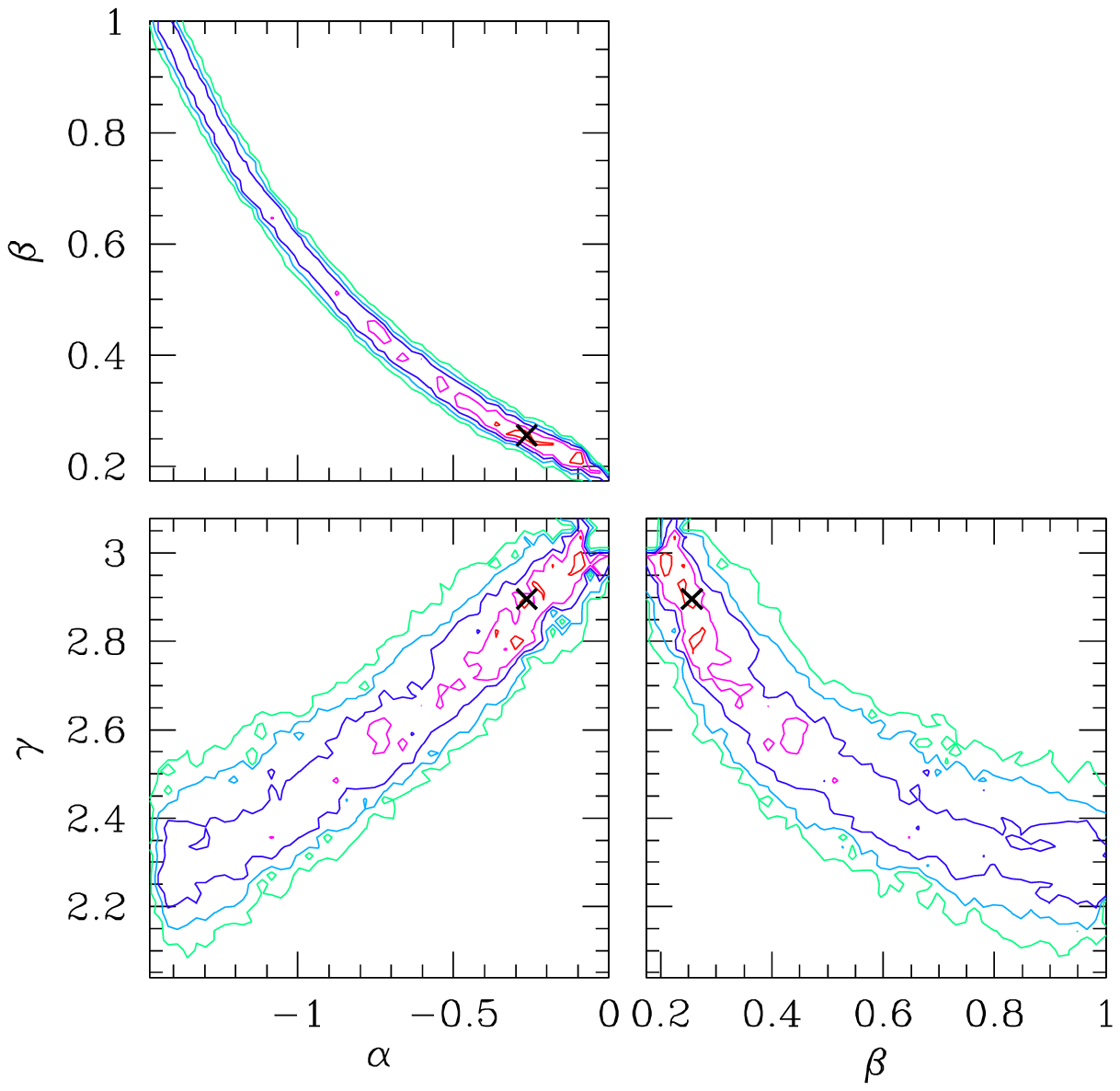}{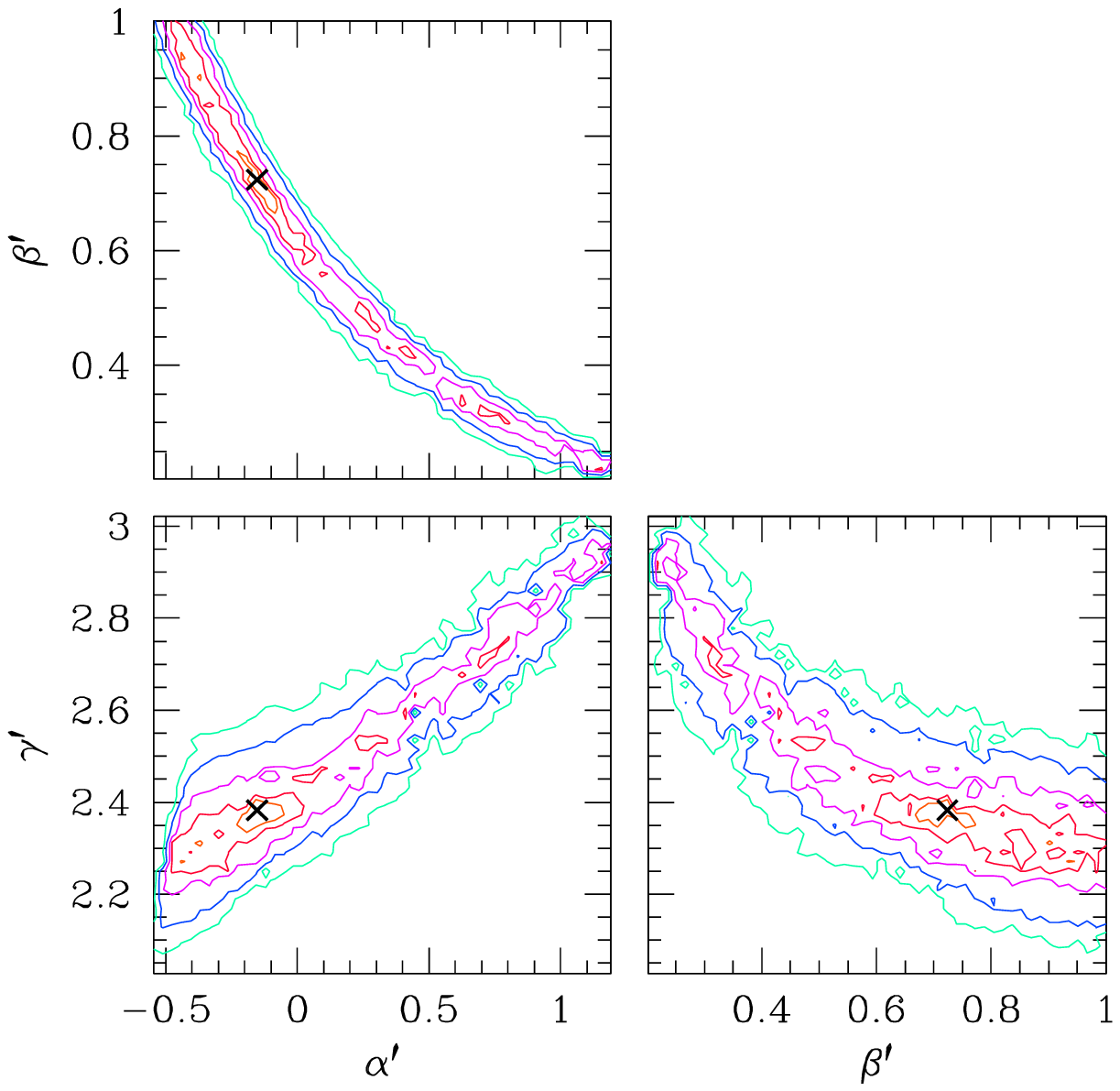}
\caption{The best fit (cross) and the projected distribution of the parameters
  on  2-D  planes. The  outer  first and  second  contours  correspond to  the
  distribution  of the  95\% and  68\% parameters  starting from  the smallest
  $\chi^2$ values.   Here results shown in  the left and right  panels are for
  stellar mass-  and optical  luminosity- X-ray luminosity  scaling relations,
  respectively. }
\label{fig:error}
\end{figure*}

\section{Groups with and without strong X-ray Emission}
\label{sec_diff}

Having discussed  various optical and  X-ray correlations for the  groups that
are linked  with X-ray clusters, we  now focus on groups  of comparable masses
($M_G$) but lacking  strong X-ray emission (i.e., for  which no measurement of
$\Lx$ is available). Although both samples  have the same sky coverage and lie
in the same redshift range, many groups fall in this category even the richest
ones. There are of cause various  survey selections, e.g. in the X-ray fluxes,
in the bright  star mask, etc., that prevent us from  getting a complete X-ray
cluster catalogue.  The  number is much larger than  those completeness values
quoted in e.g. Ebeling et al. (2000).

In this  section, we  investigate the possible  Malmquist bias induced  by the
limiting  flux in  the  RASS observation,  and  probe whether  the groups  and
clusters  with strong X-ray  emission have  different galaxy  populations from
those of the same optical component but without strong X-ray emission.

\subsection{Unbiased scaling relations} 
\label{sec:conditional}

\begin{figure*}
\center{ \includegraphics[height=13.0cm,width=17.0cm,angle=0]{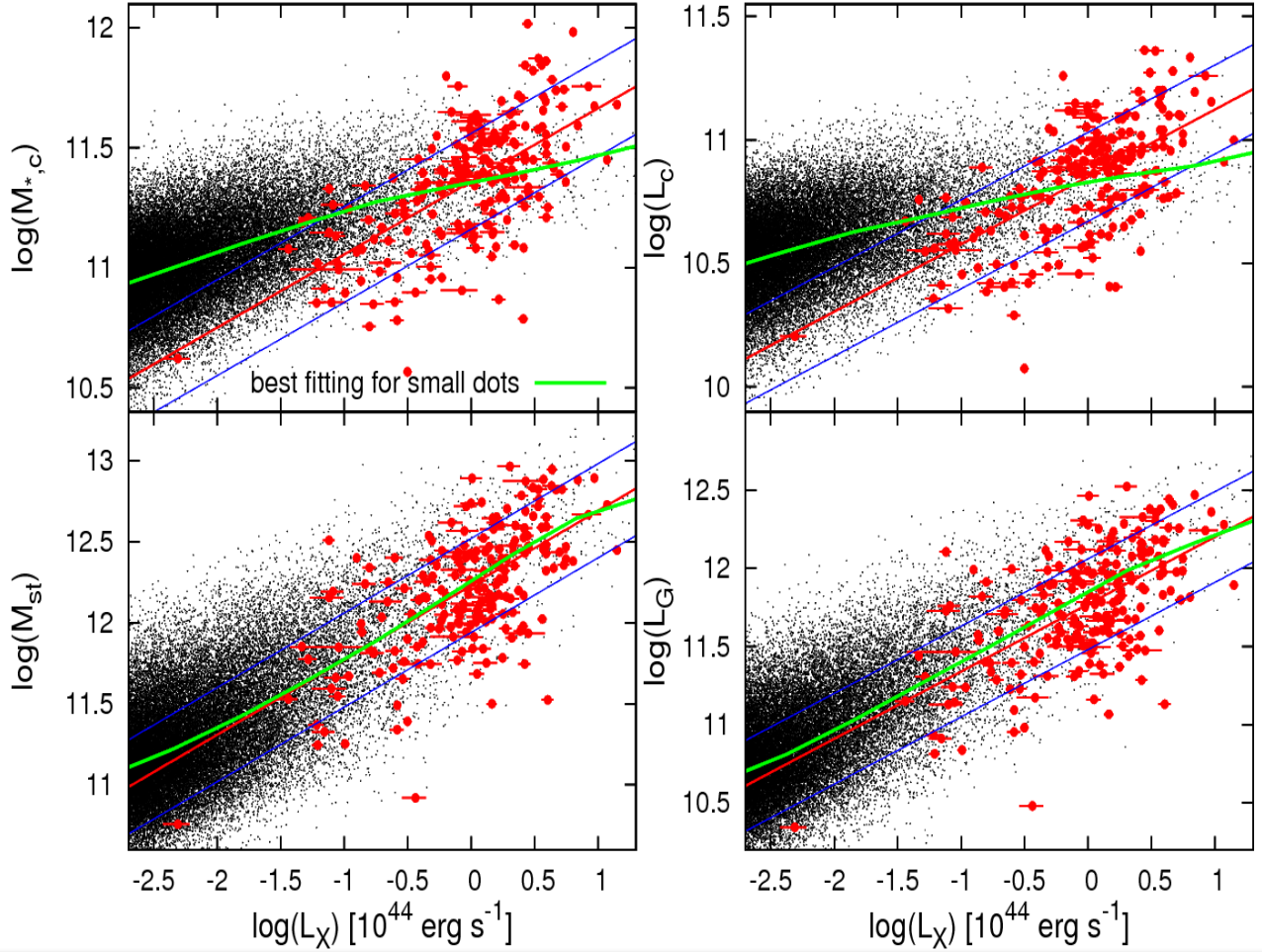}}
\caption{The   distributions  of   stellar  masses   (upper-left   panel)  and
  luminosities (upper-right panel) of central galaxies, characteristic stellar
  masses (lower-left panel) and  luminosities (lower-right panel) of groups as
  a function of X-ray cluster luminosity  $\log \Lx$.  Here big dots and solid
  straight  lines are the  results for  the observed  X-ray clusters  (same as
  those  shown   in  the  upper   panels  of  Figs   \ref{fig:central-Lx}  and
  \ref{fig:group-Lx}).  The  small dots show  the distributions of  all galaxy
  groups  whose X-ray luminosities  are assigned  using the  scaling relations
  (Eqs. \ref{eq:Lx-Mxx}  and \ref{eq:Lx-Lxx}).   The curves are  the resulting
  average  $\log  \Lx$  as  a  function  of  stellar  mass  or  luminosity  in
  consideration.}
\label{fig:Malmquist}
\end{figure*}

As shown  in Section \ref{sec:general}, $M_{\ast,c}$, $M_{\rm  st}$, $L_c$ and
$L_G$ are  all strongly  correlated with $L_X$,  albeit with  relatively large
scatter.  And  we did  not find strong  correlations between $L_X$  with other
optical properties, e.g, the color  and concentration of the central galaxies,
the magnitude  and luminosity gaps between  the first and  second most massive
(luminous)  galaxies, etc. Note  also these  relations are  probed based  on a
small set of {\it observed} X-ray  clusters. Because of the quite shallow flux
limit of the RASS, the relations  we obtained among them might suffer from the
Malmquist bias.  Here we  try to find  the unbiased scaling  relations between
$M_{\ast,c}$, $M_{\rm  st}$ (or  $L_c$ , $L_G$)  and $L_X$, assuming  that the
groups not observed in X-ray, apart from other selections like the bright star
mask, are mainly due to the flux limit in the RASS observation.

In literature, there are claims about the existence of a genuine population of
clusters that are X-ray under-luminous  (e.g., Castander \etal 1994; Balogh et
al.  2011).  However,  as pointed out in a recent paper  by Andreon \& Moretti
(2011) using Swift 1.4 Ms X-ray observations, there is no distinct populations
of  X-ray clusters, although  the scatter  in the  X-ray luminosity  is large.
Therefore, the X-ray underluminous groups  in our catalogue are expected to be
systems whose X-ray  luminosities are at the lower end of  the scatter. As the
optical group  sample is more complete than  the X-ray sample, we  can use the
observed  number  of  X-ray  clusters  to constrain  the  {\it  true}  scaling
relations and  their scatter, in  a manner that  is not affected  by Malmquist
bias. To this  end, we first measure the  differential number distributions of
X-ray  clusters with  respect to  X-ray luminosity  ($\hat  N(\Lx)$), redshift
($\hat  N(z)$) and  X-ray flux  ($\hat N(F_X)$),  respectively.   These number
distributions are obtained with the  survey completeness of the X-ray clusters
properly  taken  into account  and  with  only  X-ray clusters  brighter  than
$3.0\times10^{-12}\,{\rm  erg~cm^{-2}~s^{-1}}$ and  redshift $z\le  0.1$ being
used.  Here every X-ray cluster is  counted with a weight $1/c/f_{sky} $ where
$c$ is the completeness factor and $f_{sky}$ is the relative sky coverage with
respect  to the  SDSS DR7  in  consideration. The  results are  shown in  Fig.
\ref{fig:for_FP} as open squares  with (Poisson) errorbars. These measurements
are then used to constrain the unbiased scaling relations.

Since  the  characteristic stellar  mass  (luminosity)  and  the stellar  mass
(luminosity) of  central galaxy are  not independent variables (the  latter is
included in the  former), we use the characteristic  stellar mass (luminosity)
of  satellite  galaxies,  defined  as  $M_{\rm  sat}=M_{\rm  st}-M_{\ast,  c}$
($L_{\rm sat}=L_{\rm G}-L_{c}$), to replace $M_{\rm st}$ as the third quantity
in our scaling relation analysis.   Assume that the X-ray luminosity depend on
the stellar masses of the centrals and satellites as
\begin{equation}\label{eq:Lx-Mxx}
{\log L_X} = \alpha + \gamma [\log (M_{\ast, c} + \beta M_{\rm sat}) -12.0]\,,
\end{equation}
with lognormal scatter $\sigma$, and on the luminosities of the central and
\begin{equation}\label{eq:Lx-Lxx}
{\log L_X} = \alpha' + \gamma' [\log (L_{\rm c} + \beta' L_{\rm sat}) -12.0] \,,
\end{equation}
with lognormal scatter  $\sigma'$. We apply these models to  all of our galaxy
groups.   The resulting X-ray  luminosities are  then properly  converted into
X-ray fluxes in the observed band taking into account the luminosity distances
and  negative average $K$  corrections assuming  an average  X-ray temperature
$5.0 {\rm keV}$ (B\"ohringer et al. 2004).  From this `X-ray group catalogue',
we  calculate the  same quantities  as  shown in  Fig.  \ref{fig:for_FP}  with
respect  to X-ray  luminosity ($N(\Lx)$),  redshift ($  N(z)$) and  X-ray flux
($N(F_X)$), respectively. Here  taking into account the scatter  in $\log L_X$
of the  observed X-ray clusters, we set  $\sigma=\sigma'=0.67$.  Thus obtained
data,  together  with  those  direct  measurements  from  the  observed  X-ray
clusters, are used to constrain the scaling relations.  The goodness-of-fit of
each model is described by its $\chi^2$ value defined by
\begin{eqnarray}
\label{chisq}
\chi^2 &=& \sum
\left[ {N(\Lx) - \hat{N}(\Lx) \over \Delta \hat{N}(\Lx)} \right]^2 
+ \sum \left[ {N(z) - \hat{N}(z) \over \Delta \hat{N}(z)} \right]^2 \nonumber \\
&+& \sum \left[ {N(F_X) - \hat{N}(F_X) \over \Delta \hat{N}(F_X)} \right]^2 \,.
\end{eqnarray}
Here   $\hat{N}$  and   $\Delta\hat{N}$  are   the  observed   average  number
distribution and error of X-ray clusters, respectively.

To obtain the best fit and the freedom of the model parameters, we follow Yan,
Madgwick  \& White  (2003; see  also van  den Bosch  et al.   2005) and  use a
Monte-Carlo  Markov Chain (hereafter  MCMC) to  fully describe  the likelihood
function in our multi-dimensional parameter  space.  We start our MCMC from an
initial guess and allow a `burn-in' of 1000 random walk steps for the chain to
equilibrate in the  likelihood space. At any point in the  chain we generate a
new trial model by drawing the  shifts in its three free parameters from three
independent Gaussian  distributions.  The  probability of accepting  the trial
model is
\begin{equation}
\label{probaccept}
P_{\rm accept} = \left\{ \begin{array}{ll}
1.0 & \mbox{if $\chi^2_{\rm new} < \chi^2_{\rm old}$} \\
{\rm exp}[-(\chi^2_{\rm new}-\chi^2_{\rm old})/2] & \mbox{if
$\chi^2_{\rm new} \geq \chi^2_{\rm old}$} \end{array} \right.
\end{equation}
with the $\chi^2$ measures given by eq.~(\ref{chisq}).

We construct a  MCMC of $1$ million steps, with an  average acceptance rate of
$\sim  25$  percent.  In  order  to  suppress  the correlation  power  between
neighboring models  in the chain, we thin  the chain by a  factor $100$.  This
results in a final MCMC consisting of $10000$ independent models that properly
sample the full posterior  distribution.  The contours in Fig.~\ref{fig:error}
plot the  resulting projected  2-D distributions of  the parameters,  with the
best-fit  values  indicated by  a  cross.  The  outer  two  level of  contours
correspond to the  projected confidence regions of the  $95\%$ and $68\%$ sets
of parameters with smaller $\chi^2$.  And  the best fit values, which have the
smallest    $\chi^2$     value,    are    $[\alpha,     \beta,    \gamma]    =
[-0.26^{+0.15}_{-0.13},   0.26^{+0.04}_{-0.02},   2.90^{+0.16}_{-0.2}]$,   and
$[\alpha',  \beta', \gamma']  =  [-0.15^{+0.10}_{-0.10}, 0.72^{+0.06}_{-0.09},
  2.38^{+0.13}_{-0.14}]$,  respectively.  Here  the superscript  and subscript
indicate the 68\% confidence level of each best fit parameter while others are
fixed. Note  however, as  the satellite components  are in  general correlated
with the  central galaxy, both  of which increase  with the increasing  of the
host halo  mass, currently  we are not  able to  put tight constraints  on the
$\beta$ (or $\beta'$ ) parameter indeed, as indicated by the very extended 2-D
confidence contours.  We  note, in case one get a  more reliable constraint on
$\beta$ (or  $\beta'$ ) that significantly  deviates from our  best fit value,
one can get the updated $[\alpha, \gamma]$ (or $[\alpha', \gamma']$ ) from the
2-D confidence contour plots of the parameters.

Note  that in  constraining the  scaling relations  individual $L_X$  of X-ray
clusters are  not used,  as they might  be biased  tracers of the  total X-ray
cluster/group population  in question.  Rather  we use the observed  number of
clusters  as our  constraints.  The  best  fit differential  numbers of  X-ray
clusters are shown as the  solid and dotted lines in Fig.~\ref{fig:for_FP} for
the  cases where  scaling  relations are  based  on group  stellar masses  and
luminosities, respectively. The shaded areas  represent the 68\% ranges of the
$10000$ MCMC independent model predictions.

Thus  obtained  scaling   relations  can  be  used  to   `predict'  the  X-ray
luminosities   of   clusters.   As   an   example,   we   have  applied   Eqs.
(\ref{eq:Lx-Mxx}) and (\ref{eq:Lx-Lxx})  to all of our galaxy  groups, and use
the resulting  X-ray luminosities for  all groups to check  possible Malmquist
bias  in  the  observed   X-ray  sample  in  Figs.   \ref{fig:central-Lx}  and
\ref{fig:group-Lx}.  In Fig. \ref{fig:Malmquist}, we compare the distributions
of the  observed X-ray clusters  and the one  we `predicted' from  the optical
galaxy groups.  The  small dots in each panel  show the predicted distribution
of all  groups.  Here  the scaling relations  and the  corresponding lognormal
scatters  are applied  to the  stellar masses  (left panels)  and luminosities
(right panels),  respectively.  For  comparison, in each  panel, we  also show
using  solid  curves the  resulting  average  stellar  mass or  luminosity  in
consideration as  a function  of $\log  \Lx$.  It is  clear that  the observed
X-ray  clusters  do  suffer  significantly  from the  Malmquist  bias  in  the
$M_{\ast,c}-L_X$  (or $L_{c}-L_X$)  relation, especially  at the  low-mass end
where the overall relation is flatter than the X-ray selected groups. Contrary
to  the case  of  the $M_{\ast,  c}  - \Lx$  relation,  the predicted  $M_{\rm
  st}-\Lx$ relation for  X-ray clusters is in good  agreement with the overall
distribution of all  groups, suggesting that in this  case our best-fit linear
regression is not strongly affected by Malmquist bias.

Finally, we  note that the scaling  relations obtained above are  based on the
assumption  that there  are  no  distinct populations  of  X-ray luminous  and
underluminous  groups (see  e.g. Andreon  \& Moretti  2011). If  a significant
fraction  of the optical  groups/clusters belonged  to an  X-ray underluminous
population, then the scaling  relations for the X-ray luminous groups/clusters
would be different.

\begin{figure*}
\center{\includegraphics[height=5.9cm,width=6.0cm,angle=270]{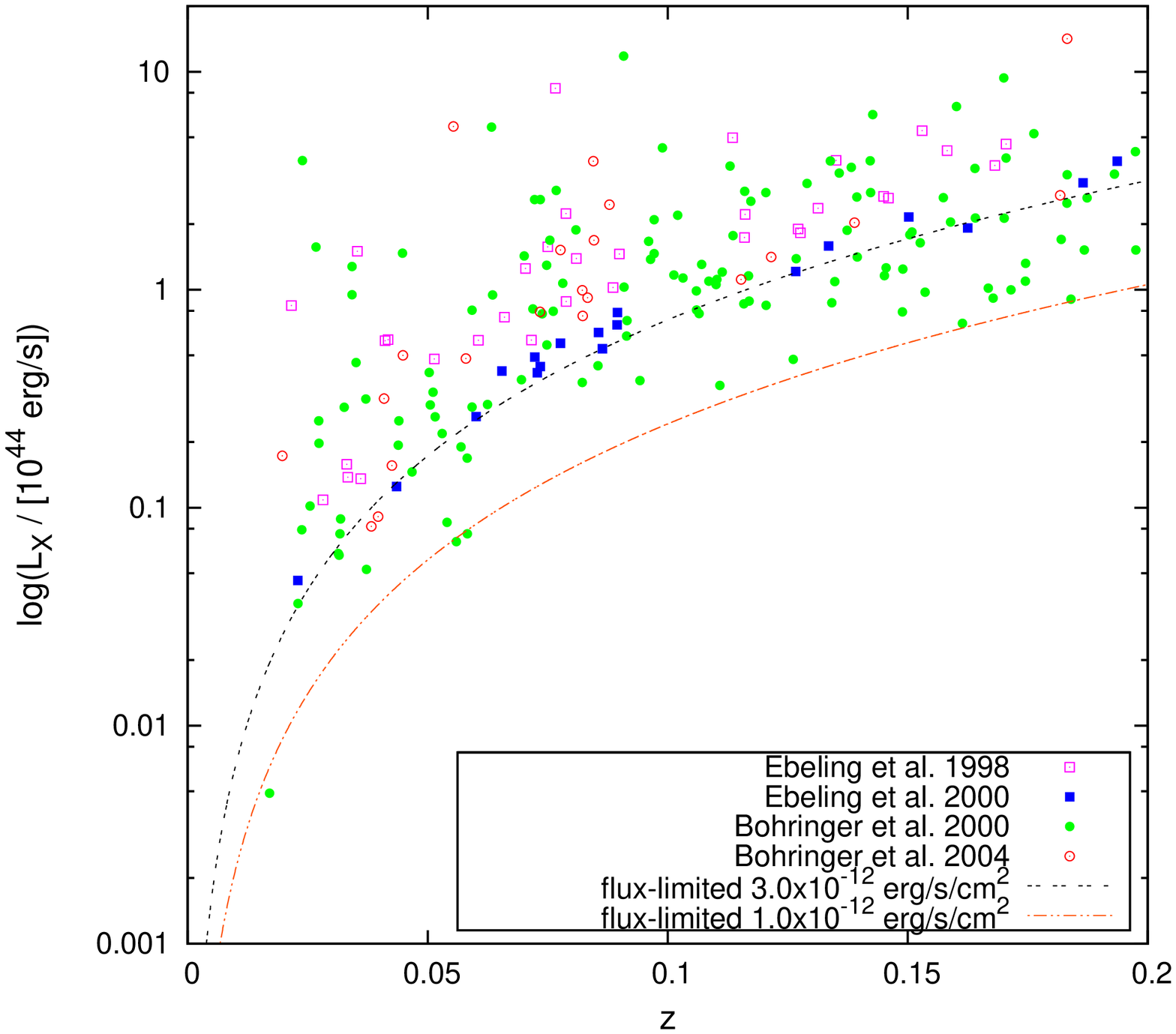}
\includegraphics[height=5.9cm,width=6cm,angle=270]{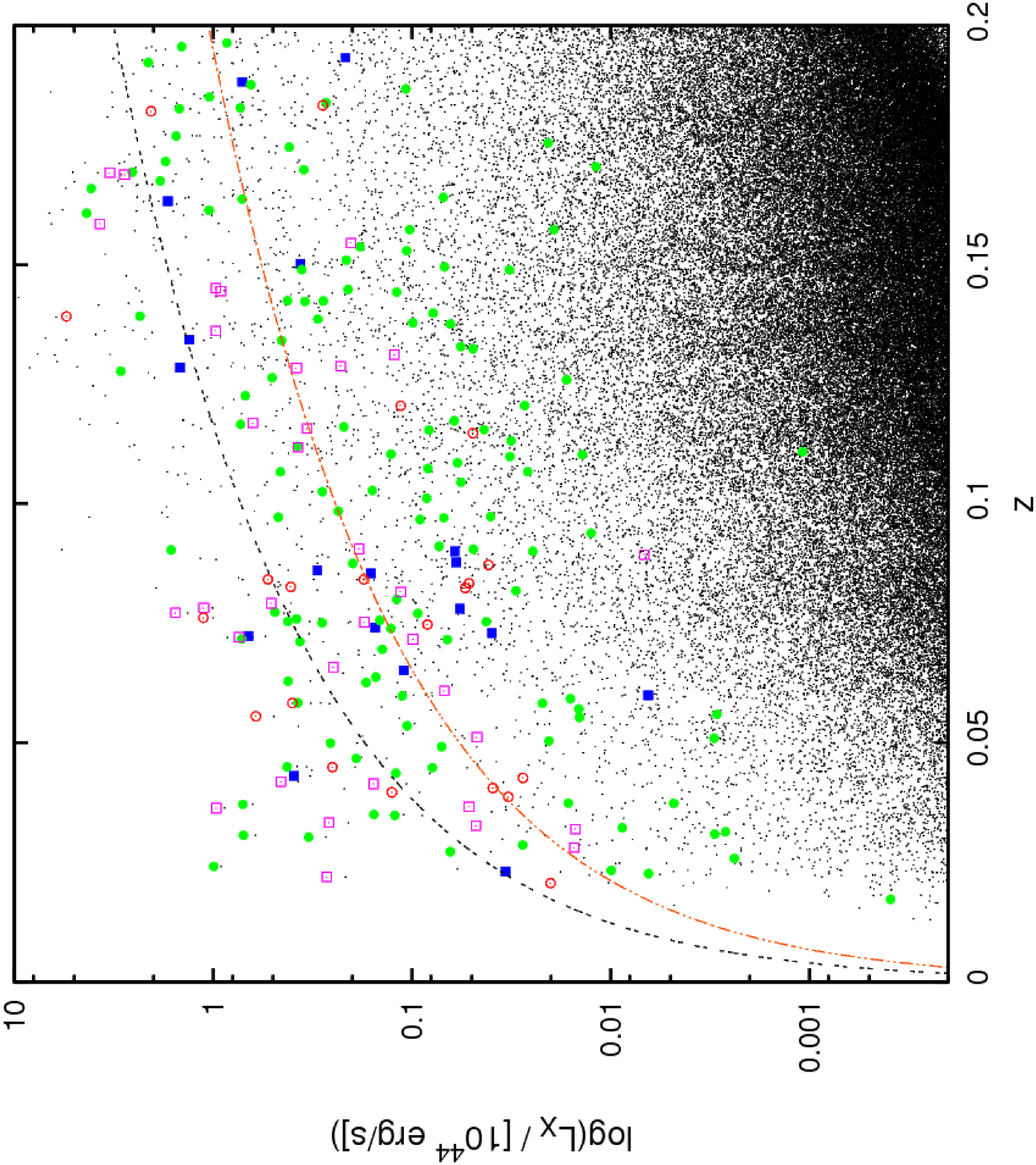}
\includegraphics[height=5.9cm,width=6.0cm,angle=270]{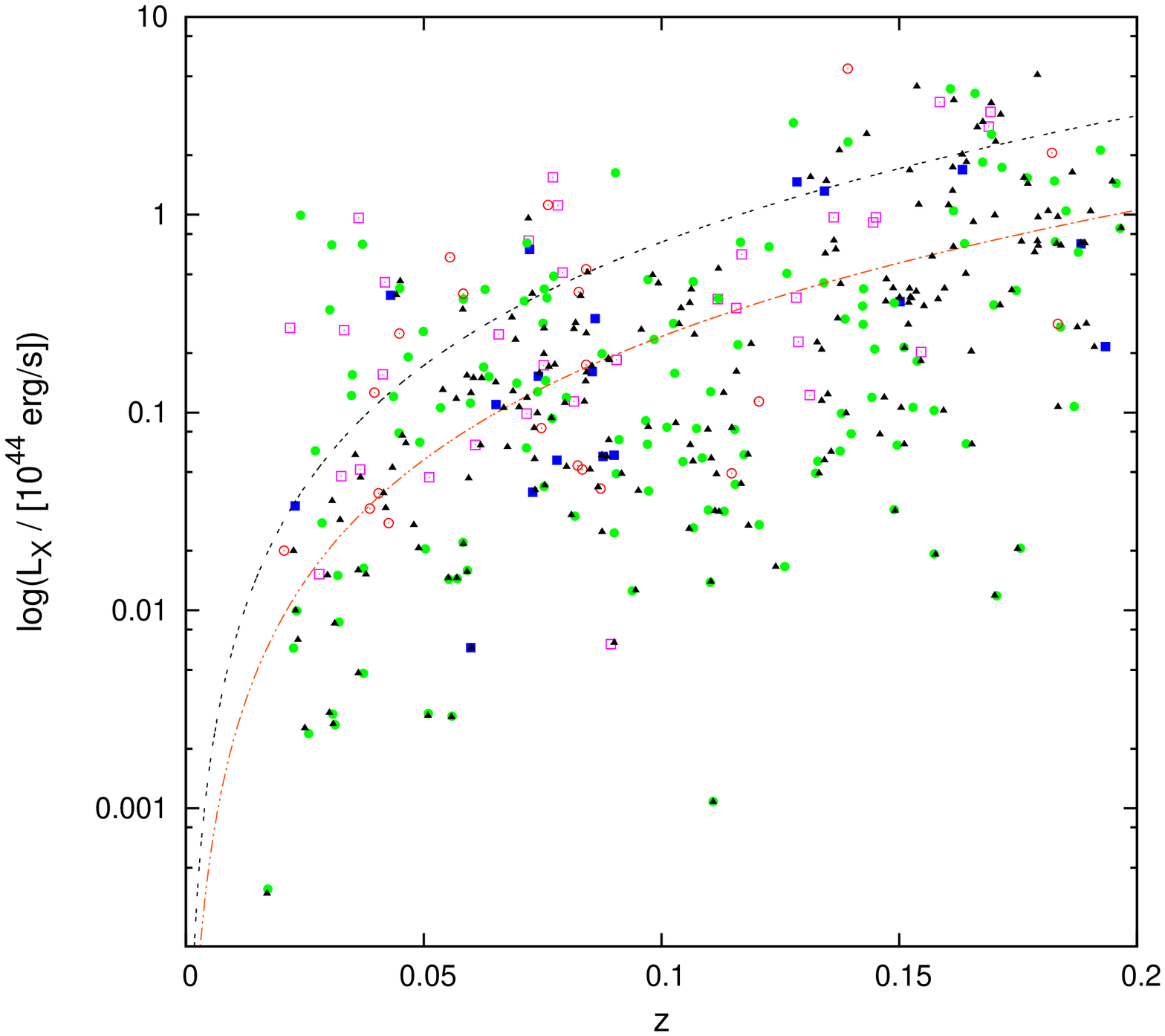}}
\caption{Left panel:  the redshift v.s.  X-ray luminosity  distribution of all
  the  X-ray  clusters  from  different  sources  (symbols).   The  two  lines
  correspond to the two flux limits, as indicated.  Middle panel: the redshift
  v.s.  X-ray luminosity distribution  of groups with their X-ray luminosities
  predicted  with  the  scaling  relation (Eq.\ref{eq:Lx-Mxx})  assuming  zero
  scatter. For  groups that are linked  to X-ray clusters, which  are shown as
  symbols, the  distribution is  quite different from  that shown in  the left
  panel.  Right panel: the same as the middle panel, but here for a controlled
  sample  of galaxy  groups, constructed  by matching  a galaxy  group without
  X-ray detection to  the one with X-ray detection,  according to its redshift
  and predicted X-ray luminosity $L_X$. }
\label{fig:dist}
\end{figure*}

\begin{figure*}
\center{\includegraphics[height=16.0cm,width=15.0cm,angle=270]{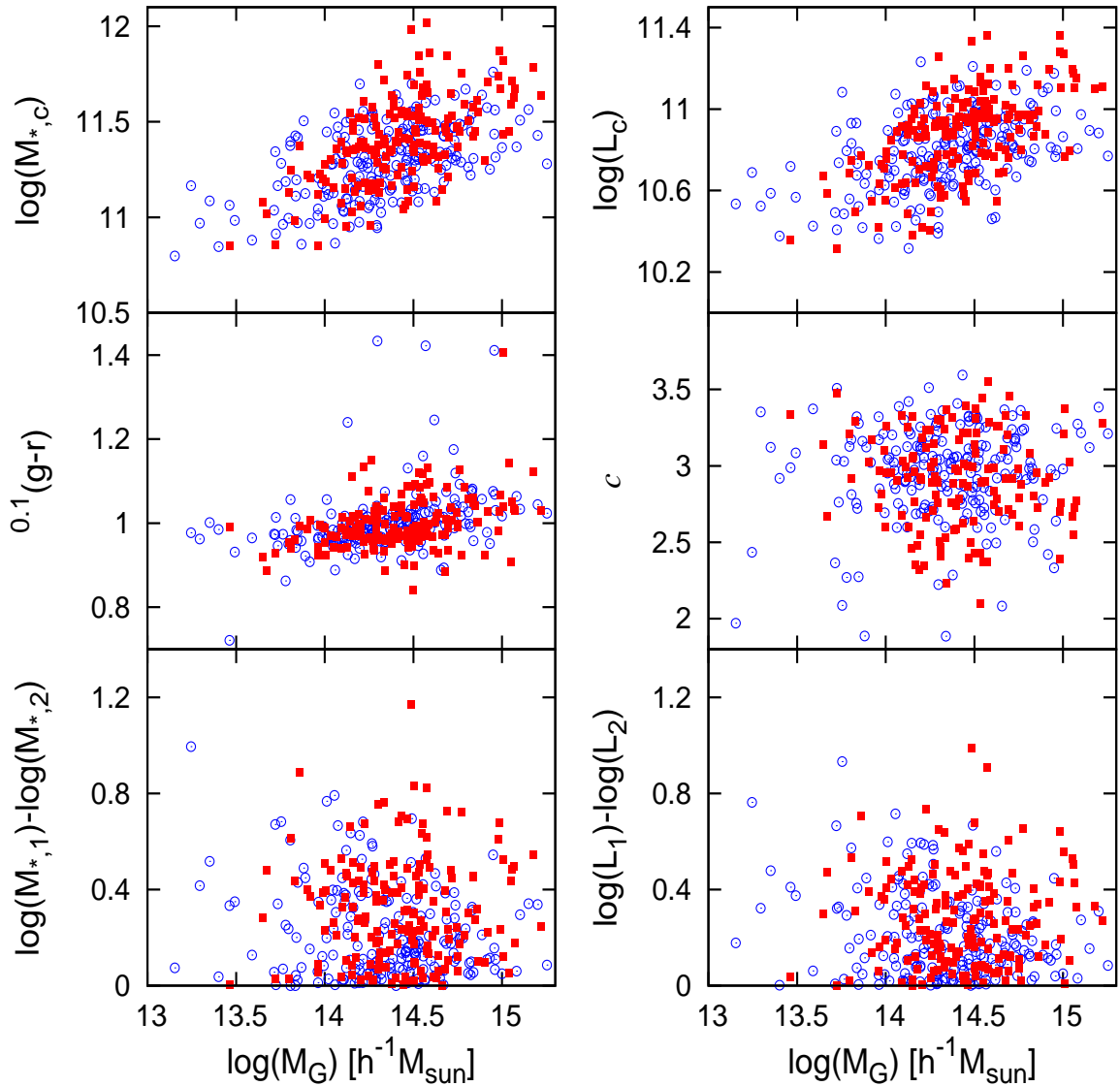}}
\caption{The optical  properties of groups  with (filled squares)  and without
  (open circles) strong X-ray emissions.  The latter are shown only for galaxy
  groups in the controlled sample.}
\label{fig:central-MG}
\end{figure*}

\begin{figure*}
\center{\includegraphics[height=14.0cm,width=14.0cm,angle=0.0]{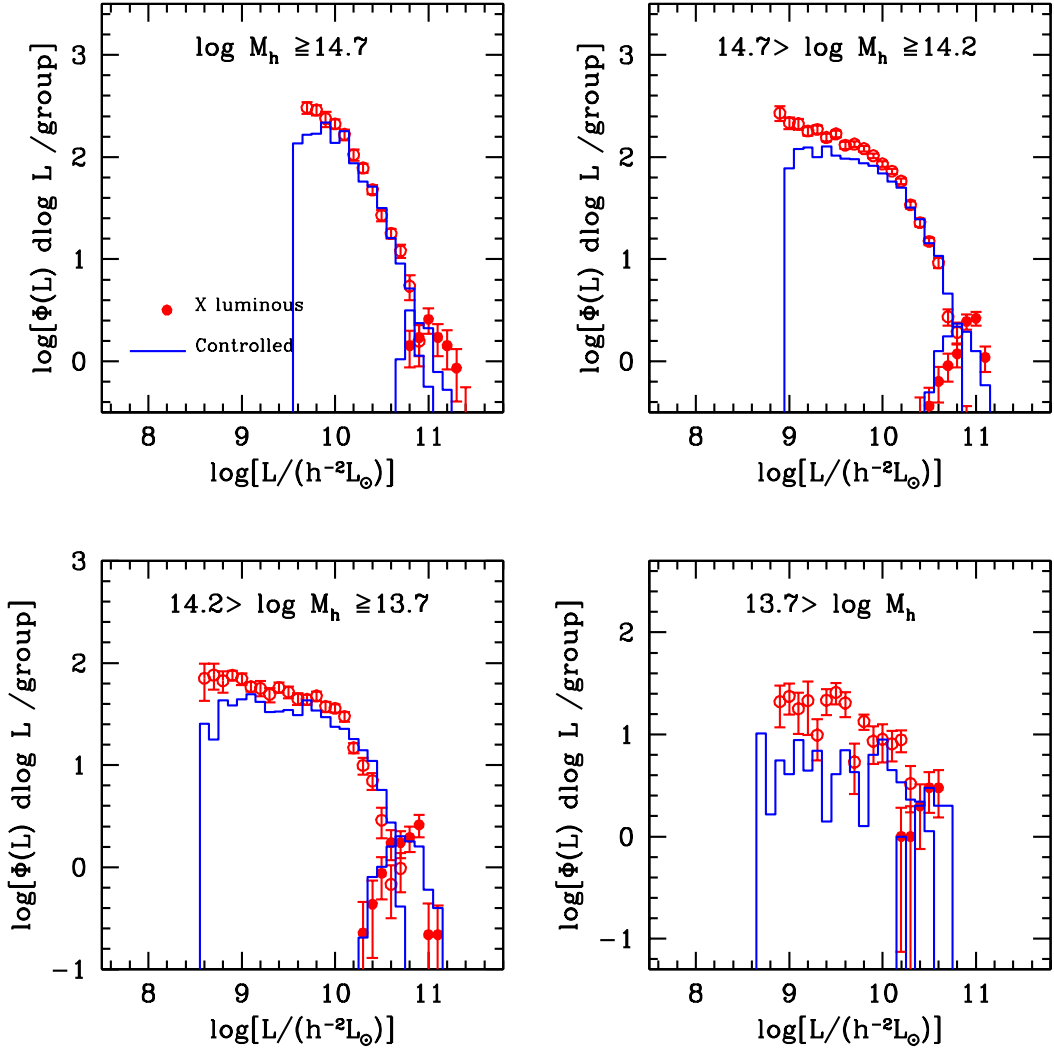}}
\caption{The conditional  stellar mass functions of groups  with (circles) and
  without (histograms)  strong X-ray  emissions.  The contribution  of central
  and satellite galaxies are plotted  separately.  The error bars are obtained
  from 200 bootstrap resampling of all the groups in consideration.}
\label{fig:CLF}
\end{figure*}

\begin{figure*}
\center{\includegraphics[height=7.0cm,width=6.5cm,angle=270]{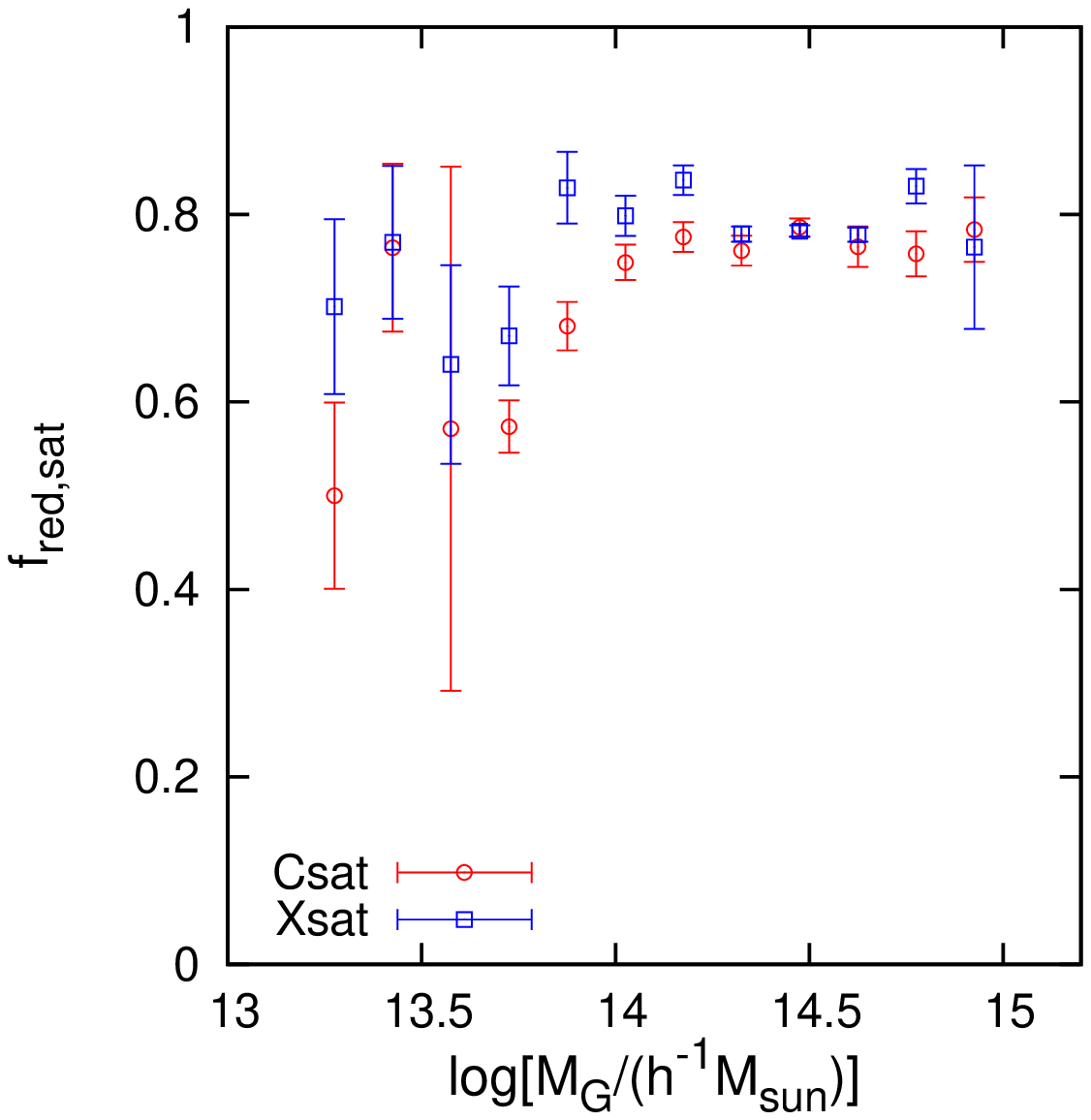}
\includegraphics[height=7.0cm,width=6.5cm,angle=270]{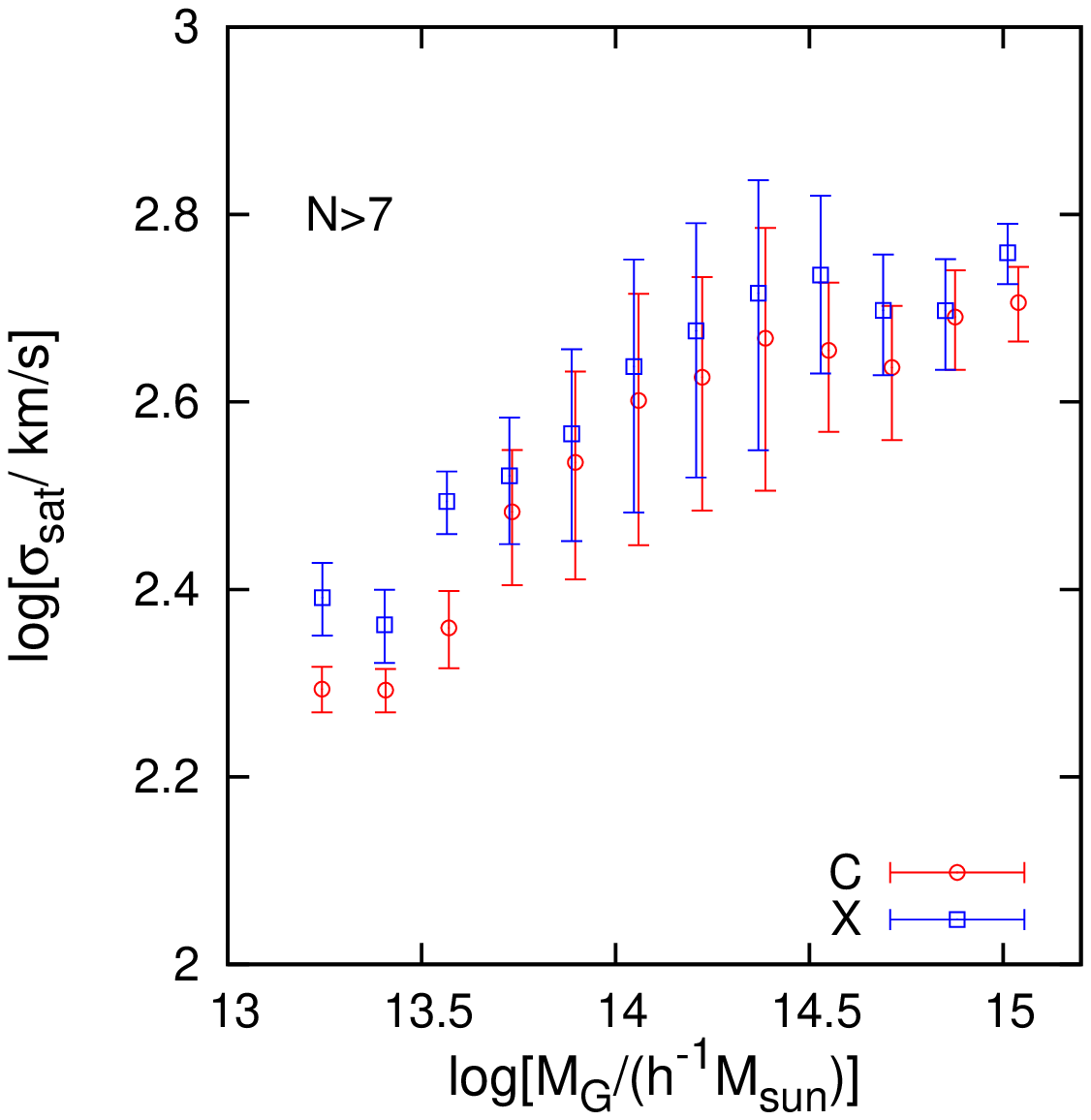}}
\caption{Left panel:  the red  fraction of satellite  galaxies in  groups with
  (squares)  and without  (circles) strong  X-ray emission.  Right  panel: the
  velocity  dispersion of  satellite  galaxies in  groups  with (squares)  and
  without (circles)  strong X-ray emission.  The error bars are  obtained from
  200 bootstrap resampling of all the groups in consideration. }
\label{fig:red_v}
\end{figure*}

\subsection{The difference between the groups with and without strong 
 X-ray emission} \label{sec:diff}

In this  subsection, we  proceed to  probe if those  groups and  clusters with
strong  X-ray emission have  different galaxy  populations from  those without
strong X-ray  emission. For this purpose,  we first check  the distribution of
X-ray clusters  with respect  to the  survey flux limits.   The left  panel of
Fig~\ref{fig:dist} plots the  X-ray luminosities of the 201  X-ray clusters in
our  sample versus  their redshifts.   The  dot-dashed and  dashed lines  mark
roughly   the    two   flux   limits   of   the    original   X-ray   samples,
$F_X=1.0\times10^{-12}\,{\rm   erg~cm^{-2}~s^{-1}}$  and   $F_X=   3.0  \times
10^{-12}\,   {\rm  erg~cm^{-2}~s^{-1}}$,   respectively.    Here  an   average
K-correction, assuming  $\Tx=5$ keV,  is used to  convert X-ray flux  to X-ray
luminosity (B\"ohringer  et al.  2004).   In order to  see whether or  not the
groups  without detected  X-ray emission  are indeed  distinct from  the X-ray
clusters,  we  plot in  the  middle  panel  of Fig.~\ref{fig:dist}  the  X-ray
luminosities,  inferred  from  the  scaling  relations with  $\sigma$  set  to
zero\footnote{As  Eqs. \ref{eq:Lx-Mxx}  and \ref{eq:Lx-Lxx}  yield  almost the
  same $L_X$,  we use the former to  make our prediction.}, for  all groups in
the SDSS  DR7 group catalog versus  the group redshifts.  The  groups that are
linked to X-ray  clusters are indicated using the same symbols  as in the left
panel, while  other groups are shown  as small dots.  As  an illustration, the
two curves indicating  the X-ray flux limits of the X-ray  data are plotted as
well.  Because  of the relatively large  scatter in the  scaling relation, the
groups with X-ray detections do not obey these `flux limits'. If there is zero
scatter in the  scaling relation, all groups above  these flux-limits would be
in  our  X-ray  cluster  sample,  while  those  below  it  would  have  evaded
detection. As  one can see,  some of our  X-ray clusters have  predicted $L_X$
that  are below  the flux  limits, and  so their  true X-ray  luminosities are
significantly  scattered upwards  relative to  the prediction.   On  the other
hand, a large number of groups  with predicted $L_X$ above the flux limits are
not detected in X-rays (by ROSAT).  This does not come as a surprise, as it is
well known  that the observed $\Lx$  contains a significant  amount of scatter
with respect  to their optical counterparts,  as modelled in  our full scaling
relations.  The  main point of  this exercise is  to demonstrate that  a large
number  of  (massive)  groups  apparently  have X-ray  luminosities  that  are
significantly below  those of the  201 X-ray clusters  in our sample,  a point
that has been  made numerous times before (e.g.,  Castander \etal 1994; Lubin,
Mulchaey \& Postman  2004; Stanek \etal 2006; Popesso  \etal 2007; Castellano
et al. 2011; Balogh et al.  2011).

To investigate the difference between the groups with and without strong X-ray
emissions  properly, we  construct a  controlled group  sample to  prevent (or
reduce) the influence of Malmquist bias. For an X-ray cluster at redshift $z$,
we first  obtain its model X-ray  luminosity $L_X$ using  the scaling relation
with zero scatter. We then  search, among all groups without X-ray detections,
the one that has the same (or similar) predicted X-ray luminosity and redshift
as the X-ray cluster  in question.  We do this for all  the X-ray clusters and
produce a  control sample of 201  groups without X-ray  detections.  Note that
because  of  the  RASS selection  effects,  not  all  the X-ray  clusters  are
detected.   And  the groups  not  linked with  known  X-ray  clusters are  not
necessary X-ray  under-luminous. To avoid  the false search of  the controlled
X-ray  under-luminous groups,  we require  that each  group in  the controlled
sample should  also fulfill the  following criteria: i)  one can not  find any
X-ray sources, on the RASS map (and  PSPC or HRI, if they are available), with
signal-to-noise  ratio $S/N >3.0$  around the  center of  this group  within a
radius of 30  arcmins or the size of its $r_{500c}$  whichever is larger.  And
ii) in the same  region one can not find any cluster  records in all published
ROSAT-based  catalogue.  Otherwise,  we reject  this group  until we  find the
control group that  fulfill those two criteria for  each X-ray cluster.  Since
the matched pairs  have similar predicted $L_X$, their  $M_{*c} + \beta M_{\rm
  sat}$ should  also be similar.  The right-hand  panel of Fig.~\ref{fig:dist}
shows the distribution  of these groups (triangles) in the  $L_X$ - $z$ plane,
compared to that of the X-ray clusters (other symbols).

Using  this   control  sample,  we   now  examine  whether  groups   that  are
under-luminous  in X-ray  emission have  a different  galaxy population  (in a
statistical sense)  from that  of X-ray luminous  groups of similar  masses at
similar  redshifts.   Fig. \ref{fig:central-MG}  shows  various optical  group
properties as  function of group  mass for both  the sample of  X-ray clusters
(filled  squares) and  our control  sample (open  circles). These  include the
stellar mass  of the  central galaxies (upper  left-hand panel),  the $r$-band
luminosity of the central galaxies (upper right-hand panel), the $^{0.1}(g-r)$
color of the  central galaxies (middle left-hand panel),  the concentration of
the  central galaxies  (middle right-hand  panels), and  the stellar  mass and
luminosity gaps  (lower left-hand and right-hand  panels, respectively).  None
of these  reveals any  indication for a  significant difference  between X-ray
luminous and X-ray under-luminous groups.
 
Next, we  compare the conditional luminosity  functions (CLF; see  Yang, Mo \&
van den Bosch 2003), of the two  samples. For this purpose we first divide the
groups in the control sample into four mass bins (A3-D3), using the same $M_G$
bins as  in A1-D1 for  the X-ray cluster  sample.  For each of  the subsamples
A1-D1 and A3-D3 we determine the CLF using the same method as outlined in Yang
\etal (2008).   The results  are shown in  Fig.~\ref{fig:CLF} as  symbols with
error  bars  for A1-D1  (filled  for centrals,  open  for  satellites) and  as
histograms  for  A3-D3, where  the  error bars  have  been  obtained from  200
bootstrap re-samplings  of all the groups in  consideration.  Different panels
correspond to  different subsamples (different  bins in halo mass,  $M_G$), as
indicated. Note that  since the halo masses of the  groups are estimated using
$M_{\rm st}$ for all member galaxies  with $\rmag \le -19.5$, as expected, the
CLFs between  the two samples at bright  end with $\rmag \le  -19.5$ are quite
similar in both the central and satellite components.  However, at the fainter
end, a significant difference between  the two samples is apparent: the groups
that are  X-ray luminous on  average have more  satellites than groups  of the
same mass that are under-luminous.

The  left-hand  panel  of  Fig.~\ref{fig:red_v}  shows the  red  fractions  of
satellite galaxies as function of group mass, $M_G$, for both group with (open
squares) and without (open circles)  strong X-ray emission.  Here we have used
the same  criteria as in Yang \etal  (2008) to separate the  galaxies into red
and blue  populations. There is a  weak indication that  groups without strong
X-ray  emission have  lower  red  fractions, especially  at  lower masses,  in
quantitative agreement with the findings by  Popesso et al. (2007) based on a
significantly  smaller sample  of  X-ray clusters.   The  right-hand panel  of
Fig.~\ref{fig:red_v} compares  the line-of-sight velocity  dispersion (see Y07
for the  detail of  this measurement)  of the satellite  galaxies for  the two
samples, as  function of the group mass  $M_G$. Here we have  only used groups
with at least 8 members.  For  haloes with $M_G \ga 10^{13.5} \msunh$ there is
no indication that $\sigma_{\rm sat}$ is different for systems with or without
strong X-ray  emission.  At the  low mass end,  however, there is a  hint that
groups  in the  X-ray under-luminous  control  sample have  smaller values  of
$\sigma_{\rm  sat}$ than  their X-ray  luminous counterparts.   However, since
this  is based  on only  a handful  of clusters,  larger samples  are required
before any definite conclusion can be reached.

Finally, we  check the large scale  environments of the groups  in our control
sample.  The  asterisks in  Fig.~\ref{fig:ratios} indicate the  average galaxy
surface number density  within $2.5\mpch <r_{\rm p} <  5\mpch$ (measured using
the method described  in Section~\ref{sec:sbgal}) as a function  of group mass
$M_G$.  A comparison with the  X-ray clusters (solid squares) shows that there
is no  indication that  groups or clusters  that are under-luminous  in X-rays
reside in  a different environment  than their X-ray luminous  counterparts of
the same mass.

\section{Conclusions}
\label{sec_summ}

Galaxy  clusters are the  largest known  gravitationally bound  objects. Apart
from their  power on the cosmological  studies, one can take  advantage of the
cross-identification between  X-ray clusters and optical  groups to understand
the  formation and  evolution  of galaxies  in  these densest  regions in  the
large-scale structure.  In this paper,  we have extracted and refined an X-ray
cluster sample from the ROSAT  broadband ($0.1$-$2.4$ keV) archive and matched
them to the  optical group catalogs constructed from the  SDSS DR7.  Since the
galaxy  groups  are  selected  from  the spectroscopic  data,  so  that  group
memberships  are   reliable  even   for  relatively  low-mass   systems,  this
cross-matched catalog  is useful  to probe galaxy  formation and  evolution in
clusters. With this cross-identified sample,  we have analyzed the optical and
X-ray properties of galaxy clusters, and the correlation between them. Our main
results are summarized as follows.

\begin{enumerate}
\item We  have made an eyeball check  of the central galaxies  that are linked
  with X-ray  clusters in the  SDSS DR7 sky  coverage. The optical  groups are
  then linked with the X-ray clusters according to their central galaxies.

\item We  have checked the general  correlation between the  optical and X-ray
  properties for all  the X-ray clusters, and found that  the stellar mass (or
  $r$-band  luminosity) of  the central  galaxy is correlated with the X-ray
  luminosity.

\item The characteristic  group stellar masses (or luminosity)  used in Y07 to
  estimate  the  halo masses  are  also in  good  correlation  with the  X-ray
  luminosity to  $\Lx^{0.46}$ (or  $\Lx^{0.43}$ for $r$-band  luminosity) with
  1-$\sigma$ scatter $\sim 0.67$ in $\log \Lx$.

\item Taking into observed X-ray flux limits and the quite large scatter
  ($\sigma\sim 0.67$) in the $\log \Lx$, We have obtained unbiased scaling
  relations between the X-ray luminosity and the group stellar masses (or
  luminosities) as: ${\log L_X} = -0.26 + 2.90 [\log (M_{\ast, c} + 0.26
    M_{\rm sat}) -12.0]$ (${\log L_X} = -0.15 + 2.38 [\log (L_{c} + 0.72
    L_{\rm sat}) -12.0]$).

\item We have compared two sets of halo masses for the X-ray clusters, and
  found that the cluster mass $M_{\rm X}$ estimated from their X-ray
  luminosity are in general agreement with the group mass estimated from the
  stellar mass, $M_{\rm G}$. Quite interestingly, the systematical difference
  between the two sets of halo mass can be used to make simple cosmological
  probes.
 
\item Dividing the clusters into  four subsamples of different $M_{\rm G}$ (or
  $M_{\rm X}$), we have investigated the surface number density of galaxies in
  and around  the X-ray  clusters.  We found  that X-ray clusters  with larger
  $M_{\rm  G}$ (or  $M_{\rm  X}$) live  in  more dense  regions  and are  more
  strongly clustered. The  strength is in general agreement  with the CDM halo
  model prediction.

\item By  comparing various  properties of groups  that are X-ray  luminous or
  under-luminous, we found that groups  linked with the X-ray clusters tend to
  have  more faint  member satellite  galaxies. The  X-ray luminous  groups in
  general have larger red fraction of satellite galaxies.

\item The last  but not the least, the  cross-identified X-ray cluster catalog
  with 201+2 entries is provided in Appendix~B to the public.

\end{enumerate}

\section*{Acknowledgements} 

We thank the anonymous referee for helpful comments that greatly improved the
presentation of this paper. This work is supported by grants from NSFC (Nos.
10821302, 10925314, 11128306, 11121062) and the CAS/SAFEA International
Partnership Program for Creative Research Teams (KJCX2-YW-T23).  HJM would
like to acknowledge the support of NSF AST-0908334. WY is supported by the
 Fundamental Research Funds for the Central Universities,NSF-10903020,
Research Foundation for Talented Scholars.


\appendix

\section{A. Check the duplications and merging pairs}

From the  total 206  X-ray cluster entries  in the  SDSS DR7 sky  coverage, we
checked their coordinates and found that the following X-ray cluster pairs are
un-resolvable in their X-ray images. In  addition, we find each of these pairs
points to the same central galaxy.

\begin{itemize}
\item  {\bf SDSS  J100031.02+440843.3}  and {\bf  RBS 0819}:  Cross-identified
  through name RX J1000.4+4409 (or RXC J1000.5+4409).
\item ${\bold{NGC\  4325}}$ and ${\bold{RX\ J1223.0+1037}}$:  The X-ray source
  RX J1223.0+1037  is probably  associated with NGC  4325 (see  Crawford \etal
  1999 and the  NED website for detail). 
\item   ${\bold{NSCS\  J145254+164255}}$   and  ${\bold{Abell\   1983}}$:  NED
  essential note shows that NSCS J145254 +164255 may be  associated with Abell
  1983.
\end{itemize}

The sources  in each of  the above pairs  are closely associated, but  are not
treated   as   the   same   X-ray   cluster  in   the   literature.    Through
cross-identification with the SDSS groups,  we argue that the clusters in each
pair actually belong to the same cluster.  To avoid double counting, we remove
the less massive cluster in each pair from our sample. 

Once X-ray clusters  are linked to central galaxies,  it is straightforward to
obtain their  group counterparts according  to the central galaxies.   In most
cases one  X-ray cluster is associated  with one optical  group.  However, the
following two  close X-ray cluster pairs  share the same  optical groups.  The
broadband  X-ray images do  show that  sources in  each of  these 2  pairs are
definitely  separated and  resolvable.  We  think  the two  pairs are  merging
clusters,  and only  keep  the bigger  one  in each  pair  for our 
investigations. However,  for completeness, the two less  massive clusters are
also included in our Appendix~B.

\begin{itemize}
\item ${\bold{NSCS\  J145254+164255}}$ and ${\bold{IC\  4516}}$: NED essential
  note shows that NSCS J145254+164255  may be associated with Abell 1983 which
  cross-identifies with IC  4516.  Note that IC 4516 is  a galaxy. The central
  galaxies of  NSCS J145254+164255 and  IC 4516 are  both located in  the same
  SDSS group (with group ID. 9).  $M_{\rm XL}/M_{\rm XS}=2.0$ (the footnote 
 'XL'/'XS' means the larger/smaller one).

\item ${\bold{NSC\ J160433+174311}}$  and ${\bold{Abell\ 2151E}}$: Abell 2151E
  is  the  subcluster  of ABELL  2151.   NED  essential  note shows  that  NSC
  J160433+174311  may be  associated with  Abell 2151.  More  importantly, the
  central galaxies of  Abell 2151E and NSC J160433+174311  are located closely
  in the same group (ID. 6).  $M_{\rm XL}/M_{\rm XS}=3.8$.
\end{itemize}

\section{B.  The catalog of X-ray cluster in the SDSS DR7 region}

\input{tab.tex}

{\noindent The contents of Table~\ref{tab} are as follows.}\hfill 
\vskip 0.2truecm

\noindent Column (1): Sequence number\\
Column (2): Cluster name (cross-identified by NED).\\
Column (3-4): Right ascension (J2000) and Declination (J2000) of X-ray
position in equatorial frame. Obtained from the literature (based on
ROSAT database).\\
Column (5): Cluster redshift. The subscript indicates the reference
(a: Ebeling et al. 1998; b: Ebeling et al. 2000; c: B\"ohringer et
al. 2000; d: B\"ohringer et al. 2004).  Column (6): Updated redshift
(see Section~\ref{sec:match}). The subscript indicates whether this
corresponds to the spectroscopic redshift of the cluster's central
galaxy (1), or whether it derives from the nearest neighbors (2).\\
Column (7): ICM gas temperature, $\Tx$ in units of ${\rm keV}$. A superscript 
`e' indicates that $\Tx$ has been estimated from the $\Tx-\Lx$ relation by the
author of the reference from which the redshift in Column (5) was taken.\\
Column (8): Intrinsic X-ray luminosity in the 0.1-2.4 keV band (in the
cluster rest frame) in units of  $10^{44} {\rm   erg~s^{-1}}$, 
taken from the same reference as the redshift in Column (5).\\
Column (9): the 1$\sigma$ fractional uncertainty for the count
rate, the X-ray flux and the X-ray luminosity.\\
Column (10): Cluster halo mass $\log [M_{\rm X}/ (h^{-1}M_{\odot})]$, which is
estimated using Eq. \ref{eq:convert}.\\
Column (11): Group ID (the ID of the optical group in the SDSS DR7
group catalog that is associated with the X-ray cluster).\\
Column (12): 10-based logarithm of the characteristic group
luminosity, $L_G$ [in $h^{-2}L_{\odot}$].\\
Column (13): 10-based logarithm of the group stellar mass, $M_{\rm
  st}$ [in $h^{-2}M_{\odot})$].\\
Column (14): 10-based logarithm of the group halo mass, $M_G$ [in $h^{-1}M_{\odot})$].\\
Column (15): ID of the cluster's central galaxy in the NYU-VAGC. For
clusters in which the central galaxy is also the most massive
galaxy, the ID has a subscript (1). A subscript (0) indicates that the
central galaxy is NOT the most massive galaxy in the cluster. \\
Column (16): Absolute magnitude of cluster's central galaxy in
$r$-band (K+E corrected to $z=0.1$, model magnitude).\\
Column (17): $^{0.1}(g-r)$ color of cluster's central galaxy. \\
Column (18): 10-based logarithm of the cluster's central galaxy,
$M_{\ast,c}$ [in $\msunhh$].\\
Column (19): Offset in arcmins between the central galaxy and the
X-ray cluster position listed in Columns (3) and (4).

\vskip 0.3truecm
{\noindent Quantities without reliable measurements are denoted by '$-$'. }

\begin{table}
\noindent
{\bf Notes on individual entries in Table~\ref{tab}:}\\
{\bf WBL 032} Two massive galaxies {\bf VV 377 NED01} and {\bf VV 382 NED01} appear to be
 equally dominant, separated in projection by 4.2 arcmin. C99 regards the latter
 as the central galaxy. However, Chandra image shows that the former galaxy is
 at the first maximum point (FMP) of the X-ray intensity of emission. We take the former
 as the central which is not the MMG.\\
{\bf ABELL 0168} Both XMM-Newton and Chandra images show {\bf ABELL 0168(N) (i.e. galaxy UGC 00797)}
 takes the place of {\bf ABELL 0168(S)} to be the X-ray FMP. C99 points that
 the X-ray image of the cluster is broad with no tight core, and the brightest
 cluster galaxy is clearly the central.\\
{\bf RXC J0736.4+3925} contains a non-MMG with QSO-like spectrum nearby the
 X-ray centroid ($\le 1$ arcmin). And the SDSS image shows the non-MMG seem to be an
 AGN or a galaxy overlapped by a star. In stead, we adopt {\bf 2MASX J07363812+3924525}
 as the central galaxy of this cluster, leading an offset 3.36 arcmin.\\
{\bf ABELL 0763} As discussed in C99, the same galaxy {\bf SDSS J091235.18+160000.6}
 is used as the central. \\
{\bf ABELL 0757} The X-ray image of the cluster has 2 maxima. The MMG (central)
 is at the FMP i.e. (138.2586,47.7059)--(in format R.A.,Dec. Hereafter the
 format like this stands for a source's position in J2000d; see also Table 6 in B\"ohringer et al. 2000)\\
{\bf SDSS J100031.02+440843.3} or {\bf RBS 0819} Both XMM-Newton and Chandra confirm
 the non-MMG {\bf SDSS J100031.00+440843.2} is the central of the cluster. \\
{\bf RX J1053.7+5450, SDSS-C4-DR3 3043} and {\bf RXC J1121.7+0249} These X-ray sources
 are all very extended and unfortunately no associated Chandra or XMM-Newton images.
 Centrals are the MMGs. \\
{\bf ABELL 1314} The central locates exactly at the X-ray centroid in Chandra and
XMM-Newton images, and is the same one adopted by C99 (173.7041,49.0776).\\ 
{\bf ABELL 1361} The X-ray image of the cluster has 2 maxima. Following C99, the
 central is at (175.9146,46.3561), also the FMP, and confirmed by
 Chandra image.\\
{\bf ABELL 1366, RXC J1229.9+1147, RBS 1198, ABELL 1612, RXC J1416.5+3045} and {\bf ABELL 1925} 
 are all very extended sources in the RASS image. The X-ray maxima for the first 3
 sources are determined by the Chandra and/or XMM-Newton high resolution images
 at (176.1535,67.4058),(187.5486,11.7444) and (194.8982,27.9596) respectively. 
Moreover, the centrals of the enrolled sources {\bf RXC J1229.9+1147} and {\bf RXC RBS 1198} in
 C99 locate at the very positions of the their (above) X-ray maxima. In the
 absence of higher resolution X-ray images, we choose the MMGs as centrals for the
 last 3 clusters.\\
{\bf ABELL 1367} is a very well-studied merging cluster. Member galaxies around the
 X-ray centroid are much less massive (at least one order of magnitude) than the
 BCG (NGC 3862, which is associated with the strong radio source 3C264, see C99).
 Although the BCG of this cluster is significantly offset $\sim 666$ kpc
 (11.61 arcmin $\ll {\rm r_{500c}}=33.61$ arcmin) from the center of the highly
 extended X-ray emission, we take it as the central galaxy of this X-ray cluster.\\
{\bf ABELL 1902, ABELL 1930} and {\bf ABELL 2033} are sources cataloged by C99.
 We use the galaxies provided by C99 as their centrals (also the MMGs), and the
 latter 2 cross-identifications are confirmed by ROSAT HRI and PSPC with a little
 offset $\le 0.5$ arcmin.\\
{\bf ABELL 2061} has a broad X-ray image. C99 suggests that the cluster is
 consist of 2 (Northern and Southern) components. And Chandra image shows {\bf ABELL 2061(N)} 
 (230.3354,30.6711) is the X-ray FMP, and also the location of its central.\\
{\bf MKW 03s} Images from 3 instruments (ROSAT/PSPC, XMM-Newton and Chandra) show
 that the X-ray FMP is at (230.4662,7.7089) where also lies the MMG of the cluster (see also C99).\\
{\bf ABELL 2067} The X-ray image from ROSAT/PSPC (230.7815,30.8718) supports the
 non-MMG {\bf 2MASX J15230842+3052387} as the cluster central.\\
{\bf ABELL 2069} Following C99 we choose {\bf 2MASX J15240741+2953203}
 (the northern component of a pair in contact) as the cluster central, which is
 also confirmed by a check according to the PSPC and Chandra images.\\
{\bf ABELL 2073} and {\bf ABELL 2169} Their RASS images are all very extended.
 Thus the MMGs are their centrals. \\
{\bf ABELL 2018} The cluster central is {\bf SDSS J154019.03+175123.3} (see C99).\\
{\bf ABELL 2124} Chandra image shows the X-ray FMP is at (236.2462,36.1097), the
 right position of the central galaxy {\bf UGC 10012} (see C99).\\
{\bf ABELL 2149} and {\bf RXC J1601.3+5354} are very close in projection but with
 very different redshifts (multiple redshift clustering in line-of-sight). And
 they have the same one X-ray image (very extended) in RASS. Here we are not
 sure whether the  foreground and background contamination on X-ray flux is
 taken into account in their reference papers.\\
{\bf ABELL 2147} C99 suggests {\bf UGC 10143} (240.5708,15.9750) as the optical
 counterpart of the X-ray centroid of the cluster, which is supported by the
 images from ROSAT/PSPC, XMM-Newton and Chandra.\\
{\bf ABELL 2148} and {\bf ABELL 2151E} For each of them, the optical
identification of the point-like object at the X-ray centroid (ROSAT/PSPC) is of a
star. In this case we use the galaxies corresponding to the second maximum points
 of X-ray intensity as the centrals. \\
{\bf ABELL 2255} Images from XMM-Newton and Chandra show that the X-ray FMP is
 at (258.1471,64.0624) where the nearest galaxy is {\bf ZwCl 1710.4+6401 A}, also
 the MMG.\\
{\bf RXC J2321.8+1505} The X-ray FMP on RASS images is at (350.4535,15.0849),
 very close to the MMG {\bf 2MASX J23214705+1504594}.\\
{\bf AWM 4} Instead of using its unreliable X-ray luminosity (suffering from
 very large uncertainty $0.979$) from B\"ohringer et al. (2000), we extract the
 record from Ebeling et al. (1998).

\end{table}

\end{document}

%% file: tab.tex
\setlength\tabcolsep{3pt}
\small{
\begin{sidetable}

\caption{The catalogue of 204 X-ray clusters and their associated groups and galaxies \label{tab}}
\begin{tabular}{llrrrrrrrrrrrrrrrrl} \hline\hline
\multicolumn{1}{c}{No.} &\multicolumn{1}{c}{X-ray cluster ID} & \multicolumn{1}{c}{R.A.}& \multicolumn{1}{c}{Dec} &
 \multicolumn{1}{c}{$z_X$} & \multicolumn{1}{c}{$z_X^c$} & \multicolumn{1}{c}{$T_X$} & \multicolumn{1}{c}{$L_X$}& \multicolumn{1}{c}{$L^{Err}_X$} &
 \multicolumn{1}{c}{$M_X$} & \multicolumn{1}{c}{Gr ID} & \multicolumn{1}{c}{$L_G$}& 
 \multicolumn{1}{c}{$M_{\rm st}$}& \multicolumn{1}{c}{$M_G$}& \multicolumn{1}{c}{Gal ID}
& \multicolumn{1}{c}{$M_r$}& \multicolumn{1}{c}{$color$}& \multicolumn{1}{c}{$M_*$}& \multicolumn{1}{c}{$Ofs$}\\ 

\multicolumn{1}{c}{(1)}&\multicolumn{1}{c}{(2)}&\multicolumn{1}{c}{(3)}&
\multicolumn{1}{c}{(4)}&\multicolumn{1}{c}{(5)}&\multicolumn{1}{c}{(6)}&
\multicolumn{1}{c}{(7)}&\multicolumn{1}{c}{(8)}&\multicolumn{1}{c}{(9)}
&\multicolumn{1}{c}{(10)}&\multicolumn{1}{c}{(11)}&\multicolumn{1}{c}{(12)}
&\multicolumn{1}{c}{(13)}&\multicolumn{1}{c}{(14)}&\multicolumn{1}{c}{(15)}
&\multicolumn{1}{c}{(16)}&\multicolumn{1}{c}{(17)}&\multicolumn{1}{c}{(18)}&\multicolumn{1}{c}{(19)}\\	\hline

  1&ABELL 0085                   & 10.4587& -9.3019&$ 0.0555_d$&$ 0.0554_1$&$ 6.90^e$& 5.581& 0.032&14.935&    11&12.062&12.468&14.575&$ 359832_1$&-22.797& 0.981&11.505& 0.12\\
  2&1RXS J011006.0+135849        & 17.5229& 13.9804&$ 0.0581_c$&$ 0.0583_1$&------& 0.076& 0.295&13.874&    93&11.729&12.156&14.291&$ 148784_1$&-21.787& 0.992&11.146& 0.60\\
  3&WBL 032                      & 18.2741& 15.5170&$ 0.0442_c$&$ 0.0469_1$&$ 2.30^e$& 0.223& 0.141&14.142&   100&11.729&12.250&14.379&$ 153588_0$&-21.971& 0.940&11.113& 2.48\\
  4&ABELL 0168                   & 18.8000&  0.3300&$ 0.0450_d$&$ 0.0448_1$&$ 2.60\ $& 0.497& 0.093&14.341&    25&11.893&12.287&14.410&$1765793_1$&-22.131& 0.974&11.223& 7.04\\
  5&SDSS CE J020.508463+00.332940& 20.4929&  0.3575&$ 0.1756_c$&$ 0.1745_1$&------& 1.095& 0.311&14.502& 20144&11.967&12.441&14.548&$2096076_1$&-22.967& 1.055&11.638& 1.76\\
  6&RXC J0137.2-0911             & 24.3140& -9.2028&$ 0.0409_d$&$ 0.0409_2$&------& 0.315& 0.084&14.229&   129&11.498&11.899&14.050&$3000001_1$&-21.773& 0.943&11.154& 0.31\\
  7&ABELL 0295                   & 30.5829& -1.1204&$ 0.0427_d$&$ 0.0425_1$&------& 0.155& 0.137&14.055&   244&11.400&11.842&13.991&$1736056_1$&-22.009& 0.991&11.200& 0.79\\
  8&MaxBCG J111.48808+41.38519   &111.5022& 41.3821&$ 0.1120_c$&$ 0.1113_1$&$ 4.30^e$& 1.204& 0.158&14.542&  1088&11.687&12.127&14.268&$1731094_1$&-22.959& 1.006&11.579& 0.87\\
  9&RXC J0736.4+3925             &114.1059& 39.4332&$ 0.1177_c$&$ 0.1180_1$&------& 2.575& 0.088&14.728&  2246&11.569&11.979&14.134&$ 230665_0$&-21.991& 0.235&10.786& 0.65\\
 10&UGCl 104                     &116.6554& 31.0136&$ 0.0579_c$&$ 0.0582_2$&------& 0.169& 0.210&14.071&   515&11.394&11.813&13.962&$ 833031_1$&-21.290& 0.943&10.848& 1.71\\
 11&1RXS J074809.3+183243        &117.0394& 18.5465&$ 0.0400_c$&$ 0.0467_1$&------& 0.146& 0.210&14.038&    34&11.819&12.239&14.367&$1140707_1$&-22.460& 0.968&11.342& 0.71\\
 12&WBL 154                      &117.8437& 50.2125&$ 0.0228_c$&$ 0.0238_2$&$ 1.80^e$& 0.079& 0.296&13.893&   213&11.129&11.594&13.722&$1783916_1$&-21.032& 0.930&10.855& 1.67\\
 13&ABELL 0598                   &117.8500& 17.5130&$ 0.1894_b$&$ 0.1865_1$&$ 6.50^e$& 3.091& 0.060&14.754&  5667&12.111&12.657&14.762&$1887146_1$&-22.237& 1.127&11.449& 0.29\\
 14&ABELL 0602                   &118.3510& 29.3660&$ 0.0621_a$&$ 0.0606_2$&$ 3.40^e$& 0.585& 0.080&14.377&   124&11.621&12.015&14.169&$ 804064_1$&-21.837& 1.014&11.136& 0.71\\
 15&ZwCl 0755.8+5408             &119.9284& 54.0016&$ 0.1032_c$&$ 0.1032_2$&$ 4.60^e$& 1.131& 0.173&14.529&   451&11.769&12.206&14.339&$3000002_1$&-22.599& 0.887&11.385& 0.64\\
 16&ABELL 0616                   &121.0915& 46.7823&$ 0.1868_c$&$ 0.1868_1$&------& 1.520& 0.367&14.579& 93613&11.664&12.159&14.293&$ 228159_1$&-22.757& 1.065&11.561& 0.33\\
 17&ABELL 0620                   &121.4304& 45.6903&$ 0.1353_c$&$ 0.1342_1$&------& 0.871& 0.209&14.456&  2623&11.543&12.117&14.260&$ 218899_1$&-21.906& 1.150&11.353& 0.36\\
 18&ABELL 0635                   &122.7599& 16.7349&$ 0.0925_c$&$ 0.0942_2$&------& 0.383& 0.285&14.264&  1883&11.173&11.746&13.889&$2131539_1$&-22.179& 1.148&11.453& 1.61\\
 19&RX J0820.9+0751              &125.2574&  7.8660&$ 0.1100_c$&$ 0.1101_2$&$ 4.40^e$& 1.055& 0.155&14.510&  1293&11.457&11.911&14.063&$1128297_1$&-21.704& 0.890&11.131& 0.21\\
 20&ZwCl 0822.8+4722             &126.3761& 47.1299&$ 0.1267_c$&$ 0.1290_1$&$ 7.10^e$& 3.070& 0.109&14.768&   263&12.379&12.833&15.004&$ 191508_1$&-23.420& 1.039&11.822& 0.37\\
 21&ABELL 0667                   &127.0190& 44.7640&$ 0.1450_a$&$ 0.1450_1$&$ 6.10^e$& 2.681& 0.070&14.731&   749&12.153&12.688&14.790&$ 436039_1$&-22.667& 1.084&11.590& 0.37\\
 22&RXC J0828.6+3025             &127.1621& 30.4280&$ 0.0503_c$&$ 0.0503_1$&$ 3.10^e$& 0.417& 0.096&14.297&    33&11.883&12.290&14.412&$1067272_1$&-22.523& 0.964&11.395& 1.83\\
 23&NSC J084254+292723           &130.7470& 29.4760&$ 0.1940_b$&$ 0.1937_1$&$ 7.00^e$& 3.889& 0.070&14.809& 43154&11.830&12.320&14.435&$1155061_1$&-22.765& 1.042&11.555& 1.39\\
 24&RXC J0844.9+4258             &131.2361& 42.9817&$ 0.0541_c$&$ 0.0540_1$&------& 0.086& 0.186&13.905&   955&11.242&11.662&13.797&$ 832301_1$&-21.957& 0.953&11.132& 0.31\\
 25&MaxBCG J136.60704+10.36365   &136.6140& 10.3450&$ 0.1328_b$&$ 0.1335_1$&$ 4.90^e$& 1.585& 0.070&14.604&   551&12.235&12.651&14.751&$1827579_1$&-22.486& 0.992&11.371& 1.19\\
 26&ABELL 0744                   &136.8570& 16.6540&$ 0.0733_b$&$ 0.0728_1$&$ 3.00^e$& 0.416& 0.060&14.290&   331&11.455&11.882&14.031&$2250692_1$&-22.365& 0.971&11.303& 1.30\\
 27&RXC J0909.1+1059             &137.2832& 10.9925&$ 0.1751_c$&$ 0.1763_1$&$ 8.10^e$& 5.198& 0.104&14.885&  6093&12.255&12.823&14.989&$2497007_1$&-22.996& 1.066&11.671& 1.60\\
 28&ABELL 0763                   &138.1240& 15.9430&$ 0.0851_a$&$ 0.0899_2$&$ 4.60^e$& 1.456& 0.090&14.595&  2594&11.067&11.500&13.616&$2305668_1$&-21.765& 0.949&11.046& 3.69\\
 29&ABELL 0757                   &138.3570& 47.6870&$ 0.0514_a$&$ 0.0513_1$&$ 3.10^e$& 0.481& 0.100&14.331&   170&11.554&11.969&14.122&$ 855554_1$&-21.453& 0.973&10.952& 4.67\\
 30&ABELL 0779                   &139.9220& 33.7630&$ 0.0230_b$&$ 0.0229_1$&$ 1.40^e$& 0.046& 0.070&13.762&    35&11.438&11.853&14.004&$1158309_1$&-22.136& 0.923&11.188& 1.61\\
 31&RXC J0920.0+0102             &140.0020&  1.0401&$ 0.0175_c$&$ 0.0170_1$&------& 0.005& 0.237&13.209&  3891&10.345&10.759&12.554&$ 184191_1$&-20.753& 0.922&10.621& 0.44\\
 32&3C 219                       &140.2857& 45.6437&$ 0.1745_c$&$ 0.1746_1$&------& 1.320& 0.216&14.548& 35702&11.326&11.752&13.895&$ 800906_1$&-22.025& 0.937&11.181& 0.33\\
 33&ABELL 0795                   &141.0238& 14.1684&$ 0.1357_c$&$ 0.1357_1$&$ 6.60^e$& 3.428& 0.096&14.794&   275&12.354&12.762&14.902&$2348261_1$&-22.429& 0.925&11.296& 0.27\\
 34&ABELL 0853                   &145.5605& 15.3865&$ 0.1664_c$&$ 0.1641_1$&$ 5.80^e$& 2.131& 0.140&14.669& 67569&11.546&12.065&14.214&$2484487_1$&-22.817& 1.065&11.594& 0.34\\
 35&ABELL 0845                   &146.0074& 64.4117&$ 0.1200_c$&$ 0.1205_1$&------& 0.848& 0.168&14.453&  3521&11.344&11.822&13.972&$1703417_1$&-22.456& 1.020&11.397& 0.60\\
 36&MaxBCG J149.55144+23.77931   &149.5430& 23.7820&$ 0.1471_c$&$ 0.1451_1$&------& 1.160& 0.262&14.524&  8110&11.692&12.107&14.250&$2239218_1$&-22.739& 0.999&11.491& 0.53\\
 37&SDSS J100031.02+440843.3     &150.1272& 44.1543&$ 0.1540_c$&$ 0.1532_1$&$ 5.20^e$& 1.655& 0.127&14.609&  2743&11.720&12.172&14.308&$ 894438_0$&-21.251& 0.973&10.868& 0.55\\
 38&NSCS J100242+324218          &150.6609& 32.6995&$ 0.0499_c$&$ 0.0505_1$&$ 2.60^e$& 0.296& 0.111&14.212&   161&11.322&11.760&13.903&$1884461_1$&-22.052& 0.985&11.222& 0.68\\
 39&ABELL 0923                   &151.6647& 25.9101&$ 0.1162_c$&$ 0.1168_1$&$ 4.40^e$& 1.158& 0.140&14.531&   851&11.809&12.261&14.387&$2173549_1$&-23.106& 0.988&11.615& 0.23\\
 40&ABELL 0961                   &154.0850& 33.6410&$ 0.1241_a$&$ 0.1272_1$&$ 5.20^e$& 1.899& 0.080&14.651&   729&11.969&12.395&14.502&$1890546_1$&-22.537& 0.962&11.359& 0.64\\
 41&ABELL 0964                   &154.1503& 24.8082&$ 0.0811_c$&$ 0.1701_2$&$ 3.00^e$& 2.129& 0.183&14.667&  4565&11.946&12.476&14.582&$3000003_1$&-22.630& 1.131&11.651& 0.45\\
 42&RX J1020.0+4100              &154.9992& 40.9873&$ 0.0922_c$&$ 0.0914_1$&$ 3.80^e$& 0.613& 0.164&14.381&   596&11.590&12.050&14.202&$1246123_1$&-22.532& 0.971&11.395& 1.93\\
 43&RX J1022.1+3830              &155.5196& 38.5120&$ 0.0491_c$&$ 0.0530_1$&$ 2.20^e$& 0.219& 0.283&14.137&   105&11.706&12.130&14.269&$1157141_0$&-21.251& 0.927&11.021& 1.45\\
 44&ABELL 0980                   &155.6170& 50.1210&$ 0.1582_a$&$ 0.1582_2$&$ 7.40^e$& 4.345& 0.090&14.846&   821&12.441&12.945&15.179&$3000004_1$&-22.986& 1.122&11.785& 0.93\\
 45&MaxBCG J155.91636+49.14401   &155.9220& 49.1329&$ 0.1440_c$&$ 0.1422_1$&$ 7.50^e$& 3.906& 0.082&14.824&  1413&11.979&12.427&14.531&$ 871988_1$&-22.838& 0.974&11.513& 0.75\\
 46&ZwCl 1023.3+1257             &156.4829& 12.6852&$ 0.1434_c$&$ 0.1423_1$&$ 6.20^e$& 2.787& 0.094&14.741&  5674&11.863&12.466&14.573&$3000005_1$&-23.646& 1.091&12.017& 0.52\\
 47&MaxBCG J157.93473+35.04138   &157.9317& 35.0495&$ 0.1259_c$&$ 0.1205_2$&$ 6.30^e$& 2.789& 0.111&14.747&   231&12.103&12.536&14.639&$1891363_1$&-22.269& 1.022&11.318& 0.52\\
 48&RXC J1032.2+4015             &158.0590& 40.2470&$ 0.0733_b$&$ 0.0776_1$&$ 3.20^e$& 0.568& 0.070&14.366&   387&11.541&11.971&14.123&$1184055_1$&-22.433& 0.972&11.347& 1.46\\
 49&ABELL 1045                   &158.7469& 30.6944&$ 0.1407_c$&$ 0.1374_1$&$ 5.40^e$& 1.873& 0.109&14.644&  8073&11.504&11.964&14.115&$2215203_1$&-22.953& 1.007&11.575& 0.23\\
 50&ABELL 1068                   &160.1829& 39.9481&$ 0.1372_c$&$ 0.1383_1$&$ 7.50^e$& 3.637& 0.083&14.808&  7318&11.602&12.023&14.177&$1182121_1$&-23.103& 0.946&11.587& 0.34\\

\end{tabular}
\end{sidetable}

\begin{sidetable}
\addtocounter{table}{-1}
\begin{center}
\caption{continued}
\begin{tabular}{llrrrrrrrrrrrrrrrll} \hline
\multicolumn{1}{c}{(1)}&\multicolumn{1}{c}{(2)}&\multicolumn{1}{c}{(3)}&
\multicolumn{1}{c}{(4)}&\multicolumn{1}{c}{(5)}&\multicolumn{1}{c}{(6)}&
\multicolumn{1}{c}{(7)}&\multicolumn{1}{c}{(8)}&\multicolumn{1}{c}{(9)}
&\multicolumn{1}{c}{(10)}&\multicolumn{1}{c}{(11)}&\multicolumn{1}{c}{(12)}
&\multicolumn{1}{c}{(13)}&\multicolumn{1}{c}{(14)}&\multicolumn{1}{c}{(15)}
&\multicolumn{1}{c}{(16)}&\multicolumn{1}{c}{(17)}&\multicolumn{1}{c}{(18)}&\multicolumn{1}{c}{(19)}\\	\hline

 51&RX J1053.7+5450              &163.4490& 54.8500&$ 0.0704_a$&$ 0.0716_1$&$ 3.30^e$& 0.588& 0.080&14.376&   146&11.712&12.114&14.256&$ 797988_1$&-21.476& 0.965&10.959& 2.99\\
 52&NSCS J105344+165124          &163.4530& 16.8420&$ 0.0856_b$&$ 0.0856_2$&$ 3.50^e$& 0.637& 0.070&14.392&   692&11.693&12.165&14.301&$3000006_1$&-23.386& 0.969&11.799& 0.64\\
 53&ABELL 1139                   &164.5434&  1.5865&$ 0.0398_d$&$ 0.0382_1$&$ 2.10^e$& 0.082& 0.200&13.897&    37&11.759&12.192&14.329&$ 278531_1$&-21.681& 1.031&11.262& 1.09\\
 54&ABELL 1132                   &164.6160& 56.7820&$ 0.1363_a$&$ 0.1351_1$&$ 7.10^e$& 3.919& 0.100&14.827&   708&12.166&12.650&14.750&$ 799723_1$&-22.674& 0.998&11.474& 1.30\\
 55&ABELL 1173                   &167.3282& 41.5624&$ 0.0763_c$&$ 0.0748_2$&$ 3.30^e$& 0.558& 0.116&14.362&   422&11.512&11.963&14.114&$1276712_1$&-21.837& 0.985&11.151& 0.86\\
 56&NGC 3551                     &167.4294& 21.7620&$ 0.0319_c$&$ 0.0318_1$&$ 1.90^e$& 0.089& 0.143&13.919&   330&11.143&11.547&13.671&$2306474_1$&-21.701& 0.886&11.016& 0.40\\
 57&ABELL 1185                   &167.6950& 28.7060&$ 0.0314_a$&$ 0.0331_1$&$ 3.90\ $& 0.158& 0.120&14.061&     7&11.911&12.339&14.453&$2205041_0$&-21.204& 0.890&10.754& 1.25\\
 58&ABELL 1190                   &167.9104& 40.8424&$ 0.0794_c$&$ 0.0781_1$&$ 3.80^e$& 1.070& 0.103&14.522&    62&11.963&12.396&14.502&$1274637_1$&-22.669& 0.899&11.343& 1.83\\
 59&ABELL 1201                   &168.2250& 13.4500&$ 0.1688_a$&$ 0.1681_1$&$ 6.90^e$& 3.714& 0.080&14.805&  1800&12.377&12.884&15.058&$1823070_1$&-23.180& 1.050&11.713& 0.86\\
 60&ABELL 1204                   &168.3324& 17.5937&$ 0.1706_c$&$ 0.1705_1$&$ 7.30^e$& 4.016& 0.108&14.823&461365&11.131&11.525&13.645&$2407813_1$&-22.316& 0.934&11.251& 0.19\\
 61&SDSS-C4-DR3 3043             &168.8865& 54.4350&$ 0.0691_c$&$ 0.0695_1$&------& 0.386& 0.138&14.273&   104&11.747&12.140&14.277&$ 965525_1$&-22.520& 0.935&11.326& 2.29\\
 62&RXC J1121.7+0249             &170.4280&  2.8184&$ 0.0468_c$&$ 0.0511_1$&------& 0.339& 0.118&14.245&    58&11.634&12.054&14.205&$ 288542_1$&-21.290& 0.991&11.061& 4.89\\
 63&SDSS-C4 3084                 &170.5604& 67.2129&$ 0.0560_c$&$ 0.0560_1$&------& 0.070& 0.203&13.854&  1253&10.911&11.328&13.408&$ 251609_1$&-21.271& 1.004&10.913& 0.59\\
 64&RBS 0976                     &170.8049& 19.5996&$ 0.1042_c$&$ 0.1103_2$&$ 4.70^e$& 1.113& 0.132&14.523& 12646&11.161&11.683&13.821&$2355104_1$&-22.398& 1.024&11.379& 0.66\\
 65&RXC J1123.9+2129             &170.9912& 21.4903&$ 0.1904_c$&$ 0.1975_2$&$ 7.50^e$& 4.296& 0.149&14.832&  3671&12.259&12.843&15.009&$2301899_1$&-22.158& 1.405&11.435& 0.69\\
 66&ABELL 1264                   &171.7530& 17.1260&$ 0.1267_b$&$ 0.1267_2$&$ 4.40^e$& 1.213& 0.060&14.540&   343&12.252&12.742&14.863&$3000007_1$&-22.715& 1.003&11.551& 0.31\\
 67&ABELL 1291                   &173.0817& 55.9789&$ 0.0527_c$&$ 0.0515_1$&$ 2.40^e$& 0.261& 0.123&14.181&   672&10.953&11.341&13.425&$ 835751_1$&-20.965& 0.982&10.781& 1.13\\
 68&ABELL 1302                   &173.3070& 66.3990&$ 0.1160_a$&$ 0.1160_2$&$ 5.10^e$& 1.737& 0.070&14.631&   299&11.921&12.396&14.503&$3000008_1$&-22.954& 1.044&11.695& 1.20\\
 69&ABELL 1314                   &173.7480& 49.0900&$ 0.0338_a$&$ 0.0333_1$&$ 5.00\ $& 0.138& 0.090&14.028&    43&11.540&11.943&14.093&$ 955566_1$&-21.820& 0.939&11.076& 2.66\\
 70&ABELL 1361                   &175.8762& 46.3845&$ 0.1167_c$&$ 0.1160_1$&$ 5.50^e$& 2.825& 0.349&14.752&  2051&11.474&11.935&14.085&$1168647_1$&-22.398& 0.948&11.336& 2.90\\
 71&ABELL 1367                   &176.1520& 19.7590&$ 0.0214_a$&$ 0.0216_1$&$ 3.50\ $& 0.846& 0.320&14.478&     3&11.917&12.327&14.441&$2330539_0$&-21.382& 0.941&10.905&11.61\\
 72&ABELL 1366                   &176.2020& 67.4130&$ 0.1159_a$&$ 0.1161_1$&$ 5.60^e$& 2.213& 0.070&14.691&   224&12.076&12.507&14.611&$ 229148_1$&-22.582& 0.936&11.371& 2.95\\
 73&ABELL 1413                   &178.8270& 23.4075&$ 0.1427_c$&$ 0.1427_2$&$ 8.90\ $& 6.364& 0.064&14.945&  5980&11.816&12.380&14.486&$3000009_1$&-23.573& 1.087&11.983& 0.19\\
 74&ZwCl 1154.2+2435             &179.2409& 24.2581&$ 0.1392_c$&$ 0.1395_2$&$ 5.20^e$& 1.414& 0.129&14.574&  3035&11.628&12.084&14.229&$2263177_1$&-22.252& 0.979&11.276& 0.52\\
 75&ABELL 1437                   &180.1057&  3.3336&$ 0.1339_c$&$ 0.1339_2$&$ 7.40^e$& 3.889& 0.086&14.826&  1256&11.971&12.481&14.589&$3000010_1$&-23.239& 1.096&11.861& 0.81\\
 76&RX J1201.9+5802              &180.4997& 58.0475&$ 0.1031_c$&$ 0.1060_2$&$ 3.70^e$& 0.807& 0.193&14.445&   714&11.564&11.994&14.152&$1200250_0$&-22.037& 0.927&11.160& 1.21\\
 77&2MASX J12025923+2836444      &180.7615& 28.6039&$ 0.1341_c$&$ 0.1348_1$&------& 1.089& 0.157&14.511&  2182&11.591&12.036&14.189&$2244173_1$&-21.544& 0.988&11.081& 1.01\\
 78&NSC J120403+280727           &181.0133& 28.1233&$ 0.1631_c$&$ 0.1668_1$&------& 1.016& 0.193&14.485&  1104&12.462&12.890&15.065&$2243047_1$&-23.005& 1.034&11.626& 0.33\\
 79&MKW 04                       &181.1049&  1.9005&$ 0.0199_d$&$ 0.0197_1$&$ 1.70\ $& 0.172& 0.060&14.085&    64&11.339&11.814&13.963&$ 265118_1$&-21.821& 0.924&11.119& 0.55\\
 80&RBS 1066                     &181.3021& 39.3493&$ 0.0381_c$&$ 0.0371_1$&$ 2.30^e$& 0.315& 0.074&14.231&  1202&10.980&11.392&13.490&$1850962_0$&-20.425& 0.992&10.566& 0.55\\
 81&NGC 4104 GROUP               &181.6470& 28.1800&$ 0.0283_a$&$ 0.0282_1$&$ 1.80^e$& 0.109& 0.090&13.970&    85&11.239&11.670&13.807&$2243092_1$&-22.105& 0.941&11.245& 0.98\\
 82&ZwCl 1207.5+0542             &182.5783&  5.3850&$ 0.0748_c$&$ 0.0762_1$&$ 3.50^e$& 0.797& 0.104&14.450&   265&11.642&12.070&14.218&$ 460946_1$&-22.623& 0.957&11.393& 0.50\\
 83&ZwCl 1215.1+0400             &184.4192&  3.6624&$ 0.0766_c$&$ 0.0768_1$&$ 5.58^e$& 2.851& 0.054&14.764&    44&12.033&12.472&14.579&$ 275461_1$&-22.192& 0.971&11.303& 0.41\\
 84&NGC 4325                     &185.7772& 10.6240&$ 0.0258_c$&$ 0.0255_1$&$ 1.80^e$& 0.102& 0.078&13.955&   555&10.838&11.254&13.308&$ 947351_1$&-21.379& 0.961&10.895& 0.17\\
 85&RXC J1225.2+3213             &186.3001& 32.2291&$ 0.0594_c$&$ 0.0592_1$&------& 0.290& 0.125&14.204&   570&11.230&11.653&13.788&$1947851_1$&-22.245& 0.983&11.270& 0.39\\
 86&ABELL 1541                   &186.8667&  8.8290&$ 0.0896_c$&$ 0.0855_2$&------& 0.448& 0.197&14.305&   220&11.830&12.251&14.380&$ 936475_1$&-22.420& 0.976&11.355& 0.73\\
 87&MaxBCG J186.96340+63.38475   &186.9603& 63.3830&$ 0.1454_c$&$ 0.1455_1$&------& 1.259& 0.155&14.544&  1557&11.863&12.319&14.434&$ 558570_1$&-21.787& 1.069&11.263& 0.22\\
 88&RXC J1229.9+1147             &187.4966& 11.7891&$ 0.0852_c$&$ 0.0915_2$&------& 0.722& 0.340&14.421&  1063&11.499&11.912&14.065&$1219366_1$&-22.462& 0.968&11.357& 4.11\\
 89&ABELL 1553                   &187.7000& 10.5560&$ 0.1652_a$&$ 0.1705_1$&$ 7.40^e$& 4.655& 0.080&14.860&  2765&12.406&12.820&14.980&$ 947413_1$&-23.436& 0.981&11.739& 0.62\\
 90&ABELL 1589                   &190.3250& 18.5510&$ 0.0718_a$&$ 0.0704_1$&$ 4.60^e$& 1.251& 0.110&14.562&    30&12.109&12.522&14.625&$2418936_1$&-22.626& 1.001&11.453& 1.42\\
 91&ABELL 1612                   &191.9300& -2.7921&$ 0.1797_d$&$ 0.1818_1$&------& 2.691& 0.340&14.721&  2617&12.327&12.873&15.037&$ 174632_1$&-22.232& 1.143&11.452& 3.09\\
 92&MACS J1255.5+3521            &193.8781& 35.3602&$ 0.1585_c$&$ 0.1614_1$&------& 0.702& 0.222&14.395&  2976&12.170&12.618&14.714&$1900466_1$&-23.035& 1.037&11.650& 0.81\\
 93&ABELL 1650                   &194.6712& -1.7569&$ 0.0845_d$&$ 0.0845_2$&$ 6.70^e$& 3.863& 0.061&14.837&    89&12.018&12.469&14.576&$3000011_1$&-23.054& 0.994&11.694& 0.29\\
 94&RBS 1198                     &194.9294& 27.9386&$ 0.0231_c$&$ 0.0239_1$&$ 8.00^e$& 3.910& 0.124&14.855&     1&12.182&12.591&14.693&$2243552_1$&-22.147& 0.935&11.210& 2.22\\
 95&ABELL 1663                   &195.7112& -2.5062&$ 0.0847_d$&$ 0.0823_1$&------& 0.754& 0.219&14.435&    84&11.980&12.408&14.514&$ 166440_1$&-22.627& 0.993&11.444& 0.77\\
 96&1RXS J130303.2+575623        &195.7614& 57.9419&$ 0.1961_c$&$ 0.1953_2$&------& 1.482& 0.184&14.570&  7144&12.140&12.645&14.746&$1201353_0$&-22.042& 1.020&11.225& 2.15\\
 97&ABELL 1668                   &195.9398& 19.2715&$ 0.0643_c$&$ 0.0635_1$&$ 3.90^e$& 0.946& 0.081&14.495&    96&11.764&12.204&14.338&$2361168_1$&-22.521& 0.997&11.393& 0.26\\
 98&ABELL 1672                   &196.1147& 33.5920&$ 0.1882_c$&$ 0.1873_1$&$ 6.30^e$& 2.635& 0.153&14.715&  5719&12.081&12.584&14.685&$1895982_1$&-22.626& 1.010&11.452& 0.31\\
 99&ABELL 1677                   &196.4778& 30.9065&$ 0.1832_c$&$ 0.1832_2$&$ 6.90^e$& 3.369& 0.116&14.776&  3581&12.182&12.717&14.835&$3000012_1$&-22.666& 1.049&11.577& 0.98\\
100&MS 1306.7-0121               &197.3208& -1.6126&$ 0.0880_d$&$ 0.0833_1$&------& 0.914& 0.151&14.482&   545&11.504&11.942&14.092&$ 169619_1$&-22.526& 0.988&11.395& 0.80\\
101&RX J1311.1+3913              &197.7710& 39.2220&$ 0.0720_b$&$ 0.0723_1$&$ 3.10^e$& 0.491& 0.060&14.331&    46&12.079&12.505&14.609&$1836950_1$&-22.855& 0.968&11.496& 0.93\\

\end{tabular}
\end{center}
\end{sidetable}

\begin{sidetable}
\addtocounter{table}{-1}
\begin{center}
\caption{continued}
\begin{tabular}{llrrrrrrrrrrrrrrrrrrrrrl} \hline
\multicolumn{1}{c}{(1)}&\multicolumn{1}{c}{(2)}&\multicolumn{1}{c}{(3)}&
\multicolumn{1}{c}{(4)}&\multicolumn{1}{c}{(5)}&\multicolumn{1}{c}{(6)}&
\multicolumn{1}{c}{(7)}&\multicolumn{1}{c}{(8)}&\multicolumn{1}{c}{(9)}
&\multicolumn{1}{c}{(10)}&\multicolumn{1}{c}{(11)}&\multicolumn{1}{c}{(12)}
&\multicolumn{1}{c}{(13)}&\multicolumn{1}{c}{(14)}&\multicolumn{1}{c}{(15)}
&\multicolumn{1}{c}{(16)}&\multicolumn{1}{c}{(17)}&\multicolumn{1}{c}{(18)}&\multicolumn{1}{c}{(19)}\\	\hline
102&ABELL 1689                   &197.8750& -1.3354&$ 0.1832_d$&$ 0.1832_2$&$ 9.23^e$&14.089& 0.080&15.130& 15382&11.892&12.448&14.554&$3000013_1$&-22.735& 1.119&11.680& 0.36\\
103&MaxBCG J197.94248+22.02702   &197.9300& 22.0267&$ 0.1716_c$&$ 0.1715_1$&------& 0.998& 0.173&14.479&  4617&12.280&12.736&14.857&$2307337_1$&-23.036& 1.003&11.613& 0.75\\
104&NSCS J132014+330824          &200.0350& 33.1430&$ 0.0362_a$&$ 0.0361_1$&$ 2.00^e$& 0.136& 0.080&14.023&    71&11.574&11.994&14.151&$1896056_0$&-21.444& 0.971&10.945& 1.58\\
105&ABELL 1716                   &200.2374& 33.9041&$ 0.1820_c$&$ 0.1820_2$&------& 1.701& 0.150&14.608&  5724&12.113&12.604&14.703&$3000014_1$&-22.395& 1.006&11.420& 0.42\\
106&RXC J1323.5+1117             &200.8760& 11.2960&$ 0.0911_b$&$ 0.0895_1$&$ 3.60^e$& 0.690& 0.080&14.411&   598&11.545&11.974&14.128&$1223762_1$&-22.539& 0.942&11.339& 0.52\\
107&NGC 5129                     &201.0497& 13.9792&$ 0.0230_c$&$ 0.0230_1$&------& 0.036& 0.171&13.701&   152&11.150&11.530&13.651&$1829896_1$&-21.918& 0.920&11.078& 0.50\\
108&ABELL 1744                   &201.4572& 59.3225&$ 0.1515_c$&$ 0.1509_1$&$ 5.30^e$& 1.841& 0.099&14.636&  3214&11.825&12.319&14.434&$ 922325_1$&-22.812& 1.025&11.541& 0.47\\
109&SDSS CE J201.573563+00.213468&201.5743&  0.2257&$ 0.0826_d$&$ 0.0822_1$&$ 4.00^e$& 0.988& 0.117&14.501&   659&11.505&11.986&14.142&$ 244060_1$&-22.626& 1.005&11.546& 0.26\\
110&ABELL 1750                   &202.7081& -1.8728&$ 0.0852_d$&$ 0.0879_1$&------& 2.440& 0.116&14.723&   703&11.482&11.948&14.098&$ 571059_1$&-22.185& 1.013&11.285& 0.67\\
111&MaxBCG J203.14997+54.31696   &203.1671& 54.3205&$ 0.1017_c$&$ 0.1066_1$&------& 0.776& 0.140&14.435&  2779&11.309&11.768&13.912&$ 969459_1$&-22.603& 0.990&11.428& 1.05\\
112&ABELL 1767                   &204.0255& 59.2079&$ 0.0701_c$&$ 0.0701_2$&$ 4.10\ $& 1.429& 0.054&14.595&    48&11.955&12.392&14.499&$3000015_1$&-22.654& 0.986&11.533& 0.56\\
113&RXC J1339.5+1830             &204.8952& 18.5122&$ 0.1140_c$&$ 0.1109_1$&------& 0.364& 0.215&14.247&450979&10.481&10.920&12.803&$2361474_1$&-21.379& 0.943&10.897& 0.79\\
114&ABELL 1775                   &205.4740& 26.3720&$ 0.0724_c$&$ 0.0755_1$&$ 4.90\ $& 1.687& 0.167&14.635&    69&11.902&12.351&14.463&$1993364_1$&-22.579& 0.961&11.531& 1.16\\
115&ABELL 1773                   &205.5228&  2.2275&$ 0.0765_d$&$ 0.0734_2$&$ 3.90^e$& 0.786& 0.129&14.447&   262&11.655&12.079&14.225&$ 268720_1$&-22.038& 0.962&11.182& 1.04\\
116&ABELL 1795                   &207.2207& 26.5956&$ 0.0622_c$&$ 0.0633_1$&$ 5.10\ $& 5.572& 0.028&14.933&    28&11.983&12.392&14.498&$2010099_1$&-22.710& 0.842&11.358& 0.20\\
117&ABELL 1804                   &207.2582& 49.3047&$ 0.1665_c$&$ 0.1678_1$&------& 0.914& 0.202&14.459&  1457&12.306&12.717&14.836&$1369451_1$&-22.411& 1.060&11.424& 0.54\\
118&NSCS J134935+280633          &207.3402& 28.1036&$ 0.0748_c$&$ 0.0748_2$&$ 5.10^e$& 1.293& 0.082&14.569&    65&11.985&12.437&14.541&$3000016_1$&-22.682& 0.969&11.523& 0.53\\
119&RXC J1351.7+4622             &207.9398& 46.3668&$ 0.0625_c$&$ 0.0625_1$&------& 0.297& 0.150&14.210&    83&11.796&12.212&14.344&$1208324_0$&-22.152& 0.967&11.239& 0.92\\
120&RXC J1353.0+0509             &208.2750&  5.1580&$ 0.0790_a$&$ 0.0789_1$&$ 3.90^e$& 0.882& 0.090&14.474&    55&12.025&12.476&14.583&$ 527307_1$&-22.739& 0.998&11.496& 0.51\\
121&ABELL 1814                   &208.5095& 14.9231&$ 0.1251_c$&$ 0.1268_1$&$ 5.00^e$& 1.389& 0.160&14.573&   459&12.032&12.469&14.576&$2380213_1$&-22.453& 0.985&11.381& 0.73\\
122&ABELL 1831                   &209.8020& 27.9780&$ 0.0612_a$&$ 0.0750_2$&$ 4.20^e$& 1.573& 0.120&14.618&   143&11.780&12.237&14.365&$3000017_1$&-22.746& 0.961&11.539& 0.66\\
123&ABELL 1885                   &213.4313& 43.6634&$ 0.0890_c$&$ 0.0909_1$&$ 4.60^e$& 1.029& 0.078&14.509&   900&11.391&11.841&13.991&$1204948_1$&-21.781& 1.002&11.118& 1.12\\
124&ABELL 1882                   &213.8092& -0.5010&$ 0.1403_d$&$ 0.1389_1$&------& 2.018& 0.200&14.662&   287&12.522&12.964&15.222&$  65728_1$&-23.009& 1.030&11.640& 1.52\\
125&SDSS CE J213.951309+00.256928&213.9650&  0.2589&$ 0.1259_c$&$ 0.1262_1$&------& 0.479& 0.224&14.311&  3770&11.312&11.745&13.887&$  75646_1$&-21.602& 0.978&11.004& 0.70\\
126&RXC J1416.5+3045             &214.1354& 30.7621&$ 0.1860_c$&$ 0.1840_1$&------& 0.904& 0.232&14.452& 20971&11.900&12.402&14.508&$1919023_1$&-22.390& 1.091&11.462& 2.20\\
127&RBS 1380                     &215.3981& 49.5519&$ 0.0716_c$&$ 0.0719_1$&$ 3.60^e$& 0.816& 0.080&14.457&   339&11.579&12.040&14.192&$1380642_1$&-22.394& 0.997&11.370& 0.08\\
128&ABELL 1902                   &215.4226& 37.2958&$ 0.1813_c$&$ 0.1574_1$&$ 6.60^e$& 2.642& 0.108&14.724& 18659&11.286&11.745&13.887&$1292787_1$&-22.221& 0.983&11.255& 2.14\\
129&RX J1423.1+2615              &215.7922& 26.2556&$ 0.0375_c$&$ 0.0372_1$&$ 1.80^e$& 0.052& 0.164&13.786&   103&11.306&11.775&13.920&$1971369_1$&-21.612& 1.006&11.207& 0.61\\
130&RBS 1385                     &215.9685& 40.2619&$ 0.0822_c$&$ 0.0822_1$&$ 3.20^e$& 0.375& 0.131&14.262&   595&11.401&11.852&14.002&$1245278_1$&-22.212& 0.989&11.279& 0.29\\
131&MaxBCG J216.34368+63.19819   &216.3447& 63.1872&$ 0.1394_c$&$ 0.1394_1$&$ 5.80^e$& 2.664& 0.112&14.731&  2250&11.878&12.431&14.533&$3000019_1$&-23.134& 1.120&11.844& 0.66\\
132&ABELL 1914                   &216.5068& 37.8271&$ 0.1712_c$&$ 0.1700_1$&$10.53^e$& 9.364& 0.052&15.032&  2411&12.359&12.891&15.073&$1296192_1$&-23.126& 1.031&11.674& 1.40\\
133&ABELL 1925                   &217.1171& 56.8829&$ 0.1051_c$&$ 0.1060_1$&$ 3.90^e$& 0.987& 0.171&14.495&   239&11.984&12.428&14.531&$ 990388_1$&-23.078& 0.960&11.573& 2.88\\
134&ABELL 1927                   &217.7794& 25.6388&$ 0.0908_c$&$ 0.0964_2$&$ 4.40^e$& 1.378& 0.105&14.580&   456&11.676&12.133&14.273&$1980477_1$&-21.997& 0.940&11.177& 0.31\\
135&ABELL 1930                   &218.1200& 31.6330&$ 0.1313_a$&$ 0.1313_2$&$ 5.80^e$& 2.362& 0.080&14.703&  2453&11.675&12.198&14.333&$3000020_1$&-22.980& 1.060&11.718& 2.43\\
136&WBL 518                      &220.1592&  3.4765&$ 0.0263_c$&$ 0.0273_1$&$ 3.29^e$& 0.198& 0.087&14.118&    22&11.597&12.034&14.188&$ 487370_1$&-22.001& 0.970&11.168& 1.33\\
137&NSC J144215+221740           &220.5768& 22.3048&$ 0.0901_c$&$ 0.0972_2$&$ 4.80^e$& 1.463& 0.090&14.594&   684&11.588&12.024&14.178&$1989751_1$&-22.400& 0.988&11.356& 0.26\\
138&ABELL 1978                   &222.7750& 14.6110&$ 0.1460_a$&$ 0.1460_1$&$ 6.00^e$& 2.631& 0.080&14.726&   861&12.162&12.579&14.678&$2385566_1$&-22.782& 0.972&11.472& 0.85\\
139&NSCS J145254+164255          &223.2449& 16.6998&$ 0.0444_c$&$ 0.0440_2$&$ 2.50^e$& 0.250& 0.155&14.172&     9&11.995&12.399&14.506&$2345992_0$&-21.297& 0.938&10.857& 0.88\\
140&ABELL 1986                   &223.2798& 21.8947&$ 0.1186_c$&$ 0.1170_2$&$ 4.10^e$& 0.886& 0.282&14.465&   162&12.120&12.566&14.664&$2005706_0$&-21.650& 1.036&11.143& 0.30\\
141&IC 4516                      &223.6166& 16.3704&$ 0.0456_c$&$ 0.0453_1$&------& 0.089& 0.415&13.917&     9&11.995&12.399&14.506&$2365986_1$&-22.459& 0.988&11.378& 1.45\\
142&ABELL 1991                   &223.6309& 18.6420&$ 0.0586_c$&$ 0.0592_1$&$ 5.40\ $& 0.804& 0.090&14.456&    42&11.961&12.372&14.479&$2290443_1$&-22.632& 0.971&11.408& 0.03\\
143&ABELL 2009                   &225.0850& 21.3620&$ 0.1530_a$&$ 0.1530_2$&$ 7.80\ $& 5.367& 0.100&14.900&  8427&11.799&12.355&14.466&$3000021_1$&-22.980& 1.087&11.744& 0.50\\
144&ABELL 2034                   &227.5489& 33.5147&$ 0.1130_c$&$ 0.1130_2$&$ 7.10^e$& 3.686& 0.064&14.818&   367&11.953&12.455&14.561&$3000022_1$&-22.839& 1.049&11.656& 1.69\\
145&ABELL 2029                   &227.7290&  5.7200&$ 0.0766_a$&$ 0.0766_2$&$ 7.80\ $& 8.391& 0.260&15.030&    12&12.242&12.666&14.772&$3000023_1$&-23.390& 0.924&11.755& 1.51\\
146&ABELL 2036                   &227.7761& 18.0437&$ 0.1161_c$&$ 0.1158_1$&$ 4.10^e$& 0.862& 0.171&14.458&  1353&11.629&12.085&14.230&$2292145_1$&-22.407& 1.010&11.365& 0.97\\
147&ABELL 2033                   &227.8480&  6.3190&$ 0.0817_a$&$ 0.0810_1$&$ 4.70^e$& 1.390& 0.100&14.586&   243&11.695&12.147&14.282&$1366833_1$&-22.562& 1.027&11.521& 1.96\\
148&SDSS-C4-DR3 1355             &227.8897&  1.7642&$ 0.0384_d$&$ 0.0396_1$&------& 0.090& 0.585&13.922&   145&11.466&11.849&14.000&$ 269480_1$&-21.625& 0.953&10.992& 0.44\\
149&ABELL 2046                   &228.1553& 34.8601&$ 0.1489_c$&$ 0.1489_2$&------& 0.792& 0.176&14.429&  4285&11.925&12.432&14.535&$3000024_1$&-23.109& 1.051&11.757& 0.13\\
150&1RXS J151247.3-012753        &228.2127& -1.4798&$ 0.1223_d$&$ 0.1216_1$&------& 1.403& 0.185&14.577&  1393&11.696&12.173&14.308&$ 577757_1$&-22.538& 1.031&11.435& 0.51\\
151&ABELL 2052                   &229.1834&  7.0185&$ 0.0353_c$&$ 0.0342_1$&$ 3.40\ $& 1.276& 0.036&14.576&    24&11.786&12.211&14.343&$1338000_1$&-22.185& 0.964&11.252& 0.22\\
152&ABELL 2055                   &229.6899&  6.2312&$ 0.1021_c$&$ 0.1021_1$&$ 6.10^e$& 2.197& 0.097&14.693&   223&11.923&12.351&14.464&$1333999_0$&-21.929& 0.803&11.094& 0.07\\
\end{tabular}
\end{center}
\end{sidetable}

\begin{sidetable}
\addtocounter{table}{-1}
\begin{center}
\caption{continued}
\begin{tabular}{llrrrrrrrrrrrrrrrrrrrrrl} \hline\hline
\multicolumn{1}{c}{(1)}&\multicolumn{1}{c}{(2)}&\multicolumn{1}{c}{(3)}&
\multicolumn{1}{c}{(4)}&\multicolumn{1}{c}{(5)}&\multicolumn{1}{c}{(6)}&
\multicolumn{1}{c}{(7)}&\multicolumn{1}{c}{(8)}&\multicolumn{1}{c}{(9)}
&\multicolumn{1}{c}{(10)}&\multicolumn{1}{c}{(11)}&\multicolumn{1}{c}{(12)}
&\multicolumn{1}{c}{(13)}&\multicolumn{1}{c}{(14)}&\multicolumn{1}{c}{(15)}
&\multicolumn{1}{c}{(16)}&\multicolumn{1}{c}{(17)}&\multicolumn{1}{c}{(18)}&\multicolumn{1}{c}{(19)}\\	\hline

153&ABELL 2064                   &230.2271& 48.6693&$ 0.1076_c$&$ 0.0738_1$&$ 5.30^e$& 0.774& 0.102&14.443&   212&11.680&12.081&14.227&$ 999416_1$&-23.021& 0.926&11.515& 0.76\\
154&ABELL 2061                   &230.3210& 30.6400&$ 0.0777_a$&$ 0.0788_1$&$ 5.60^e$& 2.237& 0.150&14.704&    29&12.196&12.599&14.697&$1323246_1$&-22.668& 0.988&11.514& 2.06\\
155&MKW 03s                      &230.4583&  7.7088&$ 0.0442_c$&$ 0.0447_1$&$ 3.00\ $& 1.472& 0.051&14.609&    51&11.653&12.104&14.248&$1302169_1$&-21.255& 1.075&11.140& 3.34\\
156&ABELL 2065                   &230.6106& 27.7095&$ 0.0723_c$&$ 0.0723_2$&$ 8.40\ $& 2.593& 0.055&14.742&    18&12.121&12.525&14.627&$1460249_1$&-21.613& 1.017&11.083& 0.67\\
157&ABELL 2063                   &230.7724&  8.6025&$ 0.0355_c$&$ 0.0342_1$&$ 4.10\ $& 0.948& 0.046&14.503&    17&11.752&12.176&14.312&$1260227_0$&-21.662& 0.962&11.179& 0.40\\
158&ABELL 2067                   &230.7830& 30.8450&$ 0.0756_b$&$ 0.0735_1$&$ 3.10^e$& 0.444& 0.070&14.306&   194&11.748&12.154&14.289&$1327738_0$&-21.844& 0.958&11.098& 1.95\\
159&ABELL 2069                   &231.0410& 29.9210&$ 0.1145_a$&$ 0.1135_1$&$ 7.90^e$& 4.978& 0.150&14.892&   259&11.976&12.435&14.539&$1318942_1$&-22.200& 1.085&11.389& 2.02\\
160&ABELL 2073                   &231.4360& 28.4280&$ 0.1515_b$&$ 0.1502_2$&$ 5.60^e$& 2.158& 0.070&14.676&  2926&11.955&12.438&14.544&$1454463_1$&-22.633& 1.060&11.524& 2.84\\
161&ABELL 2072                   &231.4770& 18.2360&$ 0.1270_a$&$ 0.1277_1$&$ 5.20^e$& 1.821& 0.070&14.640&  1110&11.852&12.271&14.397&$2016591_1$&-22.605& 0.975&11.400& 1.46\\
162&ABELL 2107                   &234.9100& 21.7890&$ 0.0411_a$&$ 0.0411_2$&$ 4.20\ $& 0.583& 0.130&14.381&    21&11.776&12.201&14.335&$3000026_1$&-22.433& 0.934&11.408& 0.41\\
163&ABELL 2110                   &234.9530& 30.7173&$ 0.0980_c$&$ 0.0972_1$&$ 5.60^e$& 2.095& 0.107&14.683&  1190&11.440&11.906&14.058&$1388342_1$&-22.524& 1.001&11.407& 0.52\\
164&ABELL 2108                   &235.0380& 17.8780&$ 0.0916_a$&$ 0.0886_1$&$ 4.30^e$& 1.022& 0.100&14.508&   182&11.828&12.248&14.375&$2020473_1$&-22.076& 0.951&11.163& 2.80\\
165&ABELL 2124                   &236.2500& 36.0660&$ 0.0654_a$&$ 0.0660_1$&$ 3.70^e$& 0.747& 0.120&14.436&    61&11.869&12.293&14.414&$1412327_1$&-22.644& 0.979&11.443& 2.62\\
166&MaxBCG J239.42665+35.50827   &239.4382& 35.5040&$ 0.1549_c$&$ 0.1589_1$&$ 5.70^e$& 2.045& 0.165&14.660&  7475&11.671&12.149&14.284&$1405455_1$&-22.541& 1.014&11.447& 0.74\\
167&ABELL 2142                   &239.5857& 27.2269&$ 0.0894_c$&$ 0.0908_1$&$11.00\ $&11.786& 0.030&15.111&    20&12.279&12.726&14.849&$1400535_1$&-22.521& 1.041&11.450& 0.42\\
168&ABELL 2149                   &240.3990& 53.9180&$ 0.0675_b$&$ 0.0654_1$&$ 3.00^e$& 0.423& 0.050&14.296&    99&11.709&12.157&14.293&$ 501846_1$&-22.175& 1.039&11.341& 2.61\\
169&RXC J1601.3+5354             &240.3474& 53.9061&$ 0.1068_c$&$ 0.1071_1$&------& 1.306& 0.079&14.563&   811&11.602&12.006&14.161&$ 250405_1$&-22.743& 0.970&11.453& 2.38\\
170&ABELL 2147                   &240.5780& 16.0200&$ 0.0353_a$&$ 0.0353_1$&$ 4.40\ $& 1.500& 0.220&14.616&     2&12.174&12.577&14.677&$2020680_0$&-22.131& 0.919&11.149& 2.76\\
171&ABELL 2148                   &240.7590& 25.4040&$ 0.0888_b$&$ 0.0895_1$&$ 3.70^e$& 0.785& 0.050&14.443&   636&11.548&12.001&14.157&$1323725_1$&-22.473& 0.973&11.353& 5.33\\
172&NSC J160433+174311           &241.1489& 17.7244&$ 0.0370_c$&$ 0.0351_1$&$ 3.50^e$& 0.464& 0.070&14.326&     6&12.115&12.536&14.638&$1972342_0$&-21.727& 0.957&11.071& 0.17\\
173&AWM 4                        &241.2380& 23.9460&$ 0.0318_a$&$ 0.0326_2$&$ 3.70\ $& 0.243& 0.080&14.168&   148&11.226&11.720&13.861&$3000027_1$&-22.166& 0.993&11.374& 0.81\\
174&ABELL 2152                   &241.3840& 16.4420&$ 0.0370_b$&$ 0.0435_1$&$ 1.70^e$& 0.125& 0.050&14.001&    10&11.992&12.399&14.505&$1994658_0$&-21.751& 0.964&11.066& 0.83\\
175&ABELL 2151E                  &241.7180& 17.7810&$ 0.0321_a$&$ 0.0390_1$&$ 1.30^e$& 0.055& 0.040&13.799&     6&12.115&12.536&14.638&$1983734_0$&-21.456& 0.986&10.965& 3.46\\
176&ABELL 2169                   &243.5400& 49.1530&$ 0.0579_b$&$ 0.0600_1$&$ 2.40^e$& 0.262& 0.060&14.179&  1027&11.093&11.491&13.606&$ 532523_1$&-21.473& 0.977&10.958& 3.63\\
177&RXC J1615.5+1927             &243.8947& 19.4600&$ 0.0308_c$&$ 0.0316_1$&------& 0.061& 0.140&13.825&   608&10.934&11.371&13.463&$1942972_1$&-21.135& 0.990&10.854& 0.18\\
178&MaxBCG J245.12969+29.89103   &245.1322& 29.8953&$ 0.0972_c$&$ 0.0960_1$&$ 5.00^e$& 1.669& 0.084&14.627&   118&12.025&12.455&14.560&$1216685_1$&-22.170& 1.003&11.292& 0.30\\
179&ABELL 2187                   &246.0591& 41.2383&$ 0.1825_c$&$ 0.1832_1$&$ 6.50^e$& 2.494& 0.129&14.702&  6978&12.239&12.784&14.921&$1011568_1$&-23.229& 1.027&11.707& 0.33\\
180&RXC J1627.3+4240             &246.8482& 42.6784&$ 0.0317_c$&$ 0.0314_1$&------& 0.062& 0.101&13.830&   381&10.815&11.246&13.296&$ 537847_1$&-21.642& 0.968&11.018& 0.42\\
181&RXC J1627.6+4055             &246.9173& 40.9197&$ 0.0301_c$&$ 0.0317_1$&------& 0.076& 0.109&13.881&     4&12.106&12.508&14.612&$ 564226_1$&-22.155& 1.063&11.329& 0.50\\
182&ABELL 2199                   &247.1582& 39.5487&$ 0.0299_c$&$ 0.0267_1$&$ 4.70\ $& 1.570& 0.026&14.629&     5&11.958&12.357&14.468&$1009286_1$&-21.856& 0.943&11.086& 0.38\\
183&NSC J164322+213144           &250.8337& 21.5261&$ 0.1535_c$&$ 0.1536_1$&------& 0.973& 0.173&14.478&  7532&11.782&12.261&14.389&$1434860_1$&-22.871& 1.040&11.583& 0.30\\
184&NSC J165252+400906           &253.2318& 40.1535&$ 0.1492_c$&$ 0.1504_2$&$ 5.00^e$& 1.786& 0.133&14.629&  6300&11.582&12.054&14.205&$ 260130_1$&-22.443& 1.011&11.375& 0.60\\
185&RXC J1654.3+2334             &253.5972& 23.5699&$ 0.0575_c$&$ 0.0570_1$&------& 0.190& 0.135&14.101&   322&11.287&11.692&13.831&$1404809_1$&-21.480& 0.961&10.984& 0.24\\
186&ABELL 2241B                  &254.9365& 32.6135&$ 0.1013_c$&$ 0.1013_2$&$ 4.30^e$& 1.169& 0.099&14.538&   949&11.598&12.041&14.194&$1013809_1$&-22.808& 1.021&11.518& 0.23\\
187&ABELL 2245                   &255.6330& 33.5130&$ 0.0843_b$&$ 0.0864_1$&$ 3.20^e$& 0.536& 0.040&14.349&   136&11.901&12.352&14.465&$ 580244_1$&-22.824& 1.014&11.550& 0.38\\
188&ABELL 2244                   &255.6786& 34.0619&$ 0.0953_c$&$ 0.0989_1$&$ 7.10\ $& 4.470& 0.038&14.869&   192&11.857&12.336&14.450&$3000028_1$&-22.630& 0.901&11.416& 0.15\\
189&ABELL 2249                   &257.4535& 34.4406&$ 0.0802_c$&$ 0.0809_2$&$ 5.60^e$& 1.883& 0.061&14.661&   208&11.729&12.161&14.296&$ 583901_1$&-22.160& 1.012&11.288& 1.10\\
190&ABELL 2255                   &258.1968& 64.0614&$ 0.0809_c$&$ 0.0734_1$&$ 7.30\ $& 2.593& 0.042&14.741&   178&11.749&12.167&14.303&$ 236440_1$&-22.596& 0.960&11.381& 4.61\\
191&NGC 6338 GROUP               &258.8414& 57.4074&$ 0.0276_c$&$ 0.0273_1$&$ 2.40^e$& 0.250& 0.154&14.176&    72&11.396&11.825&13.976&$ 199782_1$&-22.058& 0.996&11.226& 0.35\\
192&ABELL 2257                   &259.4731& 32.5860&$ 0.1054_c$&$ 0.1086_1$&------& 1.094& 0.126&14.519&  1758&11.521&12.005&14.160&$ 519321_1$&-22.678& 1.016&11.497& 0.75\\
193&RBS 1636                     &259.5410& 56.6656&$ 0.1138_c$&$ 0.1136_1$&$ 5.00^e$& 1.772& 0.085&14.637&  6235&11.370&11.783&13.930&$ 202925_1$&-22.568& 0.870&11.309& 0.52\\
194&ABELL 2259                   &260.0370& 27.6702&$ 0.1640_c$&$ 0.1640_2$&$ 7.10^e$& 3.600& 0.086&14.798&  3963&12.076&12.588&14.690&$3000029_1$&-23.240& 1.085&11.845& 0.21\\
195&SDSS-C4 3072                 &260.0386& 26.6272&$ 0.1644_c$&$ 0.1601_1$&$10.20^e$& 6.921& 0.064&14.961&  1039&12.469&12.880&15.052&$ 566931_1$&-23.224& 0.907&11.593& 0.22\\
196&MaxBCG J321.29330-06.96355   &321.3016& -6.9655&$ 0.1153_d$&$ 0.1153_2$&------& 1.107& 0.367&14.521&  1954&11.497&12.074&14.221&$3000030_1$&-22.493& 1.134&11.609& 0.51\\
197&ABELL 2396                   &328.9198& 12.5336&$ 0.1920_c$&$ 0.1930_1$&$ 6.90^e$& 3.392& 0.168&14.775&  6426&12.300&12.822&14.984&$ 350905_1$&-23.641& 1.027&11.872& 0.62\\
198&ABELL 2399                   &329.3573& -7.7946&$ 0.0579_d$&$ 0.0580_1$&------& 0.480& 0.190&14.329&    27&11.987&12.386&14.490&$ 313506_0$&-21.801& 0.950&11.052& 0.92\\
199&ABELL 2428                   &334.0645& -9.3399&$ 0.0825_d$&$ 0.0846_1$&------& 1.674& 0.101&14.631&   209&11.793&12.270&14.397&$ 732935_1$&-22.567& 1.017&11.521& 0.40\\
200&ABELL 2561                   &348.4990& 14.7440&$ 0.1627_b$&$ 0.1625_1$&$ 5.30^e$& 1.918& 0.050&14.643&  1569&12.284&12.703&14.810&$ 351366_0$&-21.863& 1.044&11.200& 0.51\\
201&RXC J2321.8+1505             &350.4671& 15.0430&$ 0.1500_c$&$ 0.1490_1$&------& 1.243& 0.150&14.540& 15567&11.382&11.840&13.990&$ 719109_1$&-22.525& 1.002&11.398& 2.73\\
202&ABELL 2593                   &351.0840& 14.6510&$ 0.0428_a$&$ 0.0417_1$&$ 3.10\ $& 0.590& 0.120&14.384&     8&12.014&12.444&14.550&$ 717807_1$&-22.322& 0.981&11.314& 0.23\\
203&ABELL 2670                   &358.5557&-10.4129&$ 0.0765_d$&$ 0.0776_1$&------& 1.511& 0.121&14.607&    23&12.194&12.618&14.716&$ 356435_1$&-22.751& 1.007&11.533& 0.38\\

\end{tabular}
\end{center}
\end{sidetable}
}